\documentclass[12pt]{article} 

\usepackage{a4}
\usepackage{times}
\usepackage[dvips]{epsfig}
\usepackage{amsmath}
\usepackage[final]{pdfpages}
\usepackage{calrsfs}
\usepackage{relsize}
\usepackage [english]{babel}
\usepackage [autostyle, english = american]{csquotes}
\usepackage{empheq}

\usepackage{url}
\MakeOuterQuote{"}

\usepackage{mathtools}
\usepackage{hyperref}
\usepackage{cleveref}
\usepackage{autonum}

\newcommand*\widefbox[1]{\fbox{\hspace{0.1em}#1\hspace{0.1em}}}

\newcommand{\bigsymbol}[1]{\mathlarger{\mathlarger{#1}}}

\newcommand{\bbigsymbol}[1]{\mathlarger{\mathlarger{\mathlarger{#1}}}}
\newcommand\balpha{\bar\alpha}
\newcommand\bbeta{\bar\beta}
\newcommand\bmu{\bar\mu}
\newcommand\An{A_n}
\newcommand\Bn{B_n}
\newcommand\Cn{C_n}
\DeclareMathOperator*{\ssum}{\textstyle\sum}

\author{Wiel Kleeven\\
Ion Beam Applications (IBA), Louvain-La-Neuve, Belgium \\
MID-111760 \\}

\title{Energy limit of compact isochronous cyclotrons}

\begin{document}

\maketitle

\newpage

\tableofcontents

\newpage





\begin{abstract}
Existing analytical models for transverse beam dynamics in isochronous cyclotrons are often not valid or not precise for relativistic energies. The main difficulty in developing such models lies in the fact that cross-terms between derivatives of the average magnetic field and the azimuthally varying components cannot be neglected at higher energies. Taking such cross-terms rigorously into account results in an even larger number of terms that need to be included in the equations.  In this paper, a method is developed which is relativistically correct and which provides results that are practical and easy to use. We derive new formulas, graphs and tables for the radial and vertical tunes in terms of the the flutter, its radial derivatives, the spiral angle and the relativistic gamma. Using this method, we study the $2\nu_r=N$ structural resonance ($N$ is number of sectors) and provide formulas and graphs for its stopband and for the modified radial tune. Combining those equations with the new equation for the verical tune, we find the stability zone and the energy limit of compact isochronous cyclotrons for any value of N. We confront the new analytical method with closed orbit simulations of the IBA C400 cyclotron for hadron therapy. 
\end{abstract}

\section{Introduction}

In this paper we derive the maximum energy that can be realized in compact isohronous cyclotrons.
This limit is determined by two competing requirements namely the need for sufficient vertical focusing on the one hand  and  the need to avoid
the stopband of the half-integer resonance $2\nu_r=N$ on the other hand ($N$ is the cyclotron rotational symmetry number; $\nu_r$ is the radial tune).
With increasing energy the isochronous field index $\bmu'$ increases rapidly and more and more azimuthal field variation $f$ is needed 
to remain vertically stable; but with higher $f$,  the stopband of the resonance broadens and the energy limit associated with it rapidly reduces. 
The energy limit depends on $N$ and on the spiral angle $\xi$ of the sectors. We derive practical formulas which are useful especially in the design phase of a new cyclotron.
Our main assumption/approximation is that $f$ is not too large. Results are derived up to $\mathcal{O}(f^2)$ (equivalent to $\mathcal{O}(F)$, where $F$ is the flutter).
For compact cyclotrons $F$ is generally well below $1$ and for these machines we expect our results to be precise. For seperate sector cyclotrons, care should be taken however.
The special case of such cyclotrons with radial sectors (no spiraling) has been studied by Gordon\cite{Gordon}, by assuming a hard-edge model where in the magnet sections 
the orbits are perfectly circular and in the empty straight sections the magnetic field is zero. In Gordon's model, there is no need to assume a small flutter, 
but on the other hand his assumptions will probably not be valid for compact cyclotrons and maybe also less accurate for coil-dominated superconducting ring cyclotrons 
where the magnetic field has the tendency to spread out more smoothly and non-uniformly. For seperate sector cyclotrons with a larger magnetic filling factor 
the flutter drops quickly ($F\approx0.25$ for a filling factor of 80\%) and we expect our results to become more accurate. Another interesting derivation of the isochronous cyclotron energy limit has been made by Danilov \textit{et\ al.} from the JINR\cite{Danilov}. In their analysis however, they take into account only the first dominant Fourier component of the field 
and they further assume that its amplitude is independent on radius and its phase increases linearly with radius. 
Also contributions due to higher order radial derivatives of the average magnetic field are ignored. 
We closely follow the Hamiltonian approach that has been firstly introduced by Hagedoorn and Verster\cite{HV-62}; in this paper we wish to pay tribute to them.

\section{Method of derivation}

We study the static (non-accelerated) motion near a
given radius $r_0$ which is related to the constant kinetic momentum $P_0$ of a particle. The reduced magnetic field $\mu(r,\theta)$
around this radius is represented by a Fourier series with respect to the azimuth
$\theta$ and the radial dependence of the average field $\bar{\mu}(r)$ and the normalized Fourier coefficients $A_n(r), B_n(r)$ are Taylor expanded relative to the same radius $r_0$. 
The magnitude of azimuthal field variation $f$ is approximately equal to the magnitude of the dominant Fourier component $C_N=(A_N^2+B_N^2)^{1/2}$ and the flutter
$F$ is approximately equal to $C_N^2/2$. We develop the general Hamiltonian $H_0$ in polar coordinates relative to the circle $r_0$ and first look for the closed orbit (CO)
which is the $N$-fold rotational symmetric solution of $H_0$. In all our derivations we use a pertubation analysis where $|f|$ serves as the measure for precission.  
In general any quantity of interest $g(\theta)$  can be split in its average part 
$\bar{g}=\tfrac{1}{2\pi}\oint g(\theta)d\theta$ and its oscillating part $osc(g)=g(\theta)-\bar{g}$. Oscillating parts of $\mathcal{O}(f)$ can be moved to the next higher order 
by a properly constructed canonical transformation. In doing so, new average contributions of $\mathcal{O}(f^2)$ are generated. 
Our goal is to derive results up to $\mathcal{O}(f)$. The reason for this is that the first significant terms in the expressions for the isochronous magnetic field 
and the radial and vertical tunes are of $\mathcal{O}(f^2)$. In line with the HV-paper\cite{HV-62}, we keep the
average part of any azimuthally varying term up to $\mathcal{O}(f^2)$, but neglect oscillating terms $\mathcal{O}(f^2)$ as they would generate new terms of $\mathcal{O}(f^3)$ when transforming them to higher order. 
However, we make one important generalization/improvement as compared to the HV-paper. In their analysis Hagedoorn and Verster  assumed that radial derivatives 
of the average magnetic field ($\bmu',\bmu'',\bmu''',\dots$) are small quantities of $\mathcal{O}(f^2)$ and therefore neglect cross-terms between those derivatives and
the Fourier content of the magnetic field in all expansions. This is a valid approach at lower energies where the radial isochronous field derivatives
are still small, but at higher energies this approximation becomes less and less accurate and ultimately breaks down completely. 
Since we are interested in the higher-energy limits of the isochronous cyclotron we cannot make this concession and therefore keep those cross-terms. This makes
the derivation and also the final results considerably more complex as many more terms need to be kept in the Hamiltonian expansion. 
The $\mathcal{O}(f^2)$ contributions to the final results all have a similar structure of the following general form:

\begin{align}
R^{(2)} = \sum_n \alpha_n(\bmu',\dots)C_n^2+\beta_n(\bmu',\dots)C_n^2\varphi_n'^2+\gamma_n(\bmu',\dots)C_nC_n'+\delta_n(\bmu',\dots)C_n'^2\ .
\end{align}

\noindent Here the summation runs over all the Fourier components ($n=kN,\hspace{0.1cm}k=1,2,\dots$) present in the magnetic field; the coefficients $\alpha_n,\dots$ 
depend on the first and higher radial derivatives $\bmu',\bmu'',\bmu''',\dots$ of the average magnetic field and the variable $\varphi'_n$ is the radial derivative of the phase $\varphi_n$
of the Fourier harmonic $n$. To 
obtain practical results we make a few assumptions and approximations that allow us to simplify this structure. Firstly it is assumed that the magnetic field is perfectly isochronous.
In this case the form-factor of the average field is completely determined
by the relativistic gamma parameter and therefore the coefficients $\alpha_n,\dots$ will depend on $\gamma$ only. Secondly we assume that the phase-derivatives $\varphi'_n$ 
do not depend on $n$. In practice this is accurately true for the first several (often up to 5) Fourier components. Since contributions of higher components rapidly drop with increasing
$n$-value, this approximation must be accurate. In this way the variable $\varphi'_n=\varphi'$ can be taken out of the series summations. Thirdly we introduce
a method where the higher Fourier harmonics ($n>N$) are expressed in terms of the dominant harmonic ($n=N$). For this we assume a hard-edge profile of the azimuthally
varying field with a symmetrical structure of equal hill and valley angle. For such a profile only the odd harmonics ($k=1,3,\dots$) are non-zero and the magnitude of the harmonics
drop with $1/n$. In this way, the n-dependence of the harmonic amplitudes $C_n$ can be included in the coefficients $\alpha_n,\dots$ and the dominant components $C_N$
can be taken outside of the series summation. The assumption of a hard-edge profile represents a certain limitation but it allows us to approximately take into account
the higher harmonic content and therefore is expected to be better than just taking into account the dominant harmonic; at the same time it allows to express the dominant Fourier coefficients $C_N$ in terms of the flutter $F$. In a final step we  sum the series analytically and express the results in elementary functions of $\gamma$ and $N$. 
The $\mathcal{O}(f^2)$ contributions to the final results are thus transformed to the following simpler form:

\begin{align}
R^{(2)} \approx F\left(a_N(\gamma)+b_N(\gamma)\varphi'^2+c_N(\gamma)\frac{F'}{F}+d_N(\gamma)(\frac{F'}{F})^2\right)\ .
\end{align}

\noindent The above method is applied in the derivation of the isochronous magnetic field, the radial and vertical tunes and the stopband of the half-integer resonance.

\section{The radial motion}\label{hmot}

The Hamiltonian for the radial motion with respect the reference circle $r_0$ (see Eq.~(\ref{Pref})) has been given in Eq.~(\ref{hamx}). 
In this paragraph we derive the expressions for the equilibrium orbit (EO) and the isochronous magnetic field. We also determine the relation between the particle
relativistic parameter $\gamma$ and the field index $\bmu'$ and express the higher derivatives $\bmu'',\bmu'''$ in terms of $\bmu'$. We derive the Hamiltonian
with respect to the EO and bring it into its normal form. Then we solve the linear motion giving us expressions for the radial tune $\nu_r$ and the stopband of the $2\nu_r=N$
resonance. In the analysis we keep oscillating terms of $\mathcal{O}(f^1)$ but neglect those of higher order. Constant ($\theta$-independent) terms are are kept up to  $\mathcal{O}(f^2)$.
Radial derivatives of the averge field such as $\bmu',\bmu'',\bmu'''$ are considered as terms of $\mathcal{O}(f^0)$ and are always kept. In the final results, summations 
over magnetic field Fourier coefficients and their radial derivatives are eliminated and replaced by expressions with flutter and spiral angle. The derivatives of the average field are
eliminated as they are considerd as functions of the relativistic parameter $\gamma$.  

\subsection{The equilibrium orbit}\label{TEO}

The EO is a closed orbit in the median plane with the same $N$-fold symmetry as the magnetic field. it can therefore can be expanded in a Fourier series:

\begin{equation}
x_e(\theta) = \gamma_e+\ssum_n \alpha_n\cos n\theta +\beta_n\sin n\theta\ . \label{eqfou}
\end{equation}

\noindent We need to find the expressions for $\alpha_n$ and $\beta_n$ up to $\mathcal{O}(f)$ and the expression for $\gamma_e$ up to $\mathcal{O}(f)$. The radial equations
of motion are obtained from Eq.~(\ref{hamx}) as:

\begin{align}
	&\frac{dx}{d\theta} \hspace{0.1cm} = \hspace{0.35cm}\frac{\partial H_x}{\partial p_x} = (1+x) p_x (1-p_x^2)^{-1/2}\ , \nonumber \\
	&\frac{dp_x}{d\theta} = -\frac{\partial H_x}{\partial x} = (1-p_x^2)^{1/2}-(1+x)\mu(\theta,x)\ . \nonumber
\end{align}

\noindent Knowing that both $x$ and $p_x$ are functions of $\mathcal{O}(f)$ we can expand the right hand sides of above equations up to $\mathcal{O}(f^2)$. From the first equation
we will get:

\begin{equation}
	\frac{dx}{d\theta} =(1+x)p_x =p_x + \mathcal{O}(f^2)\ , \nonumber
\end{equation}

\noindent because here we can neglect a term $xp_x$ as $\langle xp_x \rangle=0$. Inserting $\dot{p}_x=\ddot{x}$ and the expression for the reduced field $\mu$ from
Eq.~(\ref{Taylor}) in the second equation, we get:

\begin{equation}
\begin{aligned}
\ddot{x} = &-\tfrac{1}{2}\dot{x}^2 -(1+\bmu')x-(\bmu'+\tfrac{1}{2}\bmu'')x^2 \nonumber \\
 &-\ssum_n\left[A_n+(A_n+A'_n)x\right]\cos n\theta+ \left[B_n+(B_n+B'_n)x\right]\sin n\theta)\ .  \nonumber
\end{aligned}
\end{equation}

\noindent Note that the "dot"-operator stands for differentiation with respect to $\theta$ ($\dot{x}=\tfrac{dx}{d\theta}$). 
In the expression above we insert  the Fourier expansion of $x=x_e$ from Eq.~(\ref{eqfou}). The first order parts of the equation give us the expressions for $\alpha_n,\beta_n$.
For the second order parts we only have to keep the average values. This gives us the expression for $\gamma_e$. We find for the Fourier coefficients of the EO:

\renewcommand*\widefbox[1]{\fbox{\hspace{0.5em}#1\hspace{0.5em}}}

\begin{empheq}[box=\widefbox]{align}
	&\alpha_n =\frac{A_n}{n^2-1-\bmu'}\ ,\label{alphae} \\
	&\beta_n =\frac{B_n}{n^2-1-\bmu'}\ , \label{betae}\\
	&\gamma_e = -\frac{1}{2(1+\bmu')}\sum_n\left[\frac{3n^2-2+\bmu''}{2(n^2-1-\bmu')^2}C_n^2 +\frac{C_nC'_n}{n^2-1-\bmu'} \right]\ .  \label{gammae}
\end{empheq}

\noindent In the expression for $\gamma_e$  we used Eqs.~(\ref{AnAn},\ref{AnAnp}) to eliminate the sin/cosine coefficients $A_n,B_n$ in favour of the Fourier amplitude $C_n$.

\noindent The radial momentum of the EO is given by $p_e=\dot{x}_e$.

\subsection{Correction of the spiral angle}\label{spiralcorrection}

In paragraph~\ref{tspiral} we have defined the spiral angle as the angle between the tangent along the sector contour and the normal to the orbit. However, the EO is not exactly
a circle as there is a small angle between the normal vector of the circle and the normal vector of the orbit. This angle is equal to the arc tangent of the radial momentum of the EO. We
can therefore define a corrected spiral angle $\bar{\xi}$ as follows:

\begin{equation}
\bar{\xi} =\xi+\arctan(p_e) =\arctan(\varphi')+\arctan(p_e)\ . \nonumber
\end{equation}

\noindent Here $\xi,\varphi'$ are the uncorrected parameters. With $\bar{\varphi}'=\tan(\bar{\xi})$ we get:

\begin{equation}
\bar{\varphi}' = \frac{\varphi'+p_e}{1-\varphi'p_e}\ . \nonumber
\end{equation}

\noindent From paragraph~(\ref{TEO}) we have for $p_e$:

\begin{equation}\label{petmp}
p_e(\theta) = \dot{x_e}=-\ssum_n\frac{nC_n}{n^2-1-\bmu'}\sin n(\theta-\varphi_n)\approx -\ssum_n\frac{C_n}{n}\sin n(\theta-\varphi_n)\ . 
\end{equation}

\noindent We evaluate $p_e$ at the entrance (and exit) of the sector and assume (as we did in paragraph~\ref{RFf}) a symmetric structure where the hill angle is equal to the 
valley angle. In this case we get $\theta-\varphi_n\approx\theta-\varphi =  \pm \pi/2N$ and then get (with $n=(2k+1)N$):

\begin{equation}
\sin n(\theta-\varphi_n)\approx \pm(-1)^{k}\ , \nonumber
\end{equation}

\noindent Inserting this expression in Eq.~(\ref{petmp}), together with the expressions for the Fourier components Eq.~(\ref{fcoefs}) and the relation for the flutter  Eq.~(\ref{FCN}), 
we find the following approximation for the radial momentum at the sector edges:

\begin{equation}
p_e =\pm \frac{\pi}{2N}\sqrt{F}\ . \nonumber
\end{equation}

\noindent It is seen that the correction at the pole edges have opposite sign and we can take the average of the two as good approximation 
for the corrected paramater $\bar{\varphi'}$:

\begin{equation}\label{pcorr}
	\bar{\varphi}' =\frac{1}{2}\left[ \frac{\varphi'_{in}+\frac{\pi}{2N}\sqrt{F}}{1-\frac{\pi}{2N}\varphi'_{in}\sqrt{F}}+\frac{\varphi'_{out}-\frac{\pi}{2N}\sqrt{F}}{1+\frac{\pi}{2N}\varphi'_{out}\sqrt{F}}\right] \ .
\end{equation}

\noindent We use Eq~(\ref{pcorr}) in paragraph~(vermot), when we compare the analytical expression of the vertical tune $\nu_z$ with results from closed orbit simulations for the IBA 
C400 cyclotron.

\noindent For cyclotron design studies one normaly will start with equal pole-edge contours at the sector entrance and exit. 
In this case one can take $\varphi'_{in}=\varphi'_{out}=\varphi'$ and the expression for the corrected spiral simplifies to:

\begin{equation}\label{spiralcorrection2}
\boxed{\bar{\varphi}' = \varphi'_{geom}\left(1+\frac{\pi^2F}{4N^2}(1+ \varphi'^{2}_{geom})\right)+\mathcal{O}(f^4)\ .}
\end{equation}

\noindent Here $\varphi_{geom}$ represents the geometrical pole-edge contour.

\subsection{The isochronous magnetic field}

The shape of the isochronous magnetic field $B_{iso}(r)$ has been given in Eq.~(5.5) of the HV-paper\cite{HV-62} as:

\begin{equation}
	B_{iso}(r)=B_0\frac{R_e}{r_0}\left[1-\left(\frac{R_e}{\lambda}\right)^2\right]^{-1/2}\ ,
\end{equation}

\noindent where $B_0=m_0\omega/q$ is the center magnetic field and $\omega$ is the (constant) angular revolution frequency of a particle (with restmass $m_0$ and charge $q$)
and $\lambda=c/\omega$, with $c$ the speed of light.
The radius $R_e$ is the effective radius of the EO and is defined as its length divided by $2\pi$:

\begin{equation}
	R_e=\frac{1}{2\pi}\oint_{EO} ds = r_0\langle (1+x_e)(1-p_e^2)^{-1/2}\rangle\ .
\end{equation}

\noindent We write:

\begin{equation}
	R_e=r_0(1+\epsilon_e)\ .
\end{equation}

\noindent Up to $\mathcal{O}(f^2)$ we find for $\epsilon_e$:

\begin{equation}
\begin{aligned}
	\epsilon_e&=\langle x_e+\tfrac{1}{2}p_e^2\rangle =\gamma_e+\tfrac{1}{4}\sum_n \frac{n^2C_n^2}{(n^2-1-\bmu')^2}\ , \\ 
	&= -\frac{1}{2(1+\bmu')}\sum_n\left[\frac{2(n^2-1)-n^2\bmu'+\bmu''}{2(n^2-1-\bmu')^2}C_n^2 +\frac{C_nC'_n}{n^2-1-\bmu'} \right]\ ,\label{epse}
\end{aligned}
\end{equation}

\noindent and for $B_{iso}(r)$:

\begin{equation}
	B_{iso}(r)= \frac{B_0}{\sqrt{1-r^2/\lambda^2}}\left(1+ \frac{\epsilon_e}{1-r^2/\lambda^2}\right)\ . \label{Biso}
\end{equation}

\noindent We also calculate the field-index $\bmu'_{iso}$ of the iscochronous field and find:

\begin{equation}
	\bmu'_{iso} = \frac {r}{B_{iso}}\frac{dB_{iso}}{dr} 
			=\frac{r^2/\lambda^2}{1-r^2/\lambda^2}\left[ 1+\frac{2\epsilon_e}{1-r^2/\lambda^2}+\frac{\lambda^2}{r^2}\epsilon'_e\right]\ ,\label{iso2}
\end{equation}

\noindent where $\epsilon'_e=rd\epsilon_e/dr$. 

\noindent We now look for a relation between the field index and the relativistic parameter $\gamma$. For this we use the definition of our 
reference momentum $P_0$ from eq.~(\ref{Pref}) which now is applied for the isochronous field Eq.~(\ref{Biso}):

\begin{equation}
\frac{P_0}{m_0 c} =\beta\gamma=\frac{qrB_{iso}(r)}{m_0c}=\frac{r/\lambda}{\sqrt{1-r^2/\lambda^2}}\left(1+\frac{\epsilon}{1-r^2/\lambda^2}\right)\ .
\end{equation}

\noindent from which we get:

\begin{equation}
	\gamma^2-1=\frac{r^2/\lambda^2}{1-r^2/\lambda^2}\left[ 1+\frac{2\epsilon_e}{1-r^2/\lambda^2}\right]\ . \label{gamma2}
\end{equation}

\noindent Comparing the right hand side of this equation with the right hand side of Eq.~(\ref{iso2}), we can write $\bmu'_{iso}$ as follows:

\begin{equation}
	\bmu'_{iso}=\gamma^2-1 +\gamma^2 \epsilon'_e. \label{miso}\ .
\end{equation}

\noindent So, since we assume that the magnetic field is iscochronous, we can split the field index $\bmu'=\bmu'_{iso}$ in a relativistic part and a flutter part as follows:

\begin{align}
	&\bmu' = \bmu'_{rel}+\bmu'_{fl}\ , \\
	&\bmu'_{rel} = \gamma^2-1 \ , \label{bmufl1}\\
	&\bmu'_{fl} = (1+\bmu')\epsilon'_e \ . 
\end{align}

\noindent We further note that, in expressions which are already of $\mathcal{O}(f^2)$ (such as the expression for $\gamma_e$ in Eq.~(\ref{gammae}),
the expression for $\epsilon_e$ in Eq.~(\ref{epse}) and also in the expression on the right hand side of Eq.~(\ref{bmufl1})), 
we can ignore the difference between $\bmu'$ and $\bmu'_{rel}$ since 
the difference will generate terms of $\mathcal{O}(f^4)$. For the same reason we can, for such expressions, calculate the higher derivatives of the average field $\bmu'',\bmu''',\dots$ 
by differentiation of the function $b(r)=b_0/\sqrt{1-r^2/\lambda^2}$. In this way the higher derivatives can expressed in $\bmu'$. We find in this way:

\begin{empheq}[box=\widefbox]{align}
	&\bmu'' = \bmu'(1+3\bmu') + \mathcal{O}(f^2)\ , \label{bmupp}\\
 	&\bmu''' =  3\bmu'^2(3+5\bmu') + \mathcal{O}(f^2)\ . \label{bmuppp}
\end{empheq}

\begin{figure}[!bht]
   \vspace*{-.5\baselineskip}
   \centering
   \includegraphics[width=9cm]{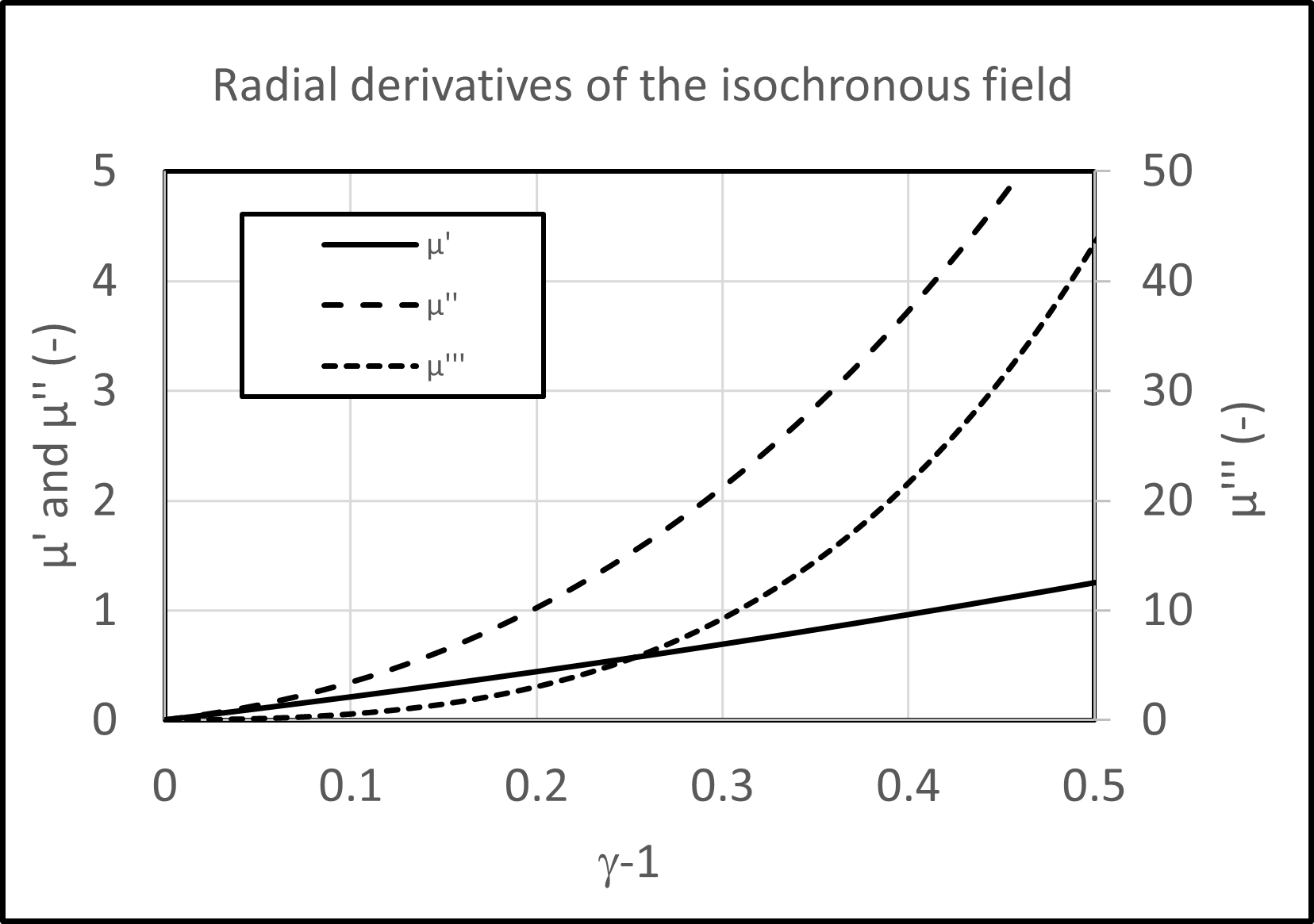}
    \caption[Derivatives of the isochronous magnetic field]{\it Normalized first and second derivatives (left scale) and third derivative (right scale) of an isochronous magnetic field.}
    \label{Bisoderivs}
\end{figure}

\noindent Figure~\ref{Bisoderivs} shows the (relativistic part) of the field-derivitatives $\bmu',\bmu''$ and $\bmu'''$. It is seen that especially the second and third derivatives become 
large for relativistic energies.

\noindent In paragraphs~\ref{radmot} and \ref{vermot}, where we derive the radial and vertical tunes, we need to make the split $\bmu' = \bmu'_{rel}+\bmu'_{fl}$, 
as we want to combine the term $\bmu'_{fl}$
with other contributions arising from the azimuthally varying part of the magnetic field. For that we need to calculate the expression for $\bmu'_{fl}$. We find this by differentiation and 
by carefully taking into account all radius-dependent terms ($\bmu', \bmu'', C_n,C_n'$) in Eqs.~(\ref{bmufl1}) and~(\ref{epse}). For the radial derivatives of $\bmu', \bmu'', C_n'$ we have:

\begin{align}
&r\frac{d}{dr}C'_n = C'_n+C''_n\ , \\
&r\frac{d}{dr}\bmu' = \bmu'-\bmu'^{2}+\bmu'' = 2\bmu'(1+\bmu')\ , \\
&r\frac{d}{dr}\bmu'' = r\frac{d}{dr}[\bmu'(1+3\bmu')]=2\bmu'(1+\bmu')(1+6\bmu')\ .
\end{align}

\noindent Using these expressions we find for $\bmu'_{fl}$:

\renewcommand*\widefbox[1]{\fbox{\hspace{0.1em}#1\hspace{0.1em}}}

\begin{empheq}[box=\widefbox]{align}
\bmu'_{fl} &= \ssum_n\frac{-1}{2(n^2-1-\bmu')}\bbigsymbol{\{}\frac{-\bmu'\left[(n^2-1)\left(3n^2-7-\bmu'(11+\bmu')\right)-3\bmu'^3\right] }{(n^2-1-\bmu')^2}C_n^2\nonumber\\
&+ \frac{\left[3(n^2-1)-3\bmu'n^2+4\bmu'+7\bmu'^2\right]}{n^2-1-\bmu'}C_nC'_n+C_nC''_n+C'^{2}_n \bbigsymbol{\}} \label{bmufl2}
\end{empheq}

We now further elaborate on the expression for $\epsilon_e$ given in Eq.~(\ref{epse}) as this term is needed to calculate the precise expression for the isochronous 
field $B_{iso}$ (given in Eq.~(\ref{Biso})) and for obtaining a precise relation between radius $r$ and $\gamma$ as determined by the relation Eq.~(\ref{gamma2}). 
We insert $\bmu''=\bmu'(1+3\bmu')$ (see Eq.~(\ref{bmupp}))  in the expression for $\epsilon_e$ and then simplify this expression  by the method explained 
in paragraph~\ref{RFf} where we substitute for the Fourier coefficient $C_n$ their expressions in terms of the flutter $F$ as defined in Eqs.~(\ref{CtoF}).

The result for $\epsilon_e$ can now be written as follows:

\begin{align}\label{epse2}
\boxed{\epsilon_e = -\frac{F}{2\gamma^4}\left(\check{a}_N+\check{c}_N\frac{F'}{F}\right)\ .}
\end{align}

\noindent Here $F$ is the flutter and $F'$ its radial derivative. The functions $\check{a}_N,\check{c}_N$ depend only on the symmetry number  $N$ and on the 
relativistic parameter $\gamma$ via the relation $\bmu'=\bmu'_{rel}=\gamma^2-1$. The expressions for these parameters are obtained as:

\begin{align}
\check{a}_N &=\frac{8\gamma^2N^2}{\pi^2}\sum_{k=0}\frac{2-\bmu'}{(n^2-1-\bmu')^2}+\frac{-2+\bmu'+3\bmu'^2}{n^2(n^2-1-\bmu')^2}\ , \\
\check{c}_N &=\frac{8\gamma^2N^2}{\pi^2}\sum_{k=0}\frac{1}{n^2(n^2-1-\bmu')^2}\ .
\end{align}

\noindent In these equations we have to replace $n$ by $n=(2k+1)N$.

\noindent The summations in the above equations can be done analytically and the coefficients $\check{a}_N,\check{c}_N$ 
can be expressed in elementary mathematical functions. Appendix~\ref{summation} shows how this is done. We find the following result:

\renewcommand*\widefbox[1]{\fbox{\hspace{0.5em}#1\hspace{0.5em}}}
\begin{empheq}[box=\widefbox]{align}
\check{a}_N &=(4\gamma^2-6)\left[1-\frac{2N}{\pi\gamma}\tan(\frac{\pi\gamma}{2N})\right]+(\gamma^2-1)\tan^2(\frac{\pi\gamma}{2N})\ , \\
\check{c}_N &=-1+\frac{2N}{\pi\gamma}\tan(\frac{\pi\gamma}{2N})\ .
\end{empheq}

\noindent One can now obtain the isochronous field as function of radius from Eq.~(\ref{Biso}) and the relation between $\gamma$ and radius $r$ from Eq.~(\ref{gamma2}), where 
$\epsilon_e$ is calculated from Eq.~(\ref{epse2}).  We re-arrange the equations as follows:

\renewcommand*\widefbox[1]{\fbox{\hspace{0.5em}#1\hspace{0.5em}}}

\begin{empheq}[box=\widefbox]{align}
&\gamma_0\hspace{0.8cm}=1/\sqrt{1-r^2/\lambda^2}\ , \\
&\gamma\hspace{1cm}=\gamma_0\bigsymbol{[}1-\frac{\gamma_0^2-1}{2\gamma_0^4}\bigsymbol{(}\check{a}_N F+\check{c}_N F'\bigsymbol{)}\bigsymbol{]}+\mathcal{O}(f^4)\ , \\
&B_{iso}(r)=B_0\gamma_0\bigsymbol{[} 1-\frac{1}{2\gamma_0^2}\bigsymbol{(}\check{a}_N F+\check{c}_N F'\bigsymbol{)}\bigsymbol{]}+\mathcal{O}(f^4)\ .
\end{empheq}

\noindent Here $\gamma_0$ is the $\mathcal{O}(f^0)$ solution for $\gamma$ and the coefficients $\check{a}_N,\check{c}_N$  must be evaluated at $\gamma=\gamma_0$.

\begin{figure}[!bht]
   \vspace*{-.5\baselineskip}
   \centering
   \includegraphics[width=\textwidth]{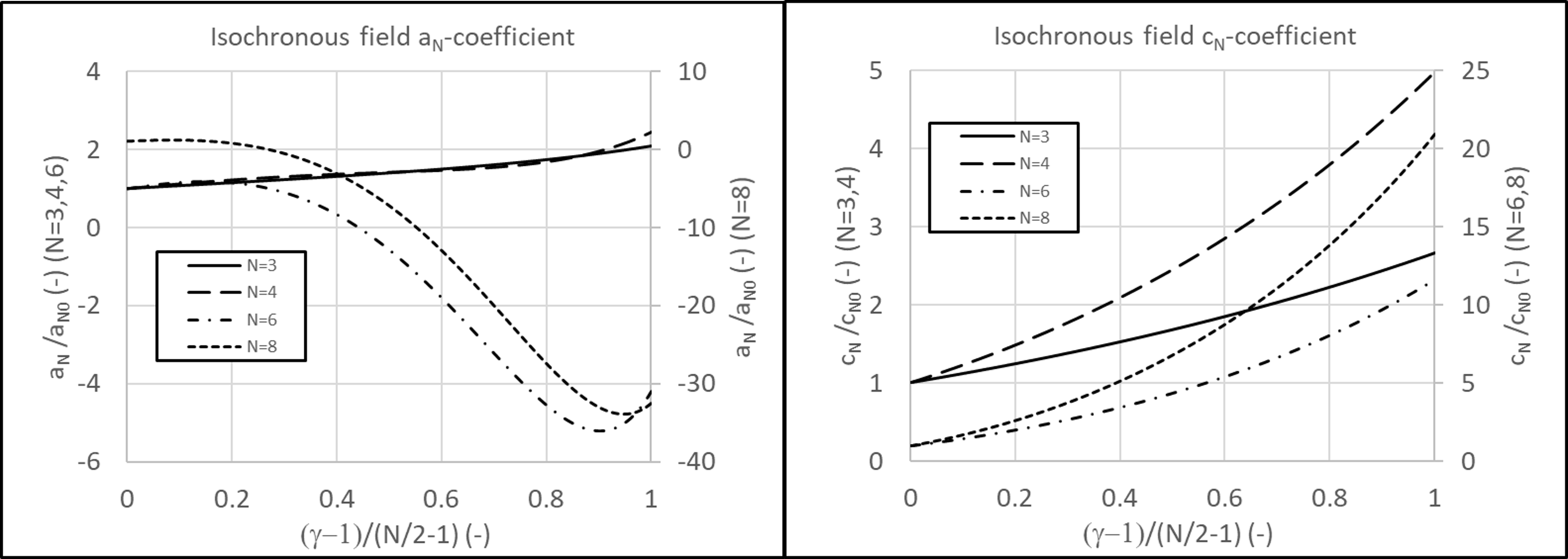}
    \caption[Energy dependence of the isochronous field coefficients]{\it Energy dependence of the isochronous field coefficients.}
    \label{Bisocoefs}
\end{figure}

\noindent Figure~\ref{Bisocoefs} shows the energy dependence of the isochronous field coefficients $\check{a}_N,\check{c}_N$ . The horizontal axis represents the relativistic 
kinetic energy $\gamma-1$, normalized by the factor $\tfrac{N}{2}-1$, i.e. the energy at which the half-integer resonance is hit (in case the stopband width equals zero, i.e. 
when there would be no  azimuthal field variation). This representation will be used several times in this report. The highest scale value of 1.0 as used in Figure~\ref{Bisocoefs}
therefore represents 100\% of the  "N/2 band-width". The vertical scale is normalized with respect to the coefficients-value  at zero kinetic energy ($\gamma=1$). 
It is seen from Figure~\ref{Bisocoefs} that the coefficient $\check{a}_N$ becomes large and negative, for high N-numbers. For such cases the required increase
of the isochronous field may be under-estimated at large energies.

\subsection{The linear radial motion}\label{radmot}

We study the linear radial motion around the EO and for this purpose introduce new canonical variables $(\pi,\xi)$ which eliminate the EO from the motion.
The method has been described in paragraph~\ref{Oneigh} and the transformation is:

\begin{equation}
\begin{aligned}
\pi &= p_x-p_e\ , \\
\xi &= x-x_e\ ,
\end{aligned}
\end{equation}

\noindent The new Hamiltonian $K_x$ is obtained as a Taylor expansion around $x_e,p_e$ (and with respect to $\pi$ and $\xi$) of the Hamiltonian $H_x$ given in Eq.~(\ref{hamx}). 
For the linear motion we only have to keep terms up to second degree in $\pi$ and $\xi$. In their coefficients we have to keep constants up to 
$\mathcal{O}(f^2)$ and oscillating terms up to $\mathcal{O}(f)$. We obtain:

\begin{align}
K_x(\xi,\pi,\theta)=\tfrac{1}{2}\left(1+x_e+\tfrac{3}{2}\dot{x}_e^2\right)\pi^2 + \dot{x}_e\pi\xi +\tfrac{1}{2}\left(\mu+(1+x_e)\frac{\partial\mu}{\partial x}\right)\xi^2\ .
\end{align}

\noindent This expression agrees with Eq.~(6.3) in the HV-paper. Note that here, we have to evaluate $\mu$ and $\frac{\partial\mu}{\partial x}$ on the EO (so at $x=x_e$).
\noindent We bring this Hamiltonian to its normal form by a canonical transformation $\pi,\xi \Rightarrow P_x,X$  using the method explained in paragraph~\ref{normform}. 
This gives us for the new Hamiltonian $\bar{K}_x(X,P_x)$:

\begin{align}
\bar{K}_x(X,P_x) = \tfrac{1}{2}P_x^2+\tfrac{1}{2}Q_x(\theta)X^2\ ,
\end{align}

\noindent where $Q_x(\theta)$ is given by:

\begin{equation}\label{Qradial1}
Q_x(\theta)=\mu+\frac{\partial\mu}{\partial x}-\tfrac{1}{2}\ddot{x}_e +x_e(\mu+2\frac{\partial\mu}{\partial x})+\frac{\partial\mu}{\partial x}x_e^2+(-\tfrac{1}{4}+\tfrac{3}{2}\mu+\tfrac{3}{2}\frac{\partial\mu}{\partial x})\dot{x}_e^2\ .
\end{equation}

\noindent We calculate the partial derivative $\partial\mu/\partial x$ from the expression for the reduced field $\mu$ given in Eq.~(\ref{Taylor}) and insert it together with the expression for $\mu$ in Eq.~(\ref{Qradial1}). 
With this we obtain:

\begin{align}
Q_x(\theta)&=1+\bmu' \nonumber \\
&+(1+3\bmu'+\bmu'')x_e+\tfrac{1}{2}(4\bmu'+5\bmu''+\bmu''')x_e^2-\tfrac{1}{2}\ddot{x_e}+\tfrac{1}{4}(5+6\bmu')\dot{x}_e^2 \nonumber \\
&+\ssum_n\left[A_n+A'_n+(A_n+3A'_n+A''_n)x_e\right]\cos n\theta \label{Qradial2} \\
&+\ssum_n\left[B_n+B'_n+(B_n+3B'_n+B''_n)x_e\right]\sin n\theta\ . \nonumber
\end{align}

\noindent We now work out this expresion in full detail. This is done with the following additional steps:   i) use the expressions for $x_e$  and $\dot{x}_e$ as defined by Eqs.~(\ref{eqfou},\ref{alphae},\ref{betae},\ref{gammae}) and  insert those in Eq.~(\ref{Qradial2}), ii) in the obtained result, split the $\mathcal{O}(f^0)$ term $1+\bmu'$ in its relativistic part and its flutter part as $1+\bmu'=1+\bmu'_{rel}+\bmu'_{fl}$ and insert for $\bmu'_{fl}$ the expression given in Eq.~(\ref{bmufl2}), iii) replace Fourier sine/cosine coefficients and their radial derivatives $A_n,A'_n,A''_n$, $B_n,B'_n,B''_n$ by Fourier amplitudes and their derivatives $C_n,C'_n,C''_n$ and phase derivative $\varphi'_n$ using Eqs.~(\ref{AnAn}-\ref{Anp2},\ref{AnAnpp}), iv) of all the $\theta$-dependent terms of $\mathcal{O}(f^2)$ keep only their average and v) substitute for the higher derivatives $\bmu'', \bmu'''$ epressions envolving the field-index $\bmu'$, using  Eqs.~(\ref{bmupp},\ref{bmuppp}).

\noindent We write the Hamiltonian in the same form as given in Eq.~(\ref{Hnorm}):

\begin{equation}\label{HPX}
\bar{K}_x(X,P_x,\theta) = \tfrac{1}{2}P_x^2 + \tfrac{1}{2}(\nu_{x0}^2+f(\theta))X^2\ .
\end{equation}

\noindent We find for $\nu_{x0}^2$:

\begin{align}
\nu_{x0}^2 &= 1+\bmu'_{rel} + 2\gamma\Delta_1 \nonumber \ , \\
\Delta_1 &=\frac{1}{\gamma}\bbigsymbol{[}\sum_n\frac{3n^2-4\bmu'(1+\bmu')(1-2\bmu')-4\bmu'^{2}(4+\bmu'(11+\bmu'))/(n^2-1-\bmu')}{16(n^2-1-\bmu')^2}C_n^2 \nonumber \\
	&- \sum_n\frac{n^2-1+\bmu'(3+4\bmu')}{4(n^2-1-\bmu')^2}C_nC'_n - \sum_n\frac{C'^{2}_n+n^2C_n^2\varphi'^{2}_n}{4(n^2-1-\bmu')}\bbigsymbol{]}\ , \label{nux0}
\end{align}

\noindent and for $f(\theta)$:

\begin{align}
f(\theta)&=\sum_n a_n \cos n\theta + b_n \sin n\theta\ , \nonumber \\
a_n &= \tfrac{3}{2} \frac{n^2+2\bmu'(1+\bmu')}{n^2-1-\bmu'}A_n+A'_n\ , \label{fxosc} \\
b_n &= \tfrac{3}{2} \frac{n^2+2\bmu'(1+\bmu')}{n^2-1-\bmu'}B_n+B'_n\ . \nonumber
\end{align}

\noindent In paragraph~\ref{remove} the general solution of a Hamiltonian with the structure of Eq.~(\ref{HPX}) has been derived by a canonical transformation 
that transforms the oscillating function $f(\theta)$ to the next higher order $\mathcal{O}(f^2)$.
The final Hamiltonian has the form:

\begin{equation}\label{HPX2}
\bar{K}_x(X,P_x,\theta) = \tfrac{1}{2}P_x^2 + \tfrac{1}{2}\nu_x^2X^2\ ,
\end{equation}

\noindent where:

\begin{equation}\label{NUX}
 \nu_x^2 =\nu_{x0}^2 + \tfrac{1}{2}\sum_n\frac{a_n^2+b_n^2}{n^2-4\nu_{x0}^2}\ .
\end{equation}

\noindent Note that in this equation we may in the summation replace $\nu_{x0}$ by $1+\bmu'$. The expression for the tune becomes:

\begin{align}
\nu_x^2 &=1+\bmu'_{rel} +\tfrac{1}{8}\sum_n\bbigsymbol{[}\frac{9\left(n^2+2\bmu'(1+\bmu')\right)^2}{n^2-4-4\bmu'}+3n^2-4\bmu'(1+\bmu')(1-2\bmu') \nonumber \\
&-\frac{4(4+\bmu'(11+\bmu'))\bmu'^{2}}{n^2-1-\bmu'}\bbigsymbol{]} \frac{C_n^2}{(n^2-1-\bmu')^2}+\tfrac{3}{2}\sum_n\frac{(1+\bmu')(C_n'^{2}+n^2C_n^2\varphi'^{2})}{(n^2-4-4\bmu')(n^2-1-\bmu')} \nonumber \\
&+\sum_n \bbigsymbol{[}\frac{n^2+(1+\bmu')(2+3\bmu')}{n^2-4-4\bmu'}-\frac{2\bmu'(1+\bmu')}{n^2-1-\bmu'}\bbigsymbol{]}\frac{C_nC'_n}{n^2-1-\bmu'} \label{NUXfinal}
\end{align} 

\begin{figure}[!bht]
   \vspace*{-.5\baselineskip}
   \centering
   \includegraphics[width=\textwidth]{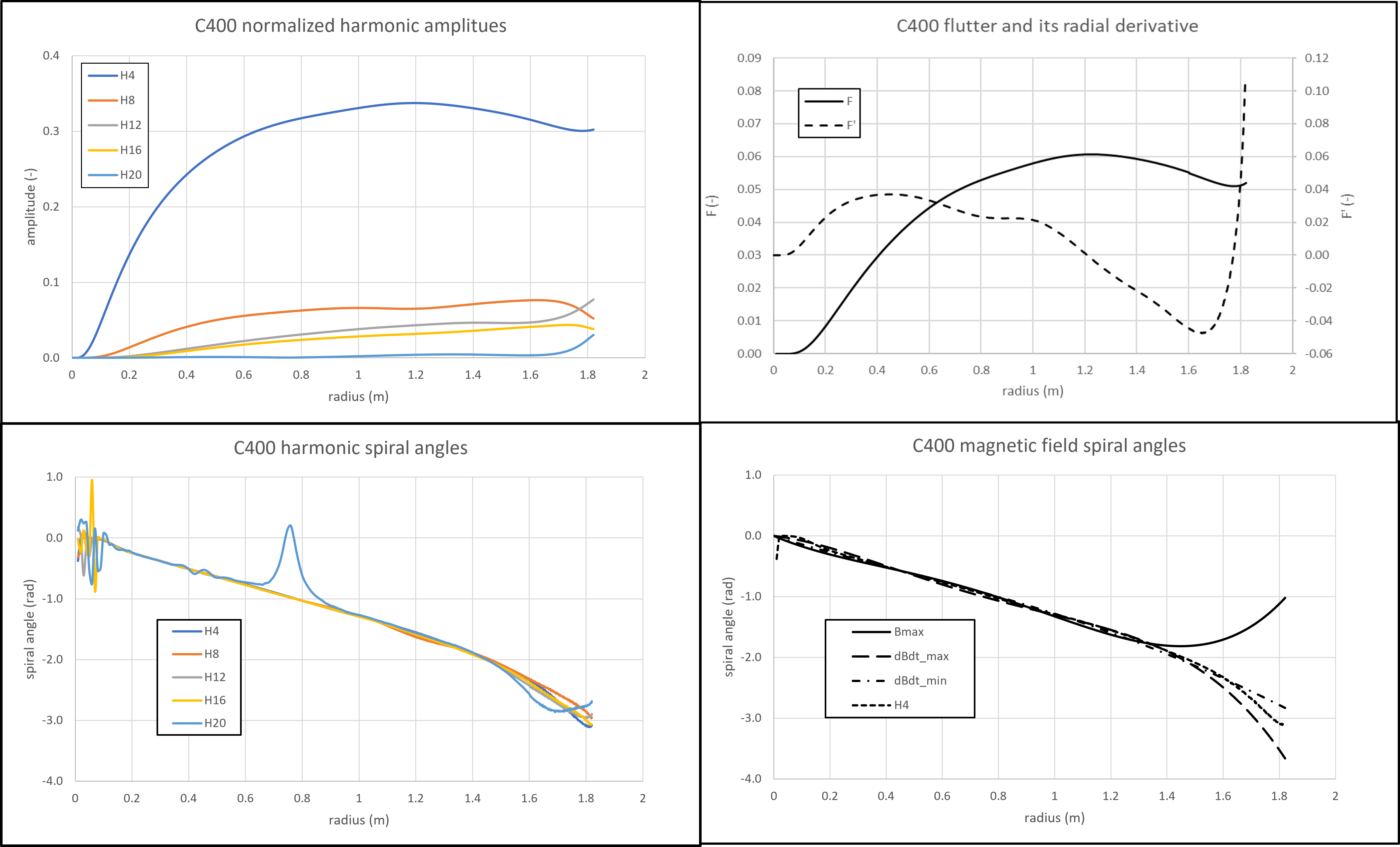}
    \caption[C400 harmonics and spiral angle comparisson]{\it C400 harmonics and spiral angle comparisson}
    \label{spiralcomp}
\end{figure}

\noindent This is a complex and rather impractical formula. We simplify it by the method explained in paragraph~\ref{RFf} and substitute for the Fourier coefficient $C_n$
their expressions in terms of the flutter $F$ as defined in Eqs.~(\ref{CtoF}). We also assume that the spiral angles $\varphi'_n$ are closely the same for the first few (up to 5)
dominant Fourier components. Figure~\ref{spiralcomp} shows that this is a valid assumption for practical cases. We therefore  replace $\varphi'_n$ by
$\varphi'$ and take this variable outside of the summations in Eq.~\ref{NUXfinal}.
The result for $\nu_x^2$ can now be written as follows:

\begin{equation}\label{NUXfinal2}
\boxed{\nu_x^2 =1+\bmu'_{rel} +\frac{8N^2F}{\pi^2}\left[\tilde{a}_N+\tilde{b}_N\varphi'^{2}+\tilde{c}_N\frac{F'}{F}+\tilde{d}_N\left(\frac{F'}{F}\right)^2\right]\ .}
\end{equation}

\noindent Here the functions $\tilde{a}_N,\tilde{b}_N,\tilde{c}_N,\tilde{d}_N$ depend only on the symmetry number  $N$ and on the relativistic parameter $\gamma$ via the
relation $\bmu'=\bmu'_{rel}=\gamma^2-1$.

\noindent The expressions for these parameters are given by:

\begin{equation}\label{rtune_coeff}
\begin{aligned}
\tilde{a}_N &= \sum_{k=0}^\infty\frac{1}{n^2(n^2-1-\bmu')^2}\bigsymbol{[}\tfrac{9}{4}\frac{\left(n^2+2\bmu'(1+\bmu')\right)^2}{n^2-4-4\bmu'}-\frac{(4+\bmu'(11+\bmu'))\bmu'^{2}}{n^2-1-\bmu'}  \\
&\ \hspace{4cm}+\tfrac{3}{4}n^2-\bmu'(1+\bmu')(1-2\bmu')\bigsymbol{]}\ , \\
\tilde{b}_N &= \sum_{k=0}^\infty\frac{3(1+\bmu')}{(n^2-4-4\bmu')(n^2-1-\bmu')}\ , \\
\tilde{c}_N &= \sum_{k=0}^\infty \frac{1}{n^2(n^2-1-\bmu')}\bigsymbol{[}\frac{n^2+(1+\bmu')(2+3\bmu')}{n^2-4-4\bmu'}-\frac{2\bmu'(1+\bmu')}{n^2-1-\bmu'}\bigsymbol{]}\ ,\\
\tilde{d}_N &= \sum_{k=0}^\infty \frac{3(1+\bmu')}{4n^2(n^2-4-4\bmu')(n^2-1-\bmu')}\ .
\end{aligned}
\end{equation}

\noindent In these equations we have to replace $n$ by $n=(2k+1)N$.

\noindent The summations in Eqs.~(\ref{rtune_coeff}) can be done analytically and the coefficients $\tilde{a}_N,\tilde{b}_N,\tilde{c}_N,\tilde{d}_N$ 
can be expressed in elementary mathematical functions. Appendix~\ref{summation} shows how this is done. We find the following result:

\renewcommand*\widefbox[1]{\fbox{\hspace{0.5em}#1\hspace{0.5em}}}
\begin{empheq}[box=\widefbox]{align}
\tilde{a}_N(\gamma) &=\tilde{q}_1\tan(\frac{\pi\gamma}{2N})+\tilde{q}_2\tan(\frac{\pi\gamma}{N})+\tilde{q}_3(1+ \tan^2(\frac{\pi\gamma}{2N}))+\tilde{q}_4 \nonumber\\
			&+\tilde{q}_5\tan(\frac{\pi\gamma}{2N})(1+ \tan^2(\frac{\pi\gamma}{2N}))\ , \nonumber\\
\tilde{b}_N(\gamma) &=  \frac{\pi}{8\gamma N}\bigsymbol{[}\tan(\frac{\pi\gamma}{N})-2\tan(\frac{\pi\gamma}{2N})\bigsymbol{]}\ ,\\
\tilde{c}_N(\gamma) &=  \frac{\pi}{96\gamma^3 N}\bigsymbol{[}(11-9\gamma^2)\frac{3\pi\gamma}{N}+3(\gamma^2+1)\tan(\frac{\pi\gamma}{N}) \label{nur-coef}\\
	&+24(2\gamma^2-3)\tan(\frac{\pi\gamma}{2N}) - \frac{12\pi\gamma}{N}(\gamma^2-1)\tan^2(\frac{\pi\gamma}{2N}) \bigsymbol{]}\ ,  \nonumber\\
\tilde{d}_N(\gamma) &=  \frac{\pi}{128\gamma^3 N}\bigsymbol{[}\frac{3\pi\gamma}{N}+ \tan(\frac{\pi\gamma}{N})-8\tan(\frac{\pi\gamma}{2N})\bigsymbol{]}\ . \nonumber
\end{empheq}

\noindent Here the coefficients $\tilde{q}_i$ are defined as:

\begin{align}
\tilde{q}_1(\gamma) &= \frac{-\pi}{8N\gamma^3}\bigsymbol{[}6-(\gamma^2+1)^2+15\tilde{q}_0\bigsymbol{]}\ , \nonumber \\
\tilde{q}_2(\gamma) &= \frac{+\pi}{32N\gamma^3}(\gamma^2+1)^2\ , \nonumber \\
\tilde{q}_3(\gamma) &= \frac{\pi^2}{16N^2\gamma^2}\bigsymbol{[}4-(\gamma^2+1)^2+7\tilde{q}_0\bigsymbol{]}\ ,\\
\tilde{q}_4(\gamma) &= \frac{+\pi^2}{32N^2\gamma^2}\bigsymbol{[}4-(\gamma^2+1)^2+16\tilde{q}_0\bigsymbol{]}\ , \nonumber \\
\tilde{q}_5(\gamma) &= \frac{-\pi^3 \tilde{q}_0}{16N^3\gamma}\ , \nonumber \\
\tilde{q}_0(\gamma) &=\frac{1}{4\gamma^4}\bigsymbol{(}4+(\gamma^2-1)(\gamma^2+10)\bigsymbol{)}(\gamma^2-1)^2 \ , \nonumber\
\end{align}

\begin{figure}[!bht]
   \vspace*{-.5\baselineskip}
   \centering
   \includegraphics[width=\textwidth]{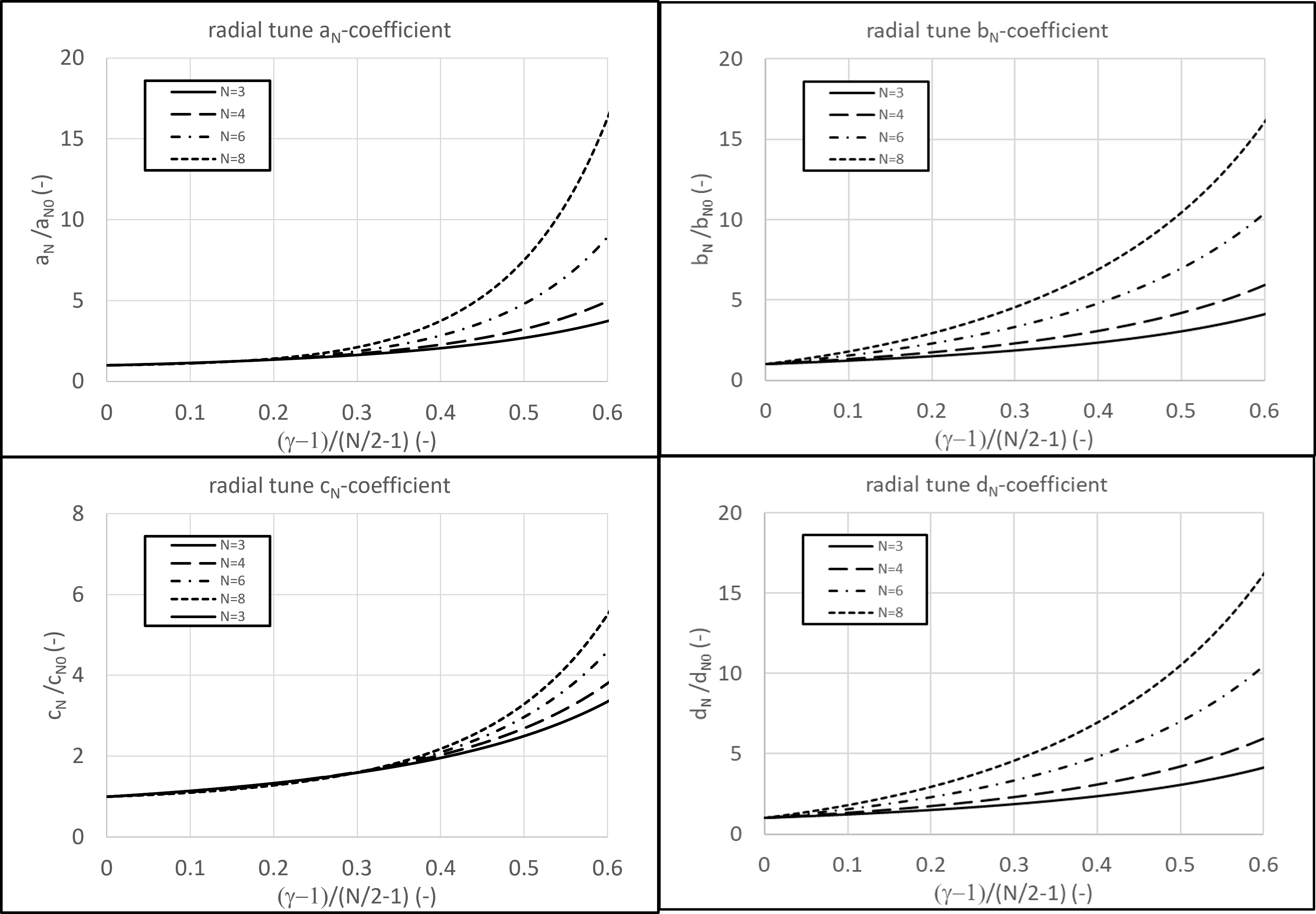}
    \caption[Energy dependence of the radial tune coefficients]{\it Energy dependence of the radial tune coefficients.}
    \label{htunecoefs}
\end{figure}

\noindent Note that the coefficients $\tilde{a}_N,\tilde{b}_N,\tilde{c}_N,\tilde{d}_N$ are singular for $\gamma=N/2$ and for $\gamma=N$. The first singularity is due to the half-integer resonance which is treated separately in paragraph~\ref{resonancestopband}. The second singularity would happen far beyond the maximum
energy that can be obtained in an isochronous cyclotron (see paragraph~\ref{isolimit}) and therefore is of no practical importance.

\noindent Figure~\ref{htunecoefs} shows the energy dependence of the horizontal tune coefficients. The representation of the axes is the same as used in Figure~\ref{Bisocoefs}. 
It is seen from Figure~\ref{htunecoefs} that the coefficients may vary more than a factor 10 in the energy range considered. 
The energy dependency is higher for higher N-values. This makes sense
because the absolue particle energy (for example at 60\% scale value) increases almost linearly with $N$ and therefore also the radial derivatives 
of the isochronous field will increase substantially (see Figure~\ref{Bisoderivs}).

\noindent In order to validate the derivations in this report, we compare results with those obtained for the C400 hadron therapy cyclotron. The design of this K=1600 
machine was initiated around the year 2004\cite{Jongen1} and finalized around the year 2010\cite{Jongen2,Jongen3}. Currently the machine is actually under construction in a collaboration between NHa and IBA. Figure~\ref{spiralcomp} shows results of the Fourier analysis of the C400 magnetic field. The upper left shows the amplitudes of the first five structural  Fourier components (normalized) a function of radius (as defined in appendix~\ref{thereduced}) and the upper right figure shows the flutter (see Eq.~(\ref{flutter})) and its radial derivative. 
The flutter is roughly equal to $C_4^2/2$ in agreement with Eq.~(\ref{flutter2}). The lower left figure shows the spiral angles of each of the first five structural Fourier components. It is
seen that they are closely the same for all five components. This property was used in the derivation of Eq.~(\ref{NUXfinal2}) and will also be used further on in the paper. 
The lower right figure shows different alternatives for the definition of the spiral angle. The first one uses the azimuth at which the magnetic field around a circle reaches its
maximum.  The second and third alternatives use the azimuth at which the azimuthal derivative of the magnetic field reaches its maximum (at sector entrance) 
or minimum (at sector exit) respectively. The fourth alternative uses the azimuth at which the basic harmonic component $C_N$ reaches its maximum. The first alternative
is not a good choice, because it deviates too much at high radii. 
For the radial tune (and also for the $\nu_r=N/2$ stopband, to be derived later) the other three alternatives give closely the same results. However, for the vertical tune we
find that the average of the second and third alternatives give the best match with the C400-tunes (see paragraph~\ref{vermot}). 
This makes sense because it is at the sector entrance and exit where the strong vertical focusing takes place. We therefore use this definition in the paper.

\begin{figure}[!bht]
   \vspace*{-.5\baselineskip}
   \centering
   \includegraphics[width=\textwidth]{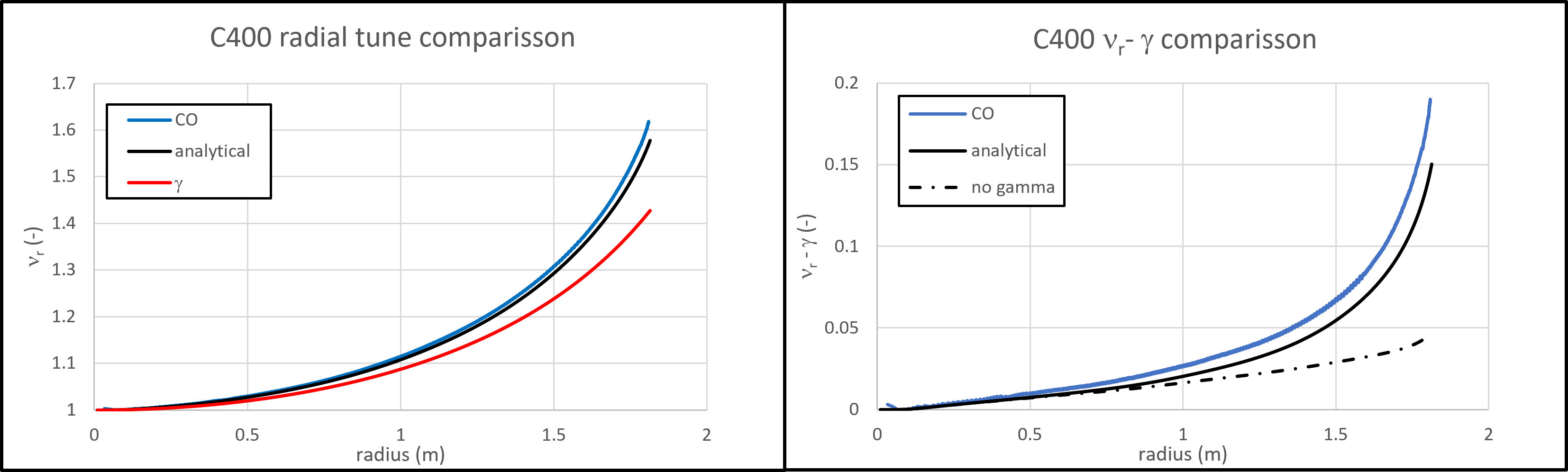}
    \caption[C400 radial tune comparisson]{\it C400 radial tune comparisson}
    \label{rtunecomp}
\end{figure}

Figure~\ref{rtunecomp} compares for the C400 our analytical radial tune  (black curve, calculated with Eq.~(\ref{NUXfinal2})) with the numerical tune obtained from a closed orbit
code (blue curve). In the left figure the relativistic contribution to the tune (=$\gamma$) is also shown seperately (red curve). At extraction, this contribution accounts for almost
75\% of the total. The right of Figure~\ref{rtunecomp} shows the part of the radial tune that is due to the flutter only. Here there is a small difference between the analytical and the closed orbit results. This difference is likely due to the fact that in the derivation of  Eq.~(\ref{NUXfinal2})), we ignore the approach towards the half-integer resonance. 
As shown in Figure~\ref{resonance stopband} the actual tune, when approaching the stopband, is higher than the "non-resonance" approximation of the tune. The dashed curve in the
right of Figure~\ref{rtunecomp} show the flutter contribution to the radial tune that is obtained if the energy-dependence of the tune-coefficients in Eq.~(\ref{NUXfinal2}) is ignored (by
evaluating these coefficients at the value $\gamma=1$). This is equivalent to a derivation in which the cross-terms between the average field radial derivatives and the magnetic
field Fourier terms are neglected. It is seen from the figure that such an approximation would have a big impact on the flutter contribution to the tune.

\section{The linear vertical motion}\label{vermot}

The derivation of the vertical motion is very much similar to that of the linear radial motion as was done in paragraph~\ref{radmot}. We start with the basic Hamiltonian for the vertical motion
derived in pargraph~\ref{Hcyclo} and given in Eq.~(\ref{hamz}). We assume that the motion in the median plane follows the EO and therefore substitute in Eq.~(\ref{hamz})
for $x,p_x$ the EO solution $x_e,p_e$. In the coefficients of this Hamiltonian we have to keep constants up to $\mathcal{O}(f^2)$ and oscillating terms up to $\mathcal{O}(f)$.
We obtain:

\begin{align}
K_z(\zeta,p_z,\theta)=\tfrac{1}{2}\left(1+x_e+\tfrac{1}{2}\dot{x}_e^2\right)p_z^2 +\tfrac{1}{2}\left(p_e\frac{\partial\mu}{\partial\theta}+(1-x_e)\frac{\partial\mu}{\partial x}\right)\zeta^2\ .
\end{align}

\noindent  Note that here, we have to evaluate $\frac{\partial\mu}{\partial\theta}$  and $\frac{\partial\mu}{\partial x}$ on the EO (so at $x=x_e$).
\noindent We bring this Hamiltonian to its normal form by a canonical transformation $p_z,\zeta \Rightarrow P_z,Z$  using the method explained in paragraph~\ref{normform}. 
This gives us for the new Hamiltonian $H_{z}(Z,P_z)$:

\begin{align}
\bar{K}_{z}(Z,P_z) = \tfrac{1}{2}P_z^2+\tfrac{1}{2}Q_z(\theta)Z^2\ ,
\end{align}

\noindent where $Q(\theta)$ is obtained as:

\begin{equation}\label{Qvertical1}
Q_z(\theta)=\dot{x}_e\frac{\partial\mu}{\partial\theta}-(1+2x_e+x_e^2+\tfrac{1}{2}\dot{x}_e^2)\frac{\partial\mu}{\partial x}+\tfrac{1}{2}\ddot{x}_e-\tfrac{1}{4}\dot{x}_e^2\ .
\end{equation}

\noindent We calculate the partial derivatives $\frac{\partial\mu}{\partial\theta}$ and  $\partial\mu/\partial x$ from the expression for the reduced field $\mu$ given in Eq.~(\ref{Taylor}) and insert them in Eq.~(\ref{Qvertical1}). We obtain:

\begin{align}
Q_z(\theta)&=-\bmu' - (2\bmu'+\bmu'')x_e-\tfrac{1}{2}(2\bmu'+4\bmu''+\bmu''')x_e^2+\tfrac{1}{2}\ddot{x_e}-\tfrac{1}{4}(1+2\bmu')\dot{x}_e^2 \nonumber \\
&-\ssum_n\left[A'_n+(2A'_n+A''_n)x_e-nB_n\dot{x}_e\right]\cos n\theta  \label{Qvertical2}\\
&-\ssum_n\left[B'_n+(2B'_n+B''_n)x_e+nA_n\dot{x}_e\right]\sin n\theta\ .  \nonumber
\end{align}

\noindent We now work out this expresion in full detail. This is done with the following additional steps:   i) use the expressions for $x_e$  and $\dot{x}_e$ as defined by Eqs.~(\ref{eqfou},\ref{alphae},\ref{betae},\ref{gammae}) and  insert those in Eq.~(\ref{Qvertical2}), ii) in the obtained result, split the $\mathcal{O}(f^0)$ term $\bmu'$ in its relativistic part and its flutter part as $\bmu'=\bmu'_{rel}+\bmu'_{fl}$ and insert for $\bmu'_{fl}$ the expression given in Eq.~(\ref{bmufl2}), iii) replace Fourier sine/cosine coefficients and their radial derivatives $A_n,A'_n,A''_n$, $B_n,B'_n,B''_n$ by Fourier amplitudes and their derivatives $C_n,C'_n,C''_n$ and phase derivative $\varphi'_n$ using Eqs.~(\ref{AnAn}-\ref{Anp2},\ref{AnAnpp}), iv) of all the $\theta$-dependent terms of $\mathcal{O}(f^2)$ keep only their average and v) substitute for the higher derivatives $\bmu'', \bmu'''$ epressions envolving the field-index $\bmu'$, using  Eqs.~(\ref{bmupp},\ref{bmuppp}).

\noindent We write the Hamiltonian in the same form as given in Eq.~(\ref{Hnorm}):

\begin{equation}\label{HPZ}
 \bar{K}_z(P_z,Z,\theta) = \tfrac{1}{2}P_z^2 + \tfrac{1}{2}(\nu_{z0}^2+f(\theta))Z^2\ .
\end{equation}

\noindent We find for $\nu_{z0}^2$:

\begin{align}
\nu_{z0}^2 = -\bmu'_{rel}+\tfrac{1}{8}&\sum_n\bbigsymbol{[}\frac{n^2(4n^2-5)+12\bmu'(n^2-(1+\bmu')(2+\bmu'))}{(n^2-1-\bmu')^2} \nonumber \\
	&-4\bmu'\frac{(n^2-1)(3n^2-7-\bmu'(11+\bmu'))-3\bmu'^{2}}{(n^2-1-\bmu')^3}\bbigsymbol{]}C_n^2 \\
	+ &\sum_n\frac{n^2-1+\bmu'(3+4\bmu')}{2(n^2-1-\bmu')^2}C_nC'_n + \sum_n\frac{C'^{2}_n+n^2C_n^2\varphi'^{2}_n}{2(n^2-1-\bmu')}\ . \nonumber
\end{align}

\noindent And $f(\theta)$ is defined by:

\begin{align}
f(\theta)&=\sum_n a_n \cos n\theta + b_n \sin n\theta\ , \nonumber \\
a_n &= -\tfrac{1}{2} \frac{n^2+6\bmu'(1+\bmu')}{n^2-1-\bmu'}A_n-A'_n\ , \\
b_n &= -\tfrac{1}{2} \frac{n^2+6\bmu'(1+\bmu')}{n^2-1-\bmu'}B_n-B'_n\ . \nonumber
\end{align}

\noindent In paragraph~\ref{remove} the general solution of a Hamiltonian with the structure of Eq.~(\ref{HPZ}) has been derived by a canonical transformation 
that transforms the oscillating function $f(\theta)$ to the next higher order $\mathcal{O}(f^2)$.
The final Hamiltonian has the form:

\begin{equation}\label{HPZ2}
 \bar{K}_z(P_z,Z,\theta) = \tfrac{1}{2}P_z^2 + \tfrac{1}{2}\nu_z^2Z^2\ ,
\end{equation}

\noindent where:

\begin{equation}\label{NUZ}
 \nu_z^2 =\nu_{z0}^2 + \tfrac{1}{2}\sum_n\frac{a_n^2+b_n^2}{n^2-4\nu_{z0}^2}\ .
\end{equation}

\noindent The expression for the tune becomes:

\begin{align}
\nu_z^2 =-\bmu'_{rel} +&\sum_n\bbigsymbol{[}\frac{\left(n^2+6\bmu'(1+\bmu')\right)^2}{n^2+4\bmu'}+n^2(4n^2-5)+12\bmu'(n^2-2-3\bmu'-\bmu'^{2}) \nonumber \\
&-4\bmu'\frac{(n^2-1)(3n^2-7-\bmu'(11+\bmu'))-3\bmu'^{3}}{n^2-1-\bmu'}\bbigsymbol{]} \frac{C_n^2}{8(n^2-1-\bmu')^2} \nonumber \\
+&\sum_n\frac{(2n^2-1+3\bmu')(C_n'^{2}+n^2C_n^2\varphi'^{2})}{2(n^2+4\bmu')(n^2-1-\bmu')} \nonumber \\
+&\sum_n \bbigsymbol{[}\frac{n^2++5\bmu'+3\bmu'^{2}}{n^2+4\bmu'}+\frac{2\bmu'(1+\bmu')}{n^2-1-\bmu'}\bbigsymbol{]}\frac{C_nC'_n}{n^2-1-\bmu'} \label{NUZfinal}
\end{align} 

\noindent This is a very complex and rather impractical formula. We simplify it by the method explained in paragraph~\ref{RFf} and substitute for the Fourier coefficient $C_n$
their expressions in terms of the flutter $F$ as defined in Eqs.~(\ref{CtoF}). We write the result as follows:

\begin{equation}\label{NUZfinal2}
\boxed{\nu_z^2 =-\bmu'_{rel} +\frac{8N^2F}{\pi^2}\left[\hat{a}_N+\hat{b}_N\varphi'^{2}+\hat{c}_N\frac{F'}{F}+\hat{d}_N\left(\frac{F'}{F}\right)^2\right]\ .}
\end{equation}

\noindent Here the functions $\hat{a}_N,\hat{b}_N,\hat{c}_N,\hat{d}_N$ depend only on the symmetry number  $N$ and on the relativistic parameter $\gamma$ via the
relation $\bmu'=\bmu'_{rel}=\gamma^2-1$.

\noindent The expressions for these parameters are given by:

\begin{align}
\hat{a}_N &= \sum_{k=0}^\infty\frac{1}{4n^2(n^2-1-\bmu')^2}\bigsymbol{[}-4\bmu'\frac{(n^2-1)(3n^2-7-\bmu'(11+\bmu'))-3\bmu'^{3}}{n^2-1-\bmu'} \nonumber \\
&\ \hspace{0.5cm}+\frac{\left(n^2+6\bmu'(1+\bmu')\right)^2}{n^2+4\bmu'}+n^2(4n^2-5)+12\bmu'(n^2-(1+\bmu')(2+\bmu'))\bigsymbol{]}\ , \nonumber \\
\hat{b}_N &= \sum_{k=0}^\infty\frac{2n^2-1+3\bmu'}{(n^2+4\bmu')(n^2-1-\bmu')}=\sum_{k=0}^\infty\frac{1}{n^2+4\bmu'}+\frac{1}{n^2-1-\bmu'}\ , \label{vtune} \\
\hat{c}_N &= \sum_{k=0}^\infty \frac{1}{n^2(n^2-1-\bmu')}\bigsymbol{[}\frac{n^2+\bmu'(5+3\bmu')}{n^2+4\bmu'}+\frac{2\bmu'(1+\bmu')}{n^2-1-\bmu'}\bigsymbol{]}\ ,\nonumber \\
\hat{d}_N &= \sum_{k=0}^\infty \frac{2n^2-1+3\bmu'}{4n^2(n^2+4\bmu')(n^2-1-\bmu')}\ .\nonumber
\end{align}

\noindent In these equations we have to replace $n$ by $n=(2k+1)N$.

\noindent The summations in Eqs.~(\ref{vtune}) can be done analytically and the coefficients $\hat{a}_N,\hat{b}_N,\hat{c}_N,\hat{d}_N$ 
can be expressed in elementary mathematical functions. Appendix~\ref{summation} shows how this is done. We find the following result:

\renewcommand*\widefbox[1]{\fbox{\hspace{0.5em}#1\hspace{0.5em}}}
\begin{empheq}[box=\widefbox]{align}
\hat{a}_N(\gamma) &=\hat{q}_1\tan(\frac{\pi\gamma}{2N})+\hat{q}_2\tanh(\frac{\pi\sqrt{\gamma^2-1}}{N})+\hat{q}_3(1+\tan^2(\frac{\pi\gamma}{2N}))+\hat{q}_4 \nonumber\\
&\ \hspace{2.45cm}+\hat{q}_5\tan(\frac{\pi\gamma}{2N})(1+ \tan^2(\frac{\pi\gamma}{2N}))\nonumber\ ,\\
\hat{b}_N(\gamma) &=\frac{\pi}{8N\gamma}\bbigsymbol{(}2\tan(\frac{\pi\gamma}{2N})+\frac{\gamma}{\sqrt{\gamma^2-1}}\tanh(\frac{\pi\sqrt{\gamma^2-1}}{N})\bbigsymbol{)}\ , \nonumber \\
\hat{c}_N(\gamma) &=\frac{\pi^2(9\gamma^2-14)}{32N^2\gamma^2}-\frac{\pi(12\gamma^4-27\gamma^2+14)}{4N\gamma^3(5\gamma^2-4)}\tan(\frac{\pi\gamma}{2N})\nonumber\\
&+\frac{\pi^2(\gamma^2-1)}{8N^2\gamma^2}\tan^2(\frac{\pi\gamma}{2N})
+\frac{\pi(3\gamma^2-2)}{32N(5\gamma^2-4)\sqrt{\gamma^2-1}}\tanh(\frac{\pi\sqrt{\gamma^2-1}}{N})\ , \nonumber \\
\hat{d}_N(\gamma) &=\frac{\pi}{128N^2\gamma^3}\bbigsymbol{(}\frac{\pi\gamma(4-3\gamma^2)}{\gamma^2-1} + 8N\tan(\frac{\pi\gamma}{2N})-\frac{N\gamma^3}{(\gamma^2-1)^{3/2}}\tanh(\frac{\pi\sqrt{\gamma^2-1}}{N})\bbigsymbol{)}\ , \nonumber
\end{empheq}

\noindent Here the coefficients $\hat{q}_i$ are defined as:

\begin{align}
\hat{q}_1(\gamma) &= \frac{\pi}{32N\gamma^3}\bigsymbol{[}84\gamma^4-176\gamma^2+101-\frac{(6\gamma^2-5)(102\gamma^4-177\gamma^2+76)\gamma^2}{(5\gamma^2-4)^2}\bigsymbol{]}\ , \nonumber \\
\hat{q}_2(\gamma) &= \frac{\pi}{256N}\frac{(9\gamma^4-12\gamma^2+32)\sqrt{\gamma^2-1}}{(5\gamma^2-4)^2}\ , \nonumber \\
\hat{q}_3(\gamma) &= \frac{\pi^2}{64N^2}\bigsymbol{[}\frac{(6\gamma^2-5)^2}{5\gamma^2-4}-\frac{36\gamma^4-80\gamma^2+45}{\gamma^2} \bigsymbol{]}\ ,\\
\hat{q}_4(\gamma) &= -\frac{\pi^2}{32N^2\gamma^2}(\gamma^2-1)(15\gamma^2-28)\ , \nonumber \\
\hat{q}_5(\gamma) &= \frac{\pi^3 }{16N^3\gamma}(\gamma^2-1)^2\ . \nonumber \
\end{align}

\noindent Note that the expressions for $\hat{b}_N,\hat{c}_N,\hat{d}_N$ are singular for $\gamma=1$ and the limits for $\gamma\downarrow 1$ need to be taken. 
These limits are as follows:

\begin{align}
\hat{a}_N(1) &=\frac{\pi}{4N}\tan(\frac{\pi}{2N})\ , \nonumber \\
\hat{b}_N(1) &=\frac{\pi}{8N}\bigsymbol{(}\frac{\pi}{N}+2\tan(\frac{\pi}{2N})\bigsymbol{)}\ , \nonumber \\
\hat{c}_N(1) &= \frac{\pi}{4N}\bigsymbol{(}-\frac{\pi}{2N}+\tan(\frac{\pi}{2N}) \bigsymbol{)}\ , \nonumber \\
\hat{d}_N(1) &=\frac{\pi}{128N^2}\bigsymbol{(}-4\pi+\frac{\pi^3}{3N^2}+8N\tan(\frac{\pi}{2N}) \bigsymbol{)}\ . \nonumber
\end{align}

\begin{figure}[!bht]
   \vspace*{-.5\baselineskip}
   \centering
   \includegraphics[width=\textwidth]{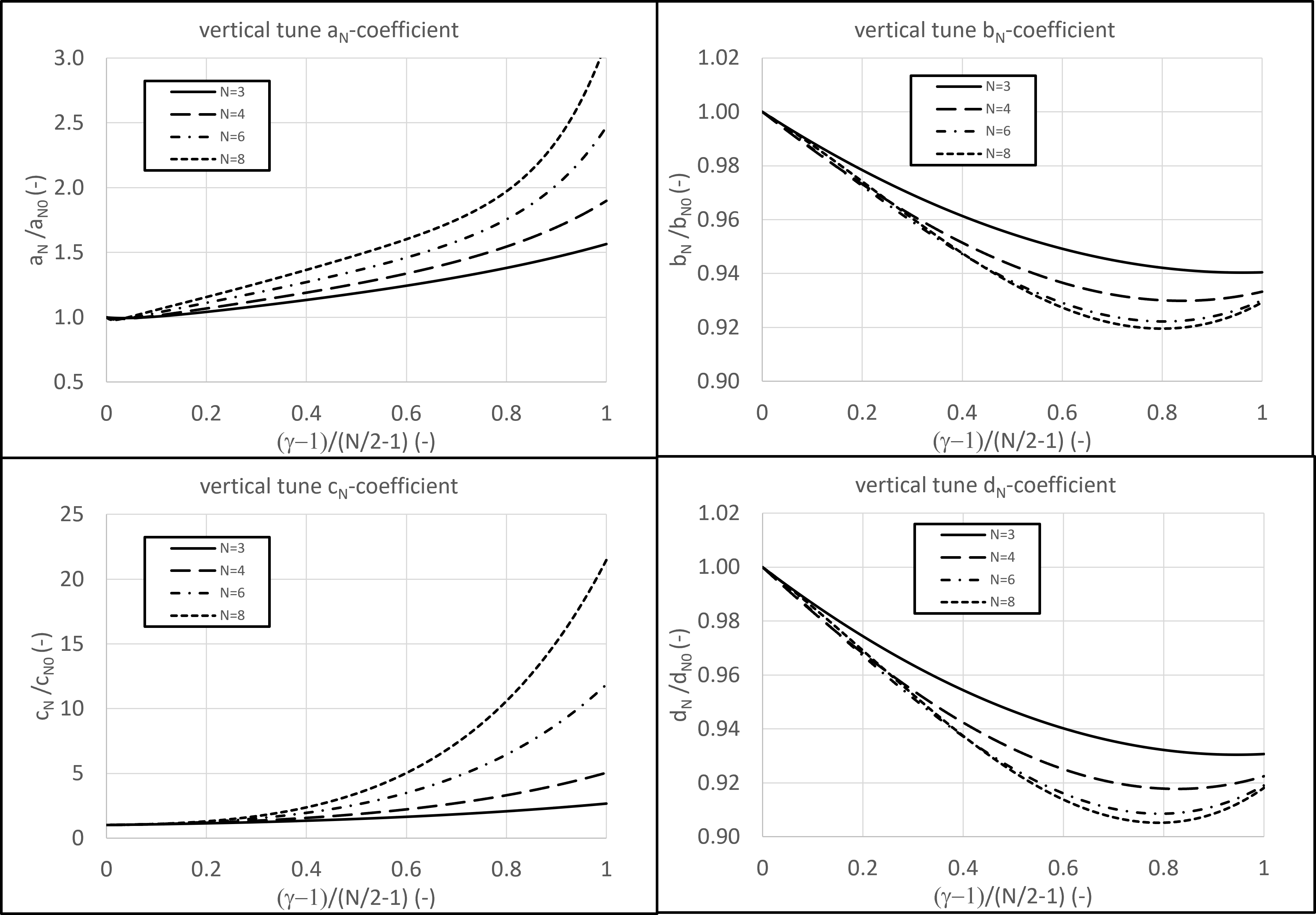}
    \caption[Energy dependence of the vertical tune coefficients]{\it Energy dependence of the vertical tune coefficients.}
    \label{vtunecoefs}
\end{figure}

\noindent Figure~\ref{vtunecoefs} shows the energy dependence of the vertical tune coefficients. The representation of the axes is the same as used in Figure~\ref{Bisocoefs}.
It is seen that the coefficient $\hat{c}_N$ may vary more than a factor 10 in the energy range considered. However, for higher energies, the most important contribution 
to the vertical tune by far comes from the spiraling of the sectors, i.e. from the coefficient $\hat{b}_N$. This coefficient depends only weakly on energy. As can be seen from
Eqs.~(\ref{vtune}) the equation for $\hat{b}_N$ contains two terms with opposite energy dependence and therefore there is some cancellation of this dependence.  For that reason the cross terms between $\bar{\mu}$-derivatives and flutter terms are less importance in the derivation of the vertical tune.

\begin{figure}[!bht]
   \vspace*{-.5\baselineskip}
   \centering
   \includegraphics[width=\textwidth]{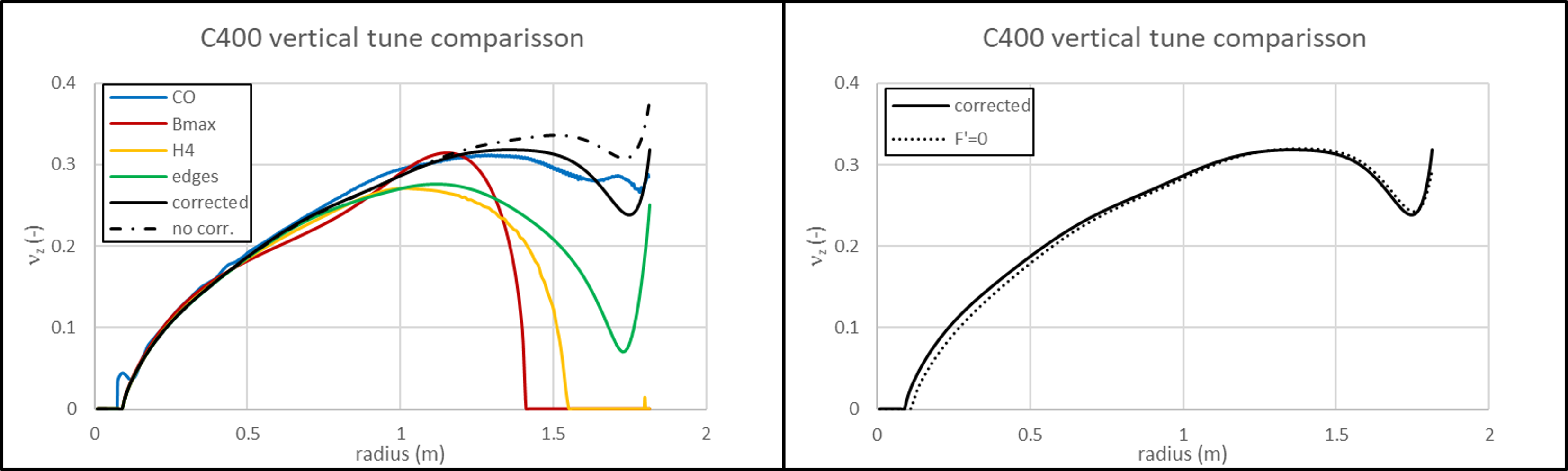}
    \caption[C400 vertical tune comparisson]{\it C400 vertical tune comparisson}
    \label{vtunecomp}
\end{figure}

Figure~\ref{vtunecomp} compares for the C400 our analytical vertical tune  (black curve, calculated with Eq.~(\ref{NUZfinal2})) with the numerical tune obtained from a closed orbit
code (CO=blue curve). 
The vertical tune depends critically on the definition of the spiral angle. The main reason for this is that the tune (squared) is obtained as the difference between two larger numbers 
(the field index $\bmu'$ as a negative contribution and the flutter as a positive contribution), which to a substantial degree cancel eached other. 

Different alternatives for the definition of the spiral angle have been introduced in paragraph~\ref{radmot} and the corresponding tunes are shown in the figure on the left.
The red curve (Bmax) uses the spiral angle obtained from the azimuth where the magnetic field around a circle is maximum. This model fits well up to a radius 
of about 1.2 m ($\approx$125 MeV/u), but beyond that immediately collapses. The orange curve (H4), based on the azimuth of the basic harmonic $C_4$, gives some improvement
but is still not satisfactory. The green curve (edges), based on the average of the sector-in and sector-out azimuths, shows a further improvement but it still deviates substantially from 
the numerical curve. The black curve shows our final result where the spiral angle (corresponding to the previous case) has been corrected for the fact that the
equilibrium orbit is not a circle and therefore enters and exits from the sector with a non-zero radial momentum. This correction is explained in paragraph~\ref{spiralcorrection} 
and Eq.~(\ref{spiralcorrection2}) was used to calculate it.

The dashed curve (corr) in Figure~\ref{vtunecomp} shows the same case but here the flutter contribution to the vertical tune is obtained by ignoring the energy-dependence of the tune-coefficients in Eq.~(\ref{NUZfinal2}) (by evaluating those at the value $\gamma=1$). 
This is equivalent to using a derivation in which the cross-terms between the average field radial derivatives and the magnetic
field Fourier terms are neglected. It is seen from the figure that such an approximation does not have such a big impact on the resulting tune.
The dotted curve in the right frame of Figure~\ref{vtunecomp} shows a case where the contribution of the flutter radial derivative to the vertical tune is ignored. 
It is seen that this contribution is small at high energies where the effect of the spiral angles dominates. For smaller machines with little or no spiralling the $F'$-contribution may
be more significant, especially in the extraction region where the flutter usually starts to drop.

\section{The stopband of the half-integer resonance}\label{resonancestopband}

In paragraph~\ref{halfint} a general treatment is given of the half-integer resonance for a Hamiltonian of the form given in Eq.~(\ref{Hnorm}). In this paragraph those results are used
to find the $\nu_r=N/2$ stopband of the isochronous cyclotron. 
\noindent The general expression for the stopband is given in Eq.~(\ref{stopband2}). In this equation we must insert expressions
for $\nu_0$ and $c_n$ as applicable for the isochronous cyclotron. These have been derived in paragraph~\ref{radmot}. For $\nu_0$ we must insert the expression 
for $\nu_{x0}$ as given in Eq.~(\ref{nux0}). With this we can write for the relativistic gamma parameters of the stopband:

\begin{equation}\label{gamma12}
\gamma_{1,2} = \frac{N}{2} \mp \frac{c_N}{2N} -\Delta_1 - \Delta_2 \ ,
\end{equation}

\noindent where $\gamma$ is the relativistic gamma and where the expression for $\Delta_1$ is given in Eq.~(\ref{nux0}) and the expression for $\Delta_2$
in Eq.~(\ref{delta2}) . The expression for $c_n$ we obtain from Eq.~(\ref{fxosc}):

\begin{align}\label{cnsquare}
c_n^2 =a_n^2+b_n^2=\left[\left(\tfrac{3}{2}\frac{n^2+2\bmu'(1+\bmu')}{n^2-1-\bmu'}+\frac{C_n'}{C_n}\right)^2+n^2\varphi_n'^2\right]C_n^2\ .
\end{align}

\noindent The parameters $c_n,\Delta_1,\Delta_2$ depend on the energy $\gamma$ through their dependence on $\bmu'$. Therefore Eq.~(\ref{gamma12}) represents
an implicit expression for the stopband limits $\gamma_{1,2}$. We can solve for $\gamma_{1,2}$ by using successive substitution in three steps. The goal is
to find the stopband limits up to order  $\mathcal{O}(f^2)$. The first step gives the stopband limits
up to order  $\mathcal{O}(f^0)$. Since $c_n$ is $\mathcal{O}(f^1)$ and $\Delta_1, \Delta_2$ are $\mathcal{O}(f^2)$, we get:

\begin{align}\label{gamma012}
\gamma^{(0)}_{1,2}&= \frac{N}{2}  \ , \\
\gamma^{(1)}_{1,2} &= \frac{N}{2} \mp \frac{c_N(\gamma^{(0)})}{2N} \ . 
\end{align}

\noindent In the second equation the term $c_N$ must be evaluated at $\gamma=N/2$. This means for  $\bmu'=\gamma^2-1=\tfrac{N^2}{4}-1$. Using this
in Eq.~(\ref{cnsquare}) one finds after the second step of successive substitution:

\begin{align}
\gamma^{(1)}_{1,2} = \frac{N}{2} \mp \frac{C_N}{2N}\sqrt{\bigsymbol{(}1+\frac{N^2}{4}+\frac{C_N'}{C_N}\bigsymbol{)}^2+N^2\varphi'^{2}_{N}} \ . \label{gamma112}
\end{align}

\noindent The third step of successive substitution can now be written as:

\begin{align}
\gamma^{(2)}_{1,2} = \frac{N}{2} \mp \frac{c_N(\gamma^{(1)})}{2N} -\Delta_1(\gamma^{(0)}) - \Delta_2(\gamma^{(0)})  \ . \label{gamma212}
\end{align}

\noindent The second term for $c_N$ in this equation must now be evaluated at $\gamma^{(1)}$ which is given in Eq.~(\ref{gamma112}).  To do this we write $c_N$ as a function
of energy $\gamma$ as follows:

\begin{align}
c_N(\gamma) = C_N\sqrt{\bigsymbol{(}\tfrac{3}{2}\frac{N^2+2\gamma^2(\gamma^2-1)}{N^2-\gamma^2}+\frac{C_N'}{C_N}\bigsymbol{)}^2+N^2\varphi_N'^2}\ ,
\end{align}

\noindent and do a Taylor expansion up to first degree:

\begin{align}
c^{(2)}_N=c_N(\gamma^{(1)})=c_N(\tfrac{N}{2}+\delta)=c_N(\tfrac{N}{2})+\delta\frac{dc_N}{d\gamma}|_{\gamma=N/2}\ ,
\end{align}

\noindent where $\delta=\gamma^{(1)}-\tfrac{N}{2}$ is obtained from Eq.~(\ref{gamma112}). We calculate $c^{(2)}_N$ up to $\mathcal{O}(f^2)$ and get:

\begin{align}
\frac{c_N^{(2)}}{2N}&=\frac{C_N}{2N}\sqrt{\bigsymbol{(}1+\frac{N^2}{4}+\frac{C_N'}{C_N}\bigsymbol{)}^2+N^2\varphi'^{2}_{N}}-\Delta_3\ ,  \\
\Delta_3  &= \frac{(7N^2-8)}{12N^3}\bigsymbol{(}1+\frac{N^2}{4}+\frac{C_N'}{C_N}\bigsymbol{)}C_N^2\ . 
\end{align}

\noindent In this way the stopband limits can be writen as:

\begin{align}\label{gamma12b}
\gamma_{1,2} &= \frac{N}{2} \mp \frac{C_N}{2N}\sqrt{\bigsymbol{(}1+\frac{N^2}{4}+\frac{C_N'}{C_N}\bigsymbol{)}^2+N^2\varphi'^{2}_{N}} -\Delta \ , \\
\Delta &= \Delta_1+\Delta_2+\Delta_3\ .
\end{align}

\noindent Here the terms $\Delta_1$ and $\Delta_2$ are both of $\mathcal{O}(f^2)$ and therefore they can be evaluated at the energy $\gamma=\gamma^{(0)}=N/2$.
We find for $\Delta_1$ and $\Delta_2$:

\begin{align}\label{delta123}
\Delta_1  &= \frac{1}{2N}\sum_{n=N}\bbigsymbol{(}\frac{12n^2+\tfrac{1}{2}N^2(N^2-4)(N^2-6)}{(4n^2-N^2)^2} -4\frac{n^2\varphi_{n}'^{2}}{4n^2-N^2} \nonumber \\
&\hspace{2cm} -\frac{(N^2-4)^2(16+(N^2-4)(\tfrac{N^2}{4}+10))}{(4n^2-N^2)^3}\bbigsymbol{)}C_n^2 \\
&-\frac{1}{2N}\sum_{n=N}\bbigsymbol{(}\frac{4(n^2-1)+(N^2-1)(N^2-4)}{n^2(4n^2-N^2)}C_nC_n'+\frac{4C_n'^2}{n^2(4n^2-N^2)}\bbigsymbol{)}\ , \\
\Delta_2 &= \frac{3C_N^2}{8N^3}\bbigsymbol{(}(1+\frac{N^2}{4}+\frac{C_N'}{C_N})^2+N^2\varphi_{N}'^{2} \bbigsymbol{)}
+\frac{3}{4N}\sum_{n>N}\frac{(8n^2+N^2(N^2-4))}{(4n^2-N^2)(n^2-N^2)}C_nC_n'\nonumber \\
&+\frac{1}{2N}\sum_{n>N}\bbigsymbol{(}\frac{9}{16}\frac{\bigsymbol{(}8n^2+N^2(N^2-4)\bigsymbol{)}^2}{(4n^2-N^2)^2(n^2-N^2)}+\frac{n^2\varphi_n'^2}{n^2-N^2}\bbigsymbol{)}C_n^2+\frac{1}{2N}\sum_{n>N}\frac{C_n'^2}{n^2-N^2}\ ,
\end{align}

\noindent As before, we eliminate the Fourier coefficients $C_n$ in favor of the flutter $F$ using  the method explained in paragraph~\ref{RFf} and assume that the spiral angles of all 
harmonics are equal ($\varphi'_n=\varphi'_N=\varphi'$). We write the stopband limits as follows:



\renewcommand*\widefbox[1]{\fbox{\hspace{0.1em}#1\hspace{0.1em}}}
\begin{empheq}[box=\widefbox]{align}
\gamma_{1,2} = \frac{N}{2} &\mp \frac{2\sqrt{F}}{\pi N}\sqrt{\bigsymbol{(}1+\frac{N^2}{4}+\frac{F'}{2F}\bigsymbol{)}^2+N^2\varphi'^{2}_{N}}\nonumber \\
&-\frac{F}{\pi^2N^3}\left(\bar{a}_N-\bar{b}_N\varphi'^2-\bar{c}_N\frac{F'}{F}+\bar{d}_N(\frac{F'}{F})^2\right)\ . \label{SBfinal}
\end{empheq}

\noindent Here $\bar{a}_N,\bar{b}_N,\bar{c}_N,\bar{d}_N$ are defined as:

\begin{align}
&\bar{a}_N = 6(1+\frac{N^2}{4})^2+\frac{1}{3}(N^2+4)(7N^2-8)+\frac{9}{32}\ssum_{k=1}\frac{(8m^2+(N^2-4))^2}{m^2(m^2-\tfrac{1}{4})^2(m^2-1)} \nonumber\\
&+\frac{1}{2}\ssum_{k=0}\bbigsymbol{(}\frac{12m^2+\tfrac{1}{2}(N^2-4)(N^2-6)}{m^2(m^2-\tfrac{1}{4})^2}-\frac{(N^2-4)^2(16+(N^2-4)(\tfrac{N^2}{4}+10))}{4N^4m^2(m^2-\tfrac{1}{4})^3}\bbigsymbol{)}\ , \nonumber \\
&\bar{b}_N = 8N^2\bbigsymbol{(}-\frac{3}{4}+\ssum_{k=0}\frac{1}{m^2-\tfrac{1}{4}}-\ssum_{k=1}\frac{1}{m^2-1}\bbigsymbol{)}\label{sblimit}\ , \\
&\bar{c}_N =\frac{4+37 N^2}{6}-\frac{1}{4}\sum_{k=0}\frac{4m^2+N^2-5}{m^2(m^2-\tfrac{1}{4})^2}+\frac{3}{2}\sum_{k=1}\frac{8m^2+N^2-4}{m^2(m^2-\tfrac{1}{4})(m^2-1)}\ ,\nonumber \\
&\bar{d}_N= -\frac{3}{2}+\ssum_{k=0}\frac{2}{m^2(m^2-\tfrac{1}{4})}-\ssum_{k=1}\frac{2}{m^2(m^2-1)}\ , \nonumber
\end{align} 

\noindent where we must substitute $m$ by $m=2k+1$, with $k=(0,)1,2,\dots$. The series  can again be summed analytically (see appendix~\ref{summation}) giving:

\begin{figure}[!bht]
   \vspace*{-.5\baselineskip}
   \centering
   \includegraphics[width=12 cm]{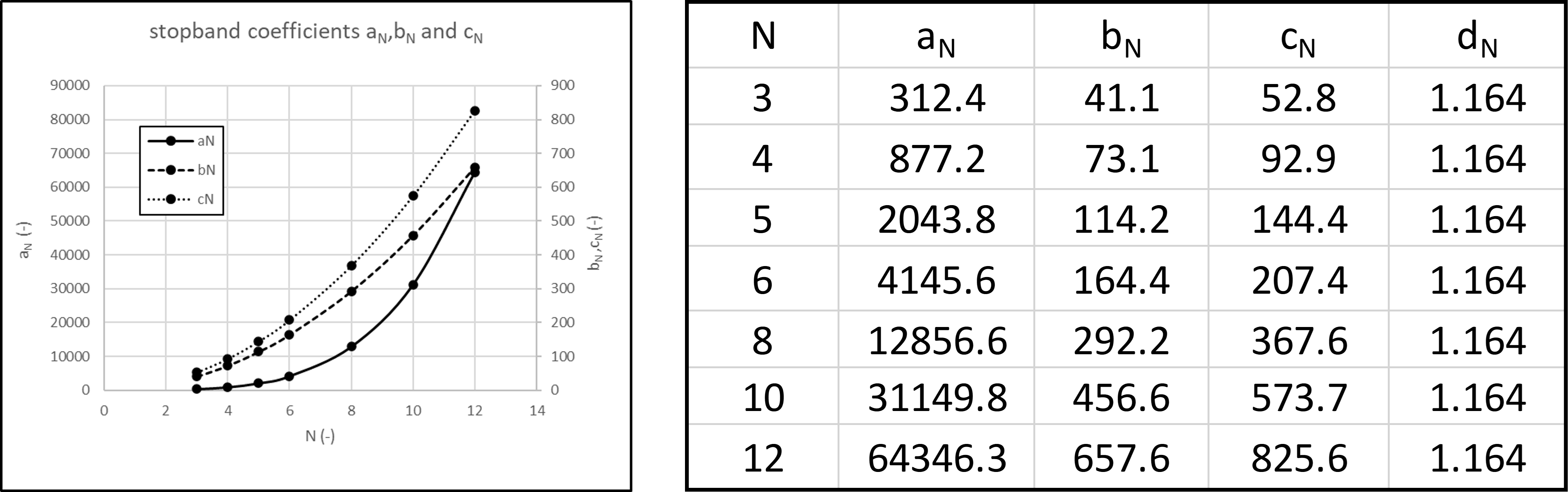}
    \caption[Stopband coefficients]{\it Stopband coefficients.}
    \label{SB-coeff}
\end{figure}


\renewcommand*\widefbox[1]{\fbox{\hspace{0.1em}#1\hspace{0.1em}}}
\begin{empheq}[box=\widefbox]{align}
\bar{a}_N &= \frac{-16+52N^2+14N^4}{3}+\pi(-40+4N^2+\frac{N^4}{2})+5\pi^2(3-\frac{N^2}{2}-\frac{N^4}{16})  \nonumber\\
&-\frac{\pi}{32N^4}(\pi^2-22\pi+60)(N^2-4)^2(N^4+36N^2-96)\ , \\
\bar{b}_N &=8N^2(\frac{\pi}{2}-1)\ ,\nonumber \\
\bar{c}_N &=\frac{16}{3}-6\pi+\frac{3\pi^2}{2}+(\frac{34}{3}-\pi-\frac{\pi^2}{4})N^2\ , \nonumber \\
\bar{d}_N&=-4+4\pi-\frac{3\pi^2}{4}\ .
\end{empheq}

\begin{figure}[!bht]
   \vspace*{-.5\baselineskip}
   \centering
   \includegraphics[width=\textwidth]{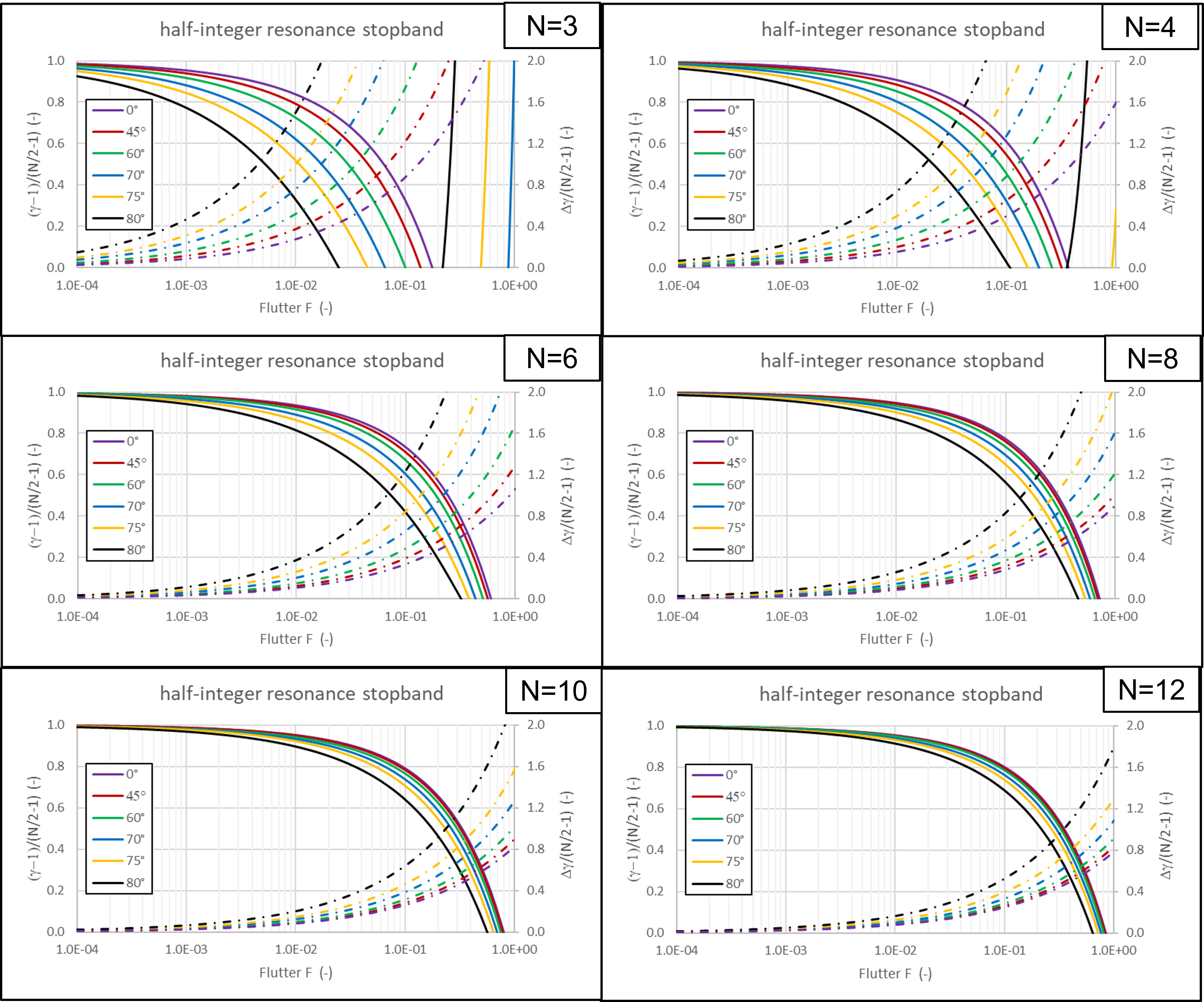}
    \caption[Stopband limit and width]{\it Stopband limit and width}
    \label{FigSB}
\end{figure}

\noindent Figure~\ref{SB-coeff} graphically shows the dependence of the coefficients $\bar{a}_N,\bar{b}_N,\bar{c}_N,\bar{d}_N$ on the cyclotron symmetry number $N$ and gives their values for a range of $N$-numbers.
\noindent Figure~\ref{FigSB} shows the stopband calculated from Eq.~(\ref{SBfinal}) for cyclotron symmetry numbers $N$  of 3,4,6,8,10,12 and for the 
case where the radial derivatives of the flutter are zero ($F'=0$).
The horizontal axis in the figures gives the flutter $F$ 
in logarithmic scale. The left axis gives the lower limit ($\gamma_1$, solid lines) and the right axis
the width ($\gamma_2-\gamma_1$, dashed lines)  of the stopband respectively. Both axes use the same scale as introduced in Figure~\ref{htunecoefs}.  Results are shown for spiral angles of 45°,60°,70°,75° and 80°.
It is seen that the lower stopband limit (i.e. the stable region) decreases monotonically with increasing flutter and increasing spiral angle. The normalized
limit $(\gamma-1)/(\tfrac{N}{2}-1)$ increases monotonically with increasing N-number. 

\noindent The stable region for a given spiral angle is the area
under the curve between $\gamma=1$ and $\gamma=\gamma_1$. We note that for low symmetry numbers, there appears a second branch of $\gamma_1$ for high values of $F$
and large spiral angles. This artefact shows that the $\mathcal{O}(f^2)$ approximation is not sufficient for very large flutter and/or spiral angle. Note that the 
maximum value of $F=1$ used in Figure~\ref{FigSB} is really large, especially if combined with a large spiral angle. We further note that the convergence of the development is 
determined not so much by the magnitude of $f$ but by the magnitude of $f/N^2$. The reason for this is that the scalloping of the equilibrium orbit is proportional to $f/N^2$.
and therefore, the EO becomes more and more circular for higher N-values. The branch therefore does not appear for the larger $N$ values. 
Note further that the apperance of this artificial branch does not compromise in any way the validity of the results because
it occurs at $F$ values that are about an order of magnitude larger than the maximum $F$-value of the corresponding stability zone. It is this value that we are interested in. 
The artificial branch can therefore be ignored completely. 
\noindent It is seen from Figure~\ref{FigSB} that the width of the stopband quickly rises to large values, for reasonable values of flutter and spiral angles. This shows that the 
half-integer resonance is extremely strong and impossible to cross by fast acceleration. It is a hard limit for the maximum energy of an isochronous cyclotron.

\subsection{Impact of different types of approximations}

In the previous paragraph we derived the stopband of the half-integer resonance in a very accurate manner namely i) up to $\mathcal{O}(f^2)$ in the magnetic field variation 
and ii) including terms that correlate derivatives of the average field $\bmu',\bmu'',\bmu'''$ with the azimuhal magnetic field variation $f$.
However, both refinements make that the derivation is complex and elaborate.
In this paragraph we investigate how these refinements impact the final result. For this purpose we re-calcuate the stopband limits by ignoring the cross-correlations. We label this 
as a non-relativistic approximation, because at low energies the average field derivatives and therefore also the correlating terms, are small. For simplicity we assume here that
the radial derivative of the flutter equals zero ($F'=0$). The approximation is found in a similar
way as used in the previous paragraph but now we insert $\bmu'=0$ in the expressions for $\Delta_1$ (in Eq.~(\ref{nux0})) and $c_n$ (in Eq.~(\ref{cnsquare})). 
The expression for $\Delta_2$ also needs to be re-calculated as it depends on $c_n$ (see Eq.~(\ref{delta2})). Further we have $\Delta_3=0$ for this case. 
We find for the stopband:

\begin{align}\label{nolrelgamma12b}
\gamma_{1,2} &= \frac{N}{2} \mp \frac{ C_N}{2}\sqrt{\frac{9N^2}{4(N^2-1)^2}+\varphi'^{2}_{N}} -\Delta_1-\Delta_2 \ , \\
\Delta_1 &=\frac{2}{N}\sum_n\left( \frac{3n^2C_n^2}{16(n^2-1)^2}-\frac{n^2C_n^2\varphi'^2_n}{4(n^2-1)}\right)\, \\
\Delta_2 &= \frac{3C_N^2}{8N}\left(\frac{9N^2}{4(N^2-1)^2}+\varphi'^{2}_{N}\right) \nonumber \\
&+\frac{1}{2N}\sum_{n>N}n^2C_n^2\left(\frac{9n^2}{4(n^2-1)^2(n^2-N^2)}+\frac{\varphi'^2_n}{n^2-N^2}\right)\ .
\end{align}

\noindent We again eliminate the Fourier coefficients $C_n$ in favor of the flutter $F$ and write the stopband limits in the non-relativistic approximation as follows:

\begin{align}
\gamma_{1,2}=\frac{N}{2} \mp \frac{2\sqrt{F}}{\pi}\sqrt{\frac{9N^2}{4(N^2-1)^2}+\varphi'^{2}_{N}}-\frac{F}{\pi^2N^3}(a_N-b_N\varphi'^2)\ ,
\end{align} 

\noindent where $a_N,b_N$ are given by:

\begin{align}
a_N&=\frac{27N^4}{2(N^2-1)^2}+6\sum_{k=0}\frac{1}{(m^2-\tfrac{1}{N^2})^2}+18\sum_{k=1}\frac{m^2}{(m^2-\tfrac{1}{N^2})^2(m^2-1)}\ , \nonumber \\
b_N&=8N^2(-\frac{3}{4}+\sum_{k=0}\frac{1}{m^2-\tfrac{1}{N^2}}-\sum_{k=1}\frac{1}{m^2-1})\ .
\end{align} 

\noindent In these equations we must substitute $m$ by $m=2k+1$, where $k=(0,)1,2,\dots$.

\noindent The series in the above two equations can be summed analytically as has been explained in appendix~\ref{summation}. We find the following epressions:

\begin{align}
a_N &=\frac{36N^6}{(N^2-1)^3}-\frac{3\pi N^3}{4}\frac{(N^4+N^2+4)}{(N^2-1)^2}\tan\frac{\pi}{2N}+\frac{3\pi^2 N^2}{8}\frac{N^2-4}{N^2-1}(1+\tan^2\frac{\pi}{2N})\ ,
\nonumber \\
b_N &=8N^2(-1+\frac{\pi N}{4}\tan\frac{\pi}{2N})\ .
\end{align} 

\begin{figure}[!bht]
   \vspace*{-.5\baselineskip}
   \centering
   \includegraphics[width=10cm]{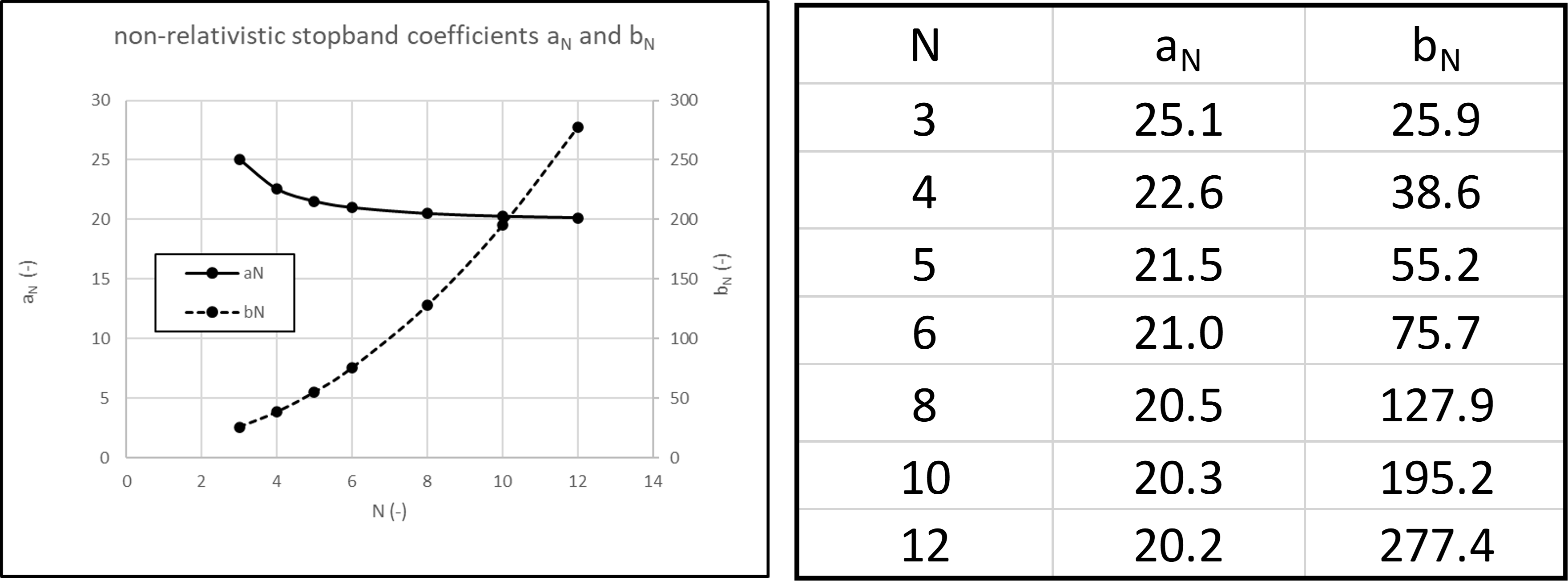}
    \caption[Stopband coefficients (non-relativistic)]{\it Stopband coefficients for non-relativistic derivation.}
    \label{SB-coeff2}
\end{figure}

\noindent Figure~\ref{SB-coeff2} shows the dependence of the coefficients $a_N$ and $b_N$ on the cyclotron symmetry number $N$.

\begin{figure}[!bht]
   \vspace*{-.5\baselineskip}
   \centering
   \includegraphics[width=\textwidth]{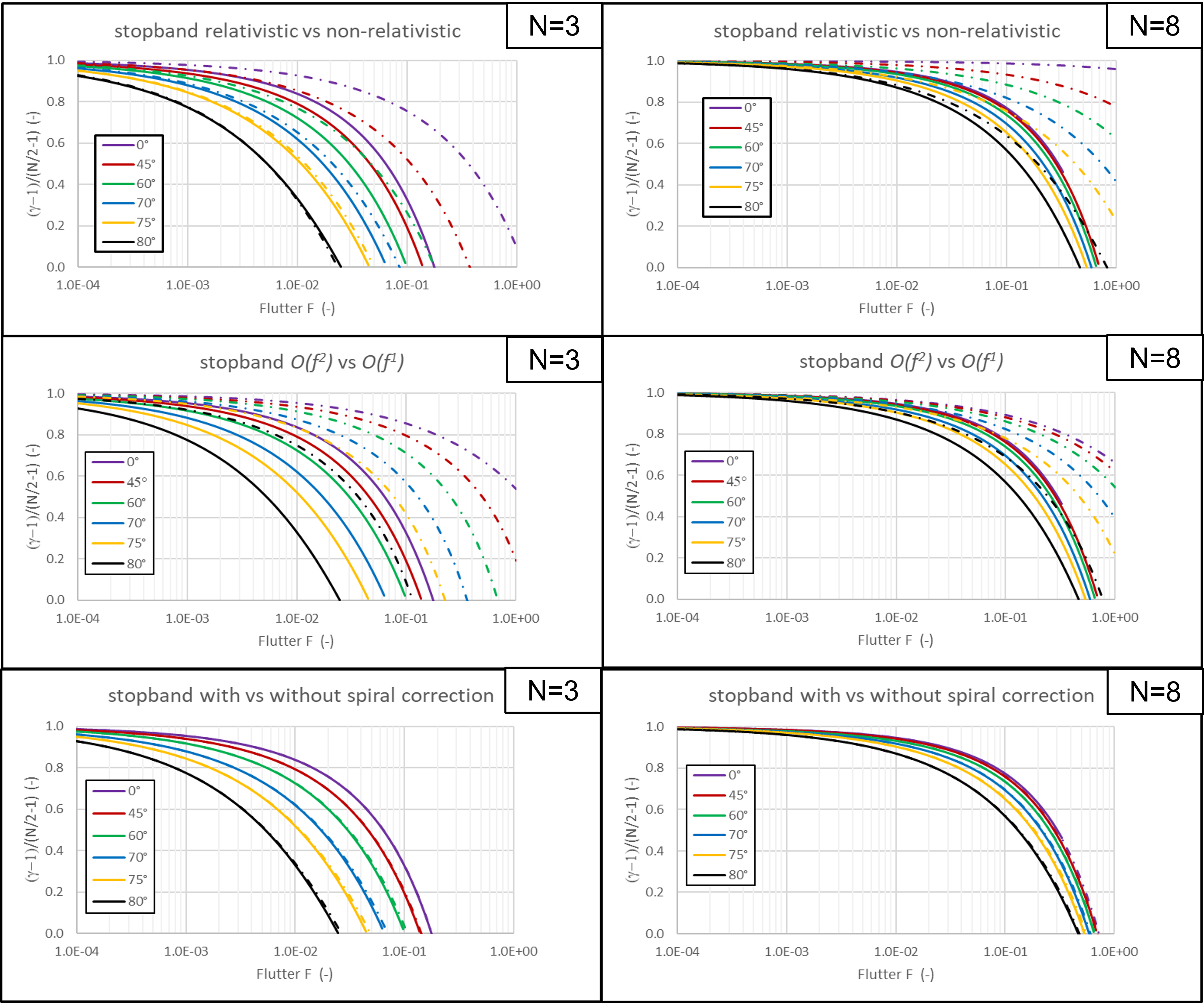}
    \caption[Impact of approximations on the limits of the half-integer resonance stopband]{\it Impact of 3 types of approximations/refinements on the calculated limits of the half-integer resonance stopband}
    \label{stopband-approximations}
\end{figure}

\noindent Figure~\ref{stopband-approximations} shows the impact of three different types of approximation on the calculated limits of the half-integer resonance stopband. The upper two figures show the differences that are due to the non-relativistic model as compared to the relativistic model. Or in other words, the improvement that is obtained by taking into account 
in the derivations the cross-terms between average field derivatives and the azimuthal field modulation. It is seen that this improvement is considerable, especially for the higher values of 
cyclotron rotational symmetry number $N$. This may be expected because higher $N$-value corresponds with higher stopband energies and thus higher values
of the field-derivatives (see Figure~\ref{derivatives}). The effect of the resonance is substantially under-estimated for the non-relativistic derivation. The middle two figures show the differences that are due to second order model ($\mathcal{O}(f^2)$) as compared to the first order model ($\mathcal{O}(f)$). Or in other words, the improvement that is obtained by taking into account terms up to $\mathcal{O}(f^2)$. It is seen that this improvement is considerable, both for lower $N$-values and higher $N$-values. It is seen that the impact of the resonance
is under-estimated if $\mathcal{O}(f^2)$ terms are ignored.

\noindent The lower two cases in Figure~\ref{stopband-approximations} show the differences that are due to use if the corrected spiral angle. This correction, as discussed in appendix~\ref{spiralcorrection}, allowed for a better agreement between the numerical C400 vertical tune function and the analytical prediction (as shown in Figure~\ref{vtunecomp}).
However, for the stopband limits this refinement only has a minor impact.

\section{Energy limit of an isochronous cyclotron}\label{isolimit}

In the previous paragraph we derived the stability zone of the isochronous cyclotron resulting from the half-integer resonance $2\nu_r=N$. It was seen that the 
stopband lower limit $\gamma_1$
can be increased by lowering the flutter $F$ or the sector spiral angle $\varphi'$. However in doing so, the vertical tune will decrease and the cyclotron may become vertically unstable.
Besides the resonance limit, there is also an energy limit due to lack of vertical focusing. This limit is determined by the condition $\nu_z=0$ and can be calculated from Eq.~(\ref{NUXfinal2})
by inserting $\nu_x^2=0$ and $\bmu'_{rel}=\gamma^2-1$. Since the tune coefficients $\hat{a}_N,\hat{b}_N$ also depend on $\gamma$, the resulting equation
is an implicit equation for $\gamma$. We solve it by the iterative method of successive substitution. The dominant tune coefficient $\hat{b}_N$ depend only weakly on $\gamma$ 
and thereforeonly a few iterations are needed (a maximum of 4 for the lowest spiral angle of 45°). 

\begin{figure}[!bht]
   \vspace*{-.5\baselineskip}
   \centering
   \includegraphics[width=\textwidth]{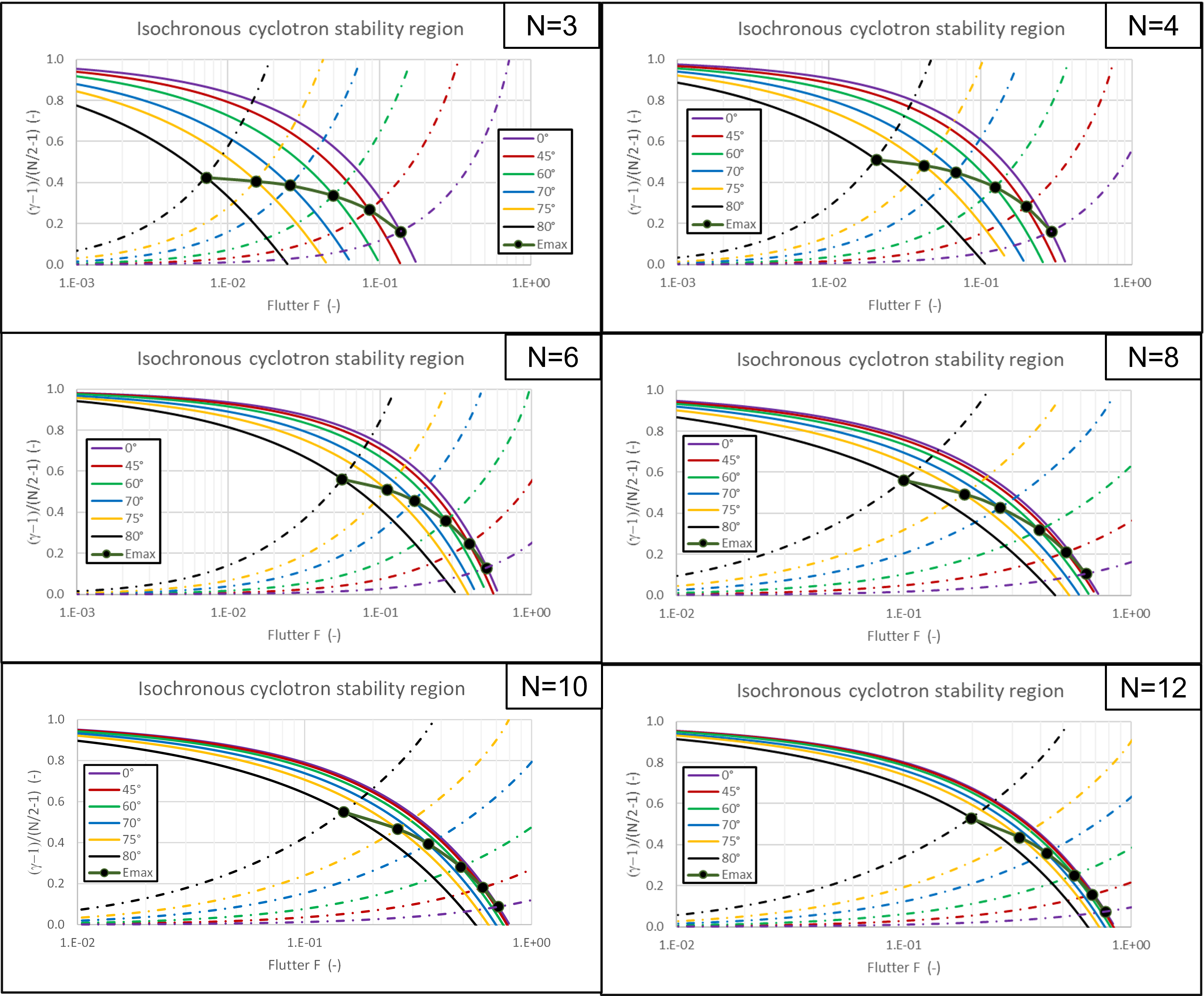}
    \caption[Stability diagram of the isochronous cyclotron]{\it Stability diagram of the isochronous cyclotron}
    \label{stability-limit}
\end{figure}

Figure~\ref{stability-limit} shows in one plot both the resonance limits (solid lines) and the vertical focusing limits (dashed lines)
as function of the flutter $F$. The different cases shown and also the axes units are the same as used in Figure~\ref{FigSB}.
It is seen that the focusing limit increases monotonically with increasing flutter and increasing spiral angle. The normalized
limit $(\gamma-1)/(\tfrac{N}{2}-1)$ decreases monotonically with increasing N-number.
In order to have a stable cyclotron, the operating point as defined by a given flutter, spiral angle and $\gamma$-value must be below  the corresponding solid lines and the corresponding dashed lines in Figure ~\ref{stability-limit}. It should be remembered that the lines itself represent extreme limits of stability and in practice sufficient distance must be taken. For
the vertical tune one could require for example a minimum value $\nu_{min}>0 $. In this case the dashed line in the plot will shift down by the amount:

\begin{align}\label{nu-correction}
	\Delta\gamma \approx -\frac{\nu_{min}^2}{2\gamma_0}\ ,
\end{align}

\noindent where $\gamma_0$ is the energy limit as given by the dashed line in Figure~\ref{stability-limit}.

\begin{figure}[!bht]
    \vspace*{-.5\baselineskip}
    \centering
    \includegraphics[width=\textwidth]{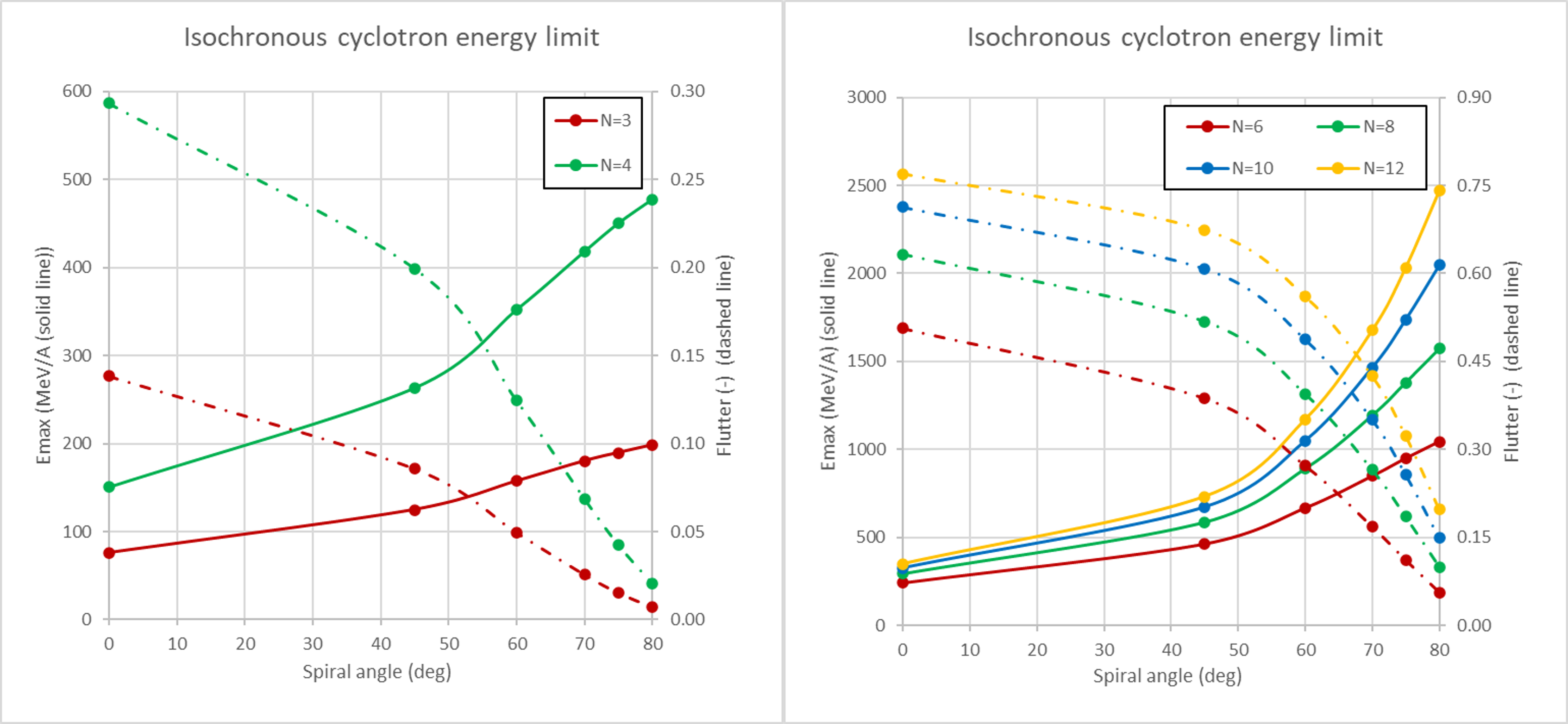}
    \caption[Energy limit of the isochronous cyclotron]{\it Energy limit of the isochronous cyclotron}
    \label{energy-limit}
\end{figure}

\begin{table}[!bht]
\centering
\caption{Energy limits of an isochronous cyclotron}
\begin{tabular}{|c|c|c|c|c|c|c|} \hline
    &\multicolumn{2}{|c|}{N=3}&\multicolumn{2}{|c|}{N=4} &\multicolumn{2}{|c|}{N=6} \\ \hline
$\xi$ (deg)&	F (-) &	T (MeV/u)&	F (-) &	T (MeV/u)&	F (-) &	T (MeV/u) \\ \hline
0&	0.1384&	75.7&	 	0.2934&	151&	0.5063&	243	 \\
45&	0.0858&	125&	0.1995&	263&	0.387&	464 \\
60&	0.0495&	157&	0.1245&	352&	0.272&	668\\
70&	0.0257&	180&	0.0686&	418&	0.168&	850 \\
75&	0.0153&	190&	0.0424&	450&	0.111&	950 \\
80&	0.0072&	198&	0.0204&	477&	0.056&	1045 \\ \hline
&\multicolumn{2}{|c|}{N=8}&\multicolumn{2}{|c|}{N=10} &\multicolumn{2}{|c|}{N=12} \\ \hline
0&		0.6324&	294&	0.7135&	326&	0.7697&	348 \\
45&		0.5180&	587&	0.6085&	673&	0.6741&	732 \\
60&		0.3945&	891&	0.4880&	1049&	0.5607&	1170 \\
70&		0.2653&	1193&	0.3510&	1465&	0.4250&	1677 \\
75&		0.1852&	1380&	0.2571&	1738&	0.3231&	2034 \\
80&		0.1000&	1572&	0.1488&	2047&	0.1975&	2471 \\ \hline
\end{tabular}
\label{table1}
\end{table}

At the intersection between solid and dashed lines the highest achievable energy is found for a given symmetry number $N$ and a given spiral angle. These points are shown as black dots
in Figure~\ref{stability-limit}. Figure~\ref{energy-limit} shows these energy limits (solid lines) as a function of the design spiral angle and  for the same N-numbers as used before. These
are kinetic energies expressed in MeV per nucleon. The graphs also show the corresponding flutter values (dashed lines) that are required to achieve these limits. 
The numerical data are given also in Table~\ref{table1}. 
\noindent The energy limits are the absolute limits for the isochronous cyclotron as dictated by the beam dynamics of these machines. In practice there are of course other limits determined by technology.

\begin{figure}[!bht]
    \vspace*{-.5\baselineskip}
    \centering
    \includegraphics[width=\textwidth]{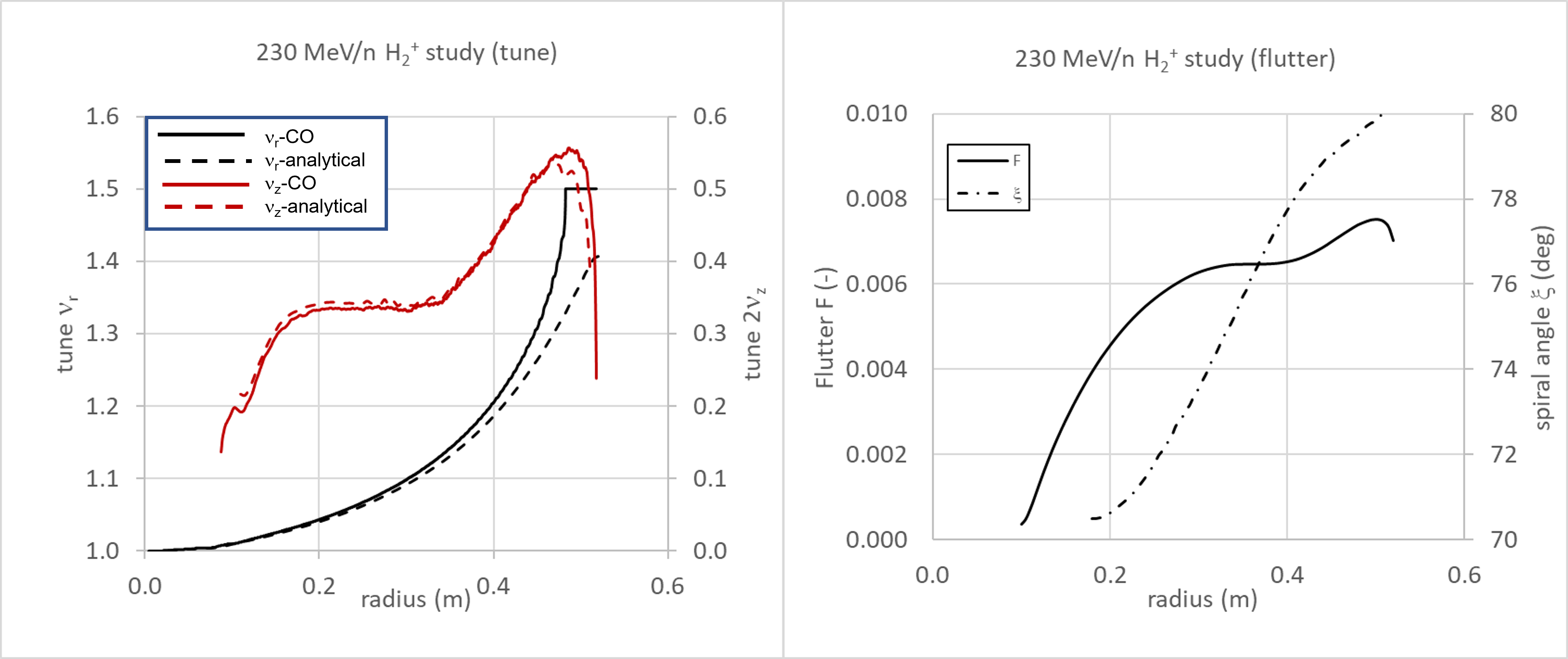}
    \caption[Example for a 230 MeV/u $H_2^+$ cyclotron study]{\it Example for a 230 MeV/u $H_2^+$ cyclotron.}
    \label{tune-H2plus}
\end{figure}

Figure~\ref{tune-H2plus} shows the tunes for a $H_2^+$ cyclotron with symmetry N=3, that has been studied at IBA. 
The left figure shows the radial tune and vertical tune (2x) obtained from a numerical closed orbit code (black-solid and red solid respectively), 
and also the radial tune (black-dashed) and vertical tune (2x, red-dashed) calculated analytically from  Eq.~(\ref{NUXfinal2}) and  Eq.~(\ref{NUZfinal2}) respectively.
In this example, the half-integer resonance hits at the radius of 48.2 cm, corresponding with an energy of 187.5 MeV/u and a vertical tune value of $\nu_z$=0.27.
The right figure shows the flutter $F$ and the spiral angle $\xi$ of the magnetic field. At the resonance energy they are F=0.0074 and $\xi$=79.5° respectively. 
Table~\ref{table1} shows an extreme energy for  $\xi$=80° of 198 MeV/u. Correcting this value for the non-zero
vertical tune (=0.27), using Eq.~(\ref{nu-correction}), we obtain the stopband energy at E=186.5 MeV/u. This is extremely close to the numerical result of 187.5 MeV/u.

\appendix

\section{The cyclotron Hamiltonian}\label{Hcyclo}

\renewcommand{\theequation}{\thesection\arabic{equation}}
\setcounter{equation}{0}

We use a polar coordinate  system ($\theta, r, z$)  that in this sequence is chosen to be right-handed. Then a positively charged particle moves in the 
positive $\theta$-direction if the average magnetic field, pointing along the z-direction, is positive. The canonical conjugate variables in polar coordinates are:

\begin{align}
	-E\hspace{2.95cm}   &; t\ ,	\\
	P_{\theta}=mrv_{\theta}+qrA_{\theta} &; \theta \ , \\
	P_{r}=mv_{r}+qA_{r}\hspace{0.45cm} &; r \ , \\
	P_{z}=mv_{z}+qA_{z}\hspace{0.45cm} &; z \ .
\end{align}

\noindent Here $E$ is the total energy of the particle, $m$ its relativistic mass, $q$ its charge, ($v_\theta, v_r, v_z$) the polar velocity components, ($P_{\theta}, P_{r}, P_{z}$)
the canonical momenta and ($A_{\theta}, A_{r}, A_{z}$)  the components of the magnetic vector potential. The magnetic field $\vec{B}$ is obtained from the vector potential via:

\begin{equation}
	\vec{B}=\nabla\times\vec{A}\ . \label{curl}
\end{equation}

\noindent The kinetic momentum $P_0$ of a particle is given by:

\begin{equation}
	P_0=mv=\sqrt{(P_{\theta}/r-qA_{\theta})^2+(P_r-qA_r)^2+(P_z-qA_z)^2}\ .	\label{Pmod}
\end{equation}

\noindent Throughout this paper, we consider the motion in a static magnetic field only (no electric fields). In this case the kinetic momentum $P_0$ is a constant of motion. 
Chosing $\theta$ as the independent variable, the Hamiltionian $\mathcal{H}$ is equal to $-P_{\theta}$. $\mathcal{H}$ can be solved from Eq.~(\ref{Pmod}) giving:

\begin{equation}\label{hamini}
	\mathcal{H} =-P_{\theta}=-r\sqrt{P_0^2-(P_r-qA_r)^2+(P_z-qA_z)^2}-qrA_{\theta}\ .
\end{equation}

We have some freedom in the choice of $\vec{A}$ and take a potential for which $A_r \equiv 0$. The expressions for $A_{\theta}$ and $A_z$ then become:

\begin{align}
A_{\theta}(\theta,r,z) &= -\frac{1}{r}\int^r r^\prime B_z(\theta,r^\prime,z) \,dr^\prime \ ,\\
A_z(\theta,r,z) &= \int^r B_{\theta}(\theta,r^\prime,z) \,dr^\prime \ .
\end{align}

\noindent Its is easily verified with Eq.~(\ref{curl}) and the divergence law $\nabla\cdot\vec{B}=0$ 
that this definition of the vector potential gives the correct result for all three magnetic field components.

\noindent We expand the magnetic field with respect to the $z$-coordinate and assume that the median plane ($z=0$) is a symmetry plane. We get:

\begin{align}
B_{\theta}(\theta,r,z) &= \frac{z}{r}\frac{\partial B}{\partial\theta} +\mathcal{O}(z^3) \ , \\
B_{r}(\theta,r,z) &= z\frac{\partial B}{\partial r} +\mathcal{O}(z^3) \ , \\
B_{z}(\theta,r,z) &= B(\theta,r)-\frac{1}{2}z^2\Delta B(\theta,r)  +\mathcal{O}(z^4) \ .
\end{align}

\noindent Here $B(\theta,r)=B_{z}(\theta,r,0)$ is the median plane field and $\Delta B$ is the 2D laplacian of $B$ in the median plane:

\begin{equation}
	\Delta B(\theta,r)=\frac{1}{r\partial r}(r\frac{\partial B}{\partial r})+\frac{1}{r^2}\frac{\partial^2 B}{\partial\theta^2} \ .
\end{equation} 

In our development of the Hamiltonian we neglect terms that envolve vertical phase space variables of higher than quadratic degree. With this simplification, the final
Hamiltonian describes linear vertical motion. For the radial motion no such simplification is made.
Almost throughout this paper the motion of the particle is analyzed in the neighborhood of a circle with radius $r_0$, where $r_0$ is related to the constant
momentum $P_0$ of the particle via:

\begin{equation}
	P_0=qr_0\bar{B}(r_0)\ .	\label{Pref}
\end{equation} 

\noindent Here $\bar{B}$ is the average magnetic field around the circle. In order to facilitate the analysis, we introduce new reduced variables
with the following normalizations:

\begin{align}
	x=\frac{r-r_0}{r_0} &;\hspace{1cm} \tilde{p}_x=\frac{P_r}{P_0}\ , \label{norm1}\\
	\zeta=\frac{z}{r_0}\hspace{0.7cm} &;\hspace{1cm} \tilde{p}_z=\frac{P_z}{P_0}\ . \label{norm2}
\end{align}

\noindent The Hamiltonian must be adjusted accordingly; using Eqs.~(\ref{jacob1},\ref{jacob2}) we find for the new Hamiltonian:

\begin{equation}
	\mathcal{K}=\frac{\mathcal{H}}{r_0 P_0}\ . \label{norm3}
\end{equation} 
 
\noindent We also define the reduced median plane magnetic field $\mu$ (around $r_0$) as follows:

\begin{equation}
	\mu(\theta,r) = \frac{B(\theta,r)}{\bar{B}(r_0)}=\frac{B(\theta,r_0+r_0 x)}{\bar{B}(r_0)} \ .
\end{equation} 

\noindent With this normalization the vector potential terms in Eq.~(\ref{hamini}) become:

\begin{align}
	rA_{\theta} &= -r_0^2\bar{B}(r_0)\left[\int^x (1+x^\prime)\mu(\theta,x^\prime)\,dx^\prime -\frac{1}{2}\zeta^2 \left((1+x) \frac{\partial\mu}{\partial x}+\int^x\frac{1}{1+x^\prime}\frac{\partial^2\mu}{\partial\theta^2}dx^\prime\right) \right]\ , \\ 
	A_z &= r_0\bar{B}(r_0)\zeta \int^x \frac{1}{1+x^\prime}\frac{\partial\mu}{\partial\theta}dx^\prime\ .
\end{align}

\noindent Inserting these expressions into Eq.~(\ref{hamini}) and applying the normalizations defined in Eqs.~(\ref{Pref}-\ref{norm3}) we find for the new Hamiltonian:

\begin{align}
	\mathcal{K} =&-(1+x)\sqrt{1-\tilde{p}_x^2-\left(\tilde{p}_z-\zeta\int^x \frac{1}{1+x^\prime}\frac{\partial\mu}{\partial\theta}dx^\prime\right)^2} \\
		&+\int^x (1+x^\prime)\mu(\theta,x^\prime)\,dx^\prime  
		-\frac{1}{2}\zeta^2 \left((1+x) \frac{\partial\mu}{\partial x}+\int^x\frac{1}{1+x^\prime}\frac{\partial^2\mu}{\partial\theta^2dx^\prime}\right)\ . \nonumber
\end{align}

\noindent Due to our choice of the vector potential the radial canonical momentum $P_r$ is equal to the radial kinetic momentum $mv_r$ and therefore the normalized momentum
$\tilde{p}_x$ is equal to the radial divergence of the particle. In order to obtain the same interpretation for the vertical momentum, we apply a canonical transformation. We use  
a type 2 generating function that depends on the original coordinates $x,\zeta$ and the new momenta $p_x,p_z$ (see Eq.~(\ref{gen2})):

\begin{align}
	&G_2(x,\zeta,p_x,p_z)=xp_x+\zeta p_z+\frac{1}{2}\zeta^2\int^x\frac{1}{1+x^\prime}\frac{\partial\mu}{\partial\theta}dx^\prime \ , \\
	&\tilde{p}_x = \frac{\partial G_2}{\partial x}=p_x+\frac{\zeta^2}{2(1+x)}\frac{\partial\mu}{\partial\theta} \ ,  \\
	&\tilde{p}_z =\frac{\partial G_2}{\partial\zeta}=p_z +\zeta\int^x\frac{1}{1+x^\prime}\frac{\partial\mu}{\partial\theta}dx^\prime \ ,  \\
	&\frac{\partial G_2}{\partial\theta}=\frac{1}{2}\zeta^2\int^x\frac{1}{1+x^\prime}\frac{\partial^2\mu}{\partial\theta^2}dx^\prime \ .
\end{align}

\noindent Keeping terms up to quadratic degree in $\zeta,p_z$, we obtain for the new Hamiltonian:

\begin{equation}
 \bar{\mathcal{K}}=-(1+x)\sqrt{1-p_x^2-p_x\frac{\zeta^2}{1+x}\frac{\partial\mu}{\partial\theta}-p_z^2}+\int^x (1+x^\prime)\mu(\theta,x^\prime)dx^\prime -\frac{\zeta^2}{2}(1+x)\frac{\partial\mu}{\partial x}\ . 
\end{equation}

\noindent We expand this Hamiltonian with respect to the vertical phase space variables and keep terms up to quadratic degree in $\zeta,p_z$. This gives:


\begin{empheq}[box=\widefbox]{align}
 \bar{\mathcal{K}}=-(1+x)(1-p_x^2)^{1/2}+\int^x (1+x^\prime)\mu(\theta,x^\prime)dx^\prime \hspace{2cm} \label{ham4d} \\
+\frac{(1+x)}{2\sqrt{1-p_x^2}}p_z^2 + \frac{1}{2}\left(\frac{p_x}{\sqrt{1-p_x^2}}\frac{\partial\mu}{\partial\theta}-(1+x)\frac{\partial\mu}{\partial x}\right)\zeta^2\ .
\end{empheq}

\noindent Since we have assumed a symmetric median plane, $\zeta=p_z=0$ is a valid solution of Eq.~(\ref{ham4d}). For this solution we can define the 2D
Hamiltonian $H_x$ describing the median plane radial motion; it is given by:

\begin{equation}
\boxed{H_x=-(1+x)(1-p_x^2)^{1/2}+\int^x (1+x^\prime)\mu(\theta,x^\prime)\,dx^\prime\ . }\label{hamx}
\end{equation} 

\noindent If at the same time the vertical excursion from the median plane $\zeta$ is small, the influence of the vertical motion on the radial motion is negligible, and we 
may consider $x,p_x$ as given functions of $\theta$ and define the 2D Hamiltonian $H_z$ for the vertical motion. 

\begin{equation}
\boxed{H_z= \frac{(1+x)}{2\sqrt{1-p_x^2}}p_z^2 + \frac{1}{2}\left(\frac{p_x}{\sqrt{1-p_x^2}}\frac{\partial\mu}{\partial\theta}-(1+x)\frac{\partial\mu}{\partial x}\right)\zeta^2\ }.\label{hamz}
\end{equation} 

\noindent The equations (\ref{hamx}) for  $H_x$ and (\ref{hamz}) for  $H_z$ agree with respectively Eq.~(4.3) and Eq.~(10.2) in the Hagedoorn-Verster paper\cite{HV-62}. 


\section{The median plane magnetic field}

\renewcommand{\theequation}{\thesection\arabic{equation}}
\setcounter{equation}{0}

The motion of the particle is dertemined by the shape of the median plane magnetic field $B(\theta,r)$. This field can be separated in an average part $\bar{B}(r)$ and an
oscillating part. This part represents the azimuthal variation of the field which we exand in a Fourier series. We  write $B(\theta,r)$ as:

\begin{equation} \label{bfield}
	B(\theta,r) = \bar{B}(r) + \ssum_n \mathcal{A}_n(r)\cos n\theta + \mathcal{B}_n(r)\sin n\theta.
\end{equation}

\noindent In our analysis we assume that the cyclotron has perfect $N$-fold symmetry. In this case only terms with $n=kN, k=1,2,\dots$ will be present in the Fourier series. 

\subsection{The reduced magnetic field}\label{thereduced}

In this paper we analyze the orbits in the vincinity of a circle with radius $r_0$ (see Eq.~(\ref{Pref})) and define the reduced magnetic field $\mu$ (around $r_0$) as follows:

\begin{equation}\label{breduced}
	\mu(\theta,r) = \frac{B(\theta,r)}{\bar{B}(r_0)}=\frac{B(\theta,r_0+r_0 x)}{\bar{B}(r_0)} \ .
\end{equation} 

\noindent Here $x$ has been defined in Eq.~(\ref{norm1}). 

\noindent Using  Eqs.~(\ref{bfield},\ref{breduced}), we can write the reduced field as:

\begin{align}
	\mu(\theta,r) = \bar{\mu}(r) + f(\theta,r)\ ,
\end{align}

\noindent where $\bar{\mu}(r)$ and $ f(\theta,r)$ are defined as :

\begin{align}
	\bar{\mu}(r) &= \bar{B}(r)/\bar{B}(r_0)\ , \\
	f(\theta,r) &= \ssum_n A_n(r)\cos n\theta + B_n(r)\sin n\theta\ , \label{Fseries}
\end{align}

\noindent and with:

\begin{equation}
	A_n(r) = \mathcal{A}_n(r)/\bar{B}(r_0)\ , \hspace{0.5cm}	B_n(r) =\mathcal{B}_n(r)/\bar{B}(r_0)\ .
\end{equation}

\noindent The Fourier series in Eq.~(\ref{Fseries}) can also be written in terms of amplitude and phase of the harmonics as:

\begin{equation}
	f(\theta,r) = \ssum_n C_n(r)\cos n(\theta-\varphi_n(r))\ ,
\end{equation}

\noindent where $C_n,\varphi_n$ relate to $A_n,B_n$ as:

\begin{align}
	\An(r) &= \Cn(r)\cos\varphi_n\ , \label{An}\\
	\Bn(r) &= \Cn(r)\sin\varphi_n\ ,  \label{Bn}
\end{align}

\noindent We expand the reduced field $\mu(\theta,x)$ in a taylor series:

\begin{align}
\mu(\theta,x) &= 1+\bmu' x +\tfrac{1}{2}\bmu'' x^2 + \tfrac{1}{6} \bmu''' x^3 + \dots \nonumber \\
&+ \ssum_n(\An+\An'x+\tfrac{1}{2}\An''x^2+\dots)\cos n\theta \nonumber \\
&+ \ssum_n(\Bn+\Bn'x+\tfrac{1}{2}\Bn''x^2+\dots)\sin n\theta \ , \label{Taylor}
\end{align}

\noindent where:

\begin{align}
	\bmu' &=\left[\frac{d}{dx}\bmu(r_0+r_0 x)\right]_{x=0} \hspace{0.4cm}= \left[\frac{r}{\bar{B}}\frac{d\bar{B}}{dr}\right]_{r=r_0}\ , \nonumber\\
	\bmu'' &=\left[\frac{d^2}{dx^2}\bmu(r_0+r_0 x)\right]_{x=0} \hspace{0.28cm}= \left[\frac{r^2}{\bar{B}}\frac{d^2\bar{B}}{dr^2}\right]_{r=r_0}\ , \nonumber\\
	\An &= \An(r_0) \hspace{2.8cm}= \hspace{3.4cm} = \left[\frac{a_n(r)}{\bar{B}(r)}\right]_{r=r_0}\ , \label{defines}\\
	\An' &=\left[\frac{d}{dx}\An(r_0+r_0 x)\right]_{x=0} \hspace{0.18cm}= \left[ r\frac{d}{dr}\An(r)\right]_{r=r_0} \hspace{0.3cm} =  \left[\frac{r}{\bar{B}}\frac{da_n}{dr}\right]_{r=r_0}\ , \nonumber\\
	\An'' &=\left[\frac{d^2}{dx^2}\An(r_0+r_0 x)\right]_{x=0} = \left[ r^2\frac{d}{dr^2}\An(r)\right]_{r=r_0} = \left[\frac{r^2}{\bar{B}}\frac{d^2a_n}{dr^2}\right]_{r=r_0}\ , 
\nonumber
\end{align}

As an important remark, we note that our definition of the field-harmonics differs with a factor $\bar{B}(r)/\bar{B}(r_0)$ from the definition used in the HV-paper\cite{HV-62}. The relation between our 
representation $A_n$ and the HV-representation $\tilde{A}_n$is as follows:

\begin{align}
&A_n = \tfrac{\bar{B}(r)}{\bar{B}(r_0)} \tilde{A}_n\ , \nonumber\\
&A_n(r_0) =\tilde{A}_n(r_0)\ , \label{compHV}\\
&A_n' = \tilde{A}_n'+\bmu' \tilde{A}_n\ , \nonumber\\
&A_n'' = \tilde{A}_n''+2\bmu' \tilde{A}_n'+\bmu''\tilde{A}_n\ , \nonumber
\end{align}

\noindent and similar equations for the sine-components.

\begin{figure}[!bht]
   \vspace*{-.5\baselineskip}
   \centering
   \includegraphics[width=9cm]{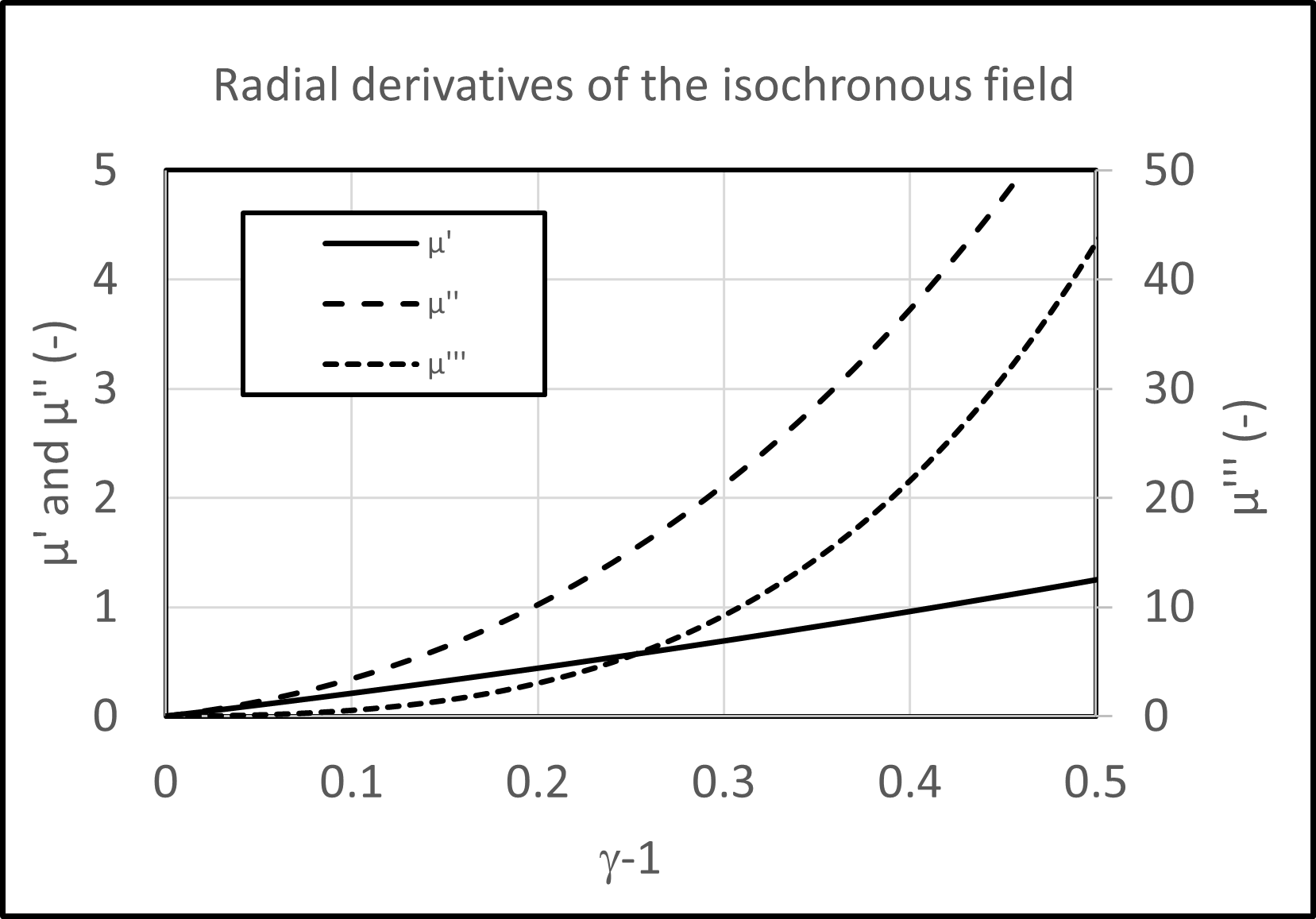}
    \caption[Derivatives of the isochronous magnetic field]{\it Normalized first and second derivatives (left scale) and third derivative (right scale) of an isochronous magnetic field.}
    \label{derivatives}
\end{figure}

\noindent The advantage of our definition is that  in the Taylor development (Eq.~(\ref{Taylor})), there are no cross-terms between derivatives of the average field and the Fourier
components. In the HV-approach, there are such cross-terms, but they have been neglected from the beginning. They were neglected not only in the magnetic field development 
but at all developments throughout their paper, with the argument 
that the average field dervatives are very small ($\mathcal{O}(f^2)$. and crossterms therefore are small up to  $\mathcal{O}(f^3)$. 
This is true for not too high particle energies but it becomes less and less valid for more relativistic energies. Figure~\ref{derivatives} shows quantities 
$\bmu',\bmu''$ and $\bmu'''$  as a function of the relativistic parameter $\gamma-1$. The value $\gamma-1=0.5$ corresponds with a kinetic energy of about 470 MeV/A. It
is seen that at this energy $\bmu'\approx 1.2$,  $\bmu''\approx 6$ and $\bmu'''\approx 43$.  Since in our study we are interested in the optics 
and stability at higher energies, we consider the derivative as functions of $\mathcal{O}(f^0)$ and therefore do not neglect the cross-terms at any moment in our development.

\subsection{Relations between magnetic field Fourier components}

Several times in our analysis we need to transform expressions using the sine/cosine representation of the azimuthal field variation into
expressions using the amplitude/phase representation. For such transformations we use the following relations:

\begin{align}
&A^2_n+B^2_n = C^2_n\ ,\label{AnAn}\\
&A_nA_n'+B_nB_n' = C_nC_n'\ ,\label{AnAnp}\\
&A_n'^2 + B_n'^2 = C_n'^2+n^2C_n^2\varphi_{n}'^{2}\ , \label{Anp2}\\
&A_nA_n''+B_nB_n''= C_n'^2-n^2C_n^2\varphi_{n}'^{2}\ . \label{AnAnpp}
\end{align}

\noindent Here $C_n',\varphi_{n}'$ are defined as shown in Eqs.~(\ref{defines}), for $A_n'$.

\subsection{The spiral angle}\label{tspiral}

\noindent Note that $\varphi'_n$ is related to the frequentlly used spiral angle $\xi_n$ as follows:

\begin{equation}
\varphi_n' =\tan\xi_n\ .
\end{equation}
 
\begin{figure}[!bht]
   \vspace*{-.5\baselineskip}
   \centering
   \includegraphics[width=9cm]{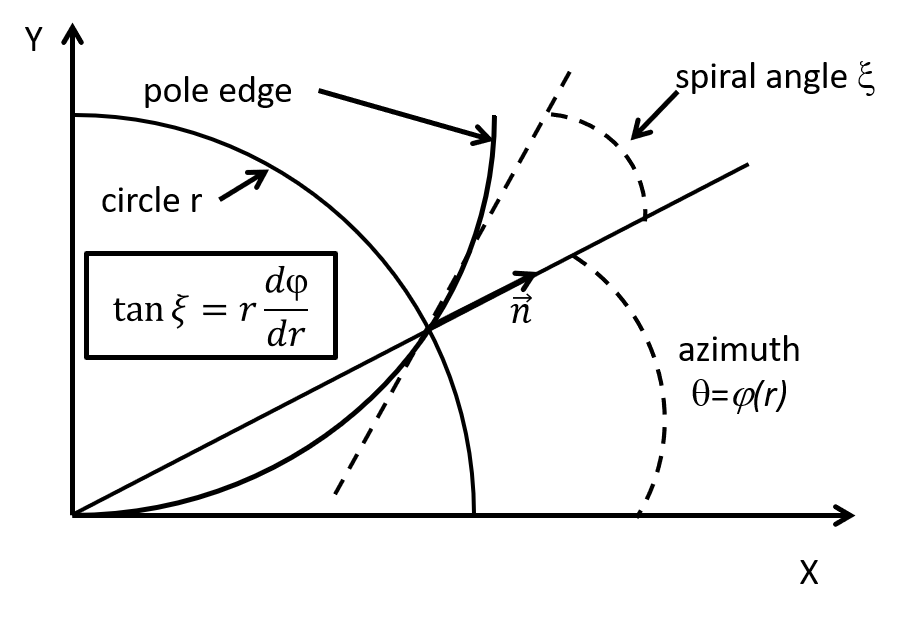}
    \caption[Definition of the spiral angle]{\it Definition of the spiral angle.}
    \label{spiral}
\end{figure}

\noindent If $\theta=\varphi_n(r)$ is the contour where the n$^{\mbox{th}}$ Fourier component is maximum, then the spiral at a given point on this contour
is defined as the angle between a radial unit vector (the vector $\vec{n}$ normal to the circle) and the tangent along the contour. In practice one often uses for $\varphi$ the contour of the entrance or exit pole edge of the sector,
or the contour of the mid-sector angle. This is illustrated in Figure~\ref{spiral}.

\subsection{Relations between Fourier harmonics and flutter}\label{RFf}

In our analysis of cyclotron optics whe derive equations for optical quantities (such as the tunes for example) 
that depend on (summations over $n$ of) the Fourier harmonics and their derivatives and on the derivatives of the average magnetic field. 
These are rather complex and also impractical equations. 
In order to simpify them we look for away to relate the higher Fourier harmonics  $(n>N)$ to the principal harmonics $(n=N) $. This envolves some approximation which however is not so significant because the optical quantities are dominantly determined by the principal harmonics and less by the higher harmonics. Making some error in the values of the higher harmonics therefore does not have a too big impact. It will lead us however to more pratical forms of the equations. In order to make such an approximation, we assume a
hard-edge profile of the azimuthal field variation. In this case the relation between the higher harmonic amplitudes $C_{kN}$ and the principal harmonic amplitude $C_N$ is given by:

\begin{equation}
C_{kN}=\frac{\sin(kN\alpha_h)}{k\sin(N\alpha_h)}C_N\ , \hspace{2cm} k=1,2,\dots \ ,
\end{equation}

\noindent where $\alpha_h$ is half of the hill angular extend.

\noindent We simplify a little bit more by assuming a symmetric structure where the hill angle is equal to the valley angle. In this case only Fourier components with $n=N,3N,5N,\dots\ ,$
are non-zero and we get:

\begin{equation}
C_{(2k+1)N} = \frac{C_N}{2k+1}(-1)^{k} \hspace{2cm} \mbox{for}\ k=0,1,2,\dots \ , \label{fcoefs} 
\end{equation}

\noindent  We can also make the relation with the frequently used flutter of the magnetic field. The flutter is defined by the equation:

\begin{equation}\label{flutter}
F(r)=\frac{\bigsymbol{\langle} B^2(\theta,r)\bigsymbol{\rangle}-\bigsymbol{\langle} B(\theta,r)\bigsymbol{\rangle}^2}{\bigsymbol{\langle} B(\theta,r)\bigsymbol{\rangle}^2}\ ,
\end{equation}

\noindent and can be expressed in the Fourier components:

\begin{equation}\label{flutter2}
F(r)=\tfrac{1}{2}\ssum_n [A^2_n(r)+B^2_n(r)] = \tfrac{1}{2}\ssum_n C^2_n(r)\ .
\end{equation}

\noindent Inserting in this expression for $F$ the Fourier components defined in Eq.~(\ref{fcoefs}) and choosing the summation coefficient as $n=(2k+1)N$ we get:

\begin{equation}
F = \tfrac{1}{2}C_N^2\ssum_{k=0}^{\infty}\frac{1}{(2k+1)^2} \ .
\end{equation}

\noindent The series in the above equation is one of the Leonard Euler series:

\begin{equation}
\ssum_{k=0}^{\infty}\frac{1}{(2k+1)^2} = \frac{\pi^2}{8}\ ,
\end{equation}

\noindent and with this we get:

\begin{equation}
F = \frac{\pi^2}{16}C_N^2\ . \label{FCN}
\end{equation}

\noindent For use elsewhere in this report, we now give the following relations between the Flutter $F$ and the (square of) the Fourier amplitudes $C_n$:

\begin{equation}
\begin{aligned}\label{CtoF}
	&\ C_n^2 &&=\frac{N^2}{n^2}C_N^2  &&=\frac{16F}{\pi^2(2k+1)^2}\ , \\
	&\ C_nC_n' &&=\frac{N^2}{n^2}C_NC_N' &&=\frac{8F'}{\pi^2(2k+1)^2}\ , \\
	&\ C_n'^2 &&=\frac{N^2}{n^2}C_N'^2 &&=\frac{4F'^2/F}{\pi^2(2k+1)^2}\ , \\
	&\ C_nC_n'' &&=\frac{N^2}{n^2}C_NC_N'' &&=\frac{8(F'+F''-F'^2/2F)}{\pi^2(2k+1)^2}\ .
\end{aligned}
\end{equation}

\noindent In the right hand side of these equations we replaced $n$ by $(2k+1)N$.

\renewcommand{\theequation}{\thesection\arabic{equation}}
\setcounter{equation}{0}

\section{Move an oscillating term to the next higher order}\label{remove}

Consider a Hamiltonian of the following form:

\begin{equation}\label{Hnorm}
 H(p,x,\theta) = \tfrac{1}{2}p^2 + \tfrac{1}{2}(\nu_0^2+f(\theta))x^2\ .
\end{equation}

\noindent Here we assume the function $f$ to be an oscillating function ($\langle f\rangle=0$) with a small oscillating amplitude. 
The parameter $\nu_0$ may be considered as the zero-order tune of the oscillation. We want to design a canonical transformation 
which removes the function $f$ from the Hamiltonian up to first order $\mathcal{O}(f)$. The new Hamiltonian may have oscillating terms of $\mathcal{O}(f^2)$, which we
consider small enough to be negligible. We will keep constant ($\theta$-independent) terms up to $\mathcal{O}(f^2)$. When this has been achieved the motion is solved 
(up to $\mathcal{O}(f^2)$) as the Hamiltonian has become a constant. We note that the transformation has been given in the HV-paper\cite{HV-62}, but only for the case where $\nu_0^2$
itself is also a small quentity of $\mathcal{O}(f^2)$. In our cases of interest, this is not always true and therefore we generalize the transformation.
We write for the generating function:

\begin{equation} \label{transfo1}
G=G_3(p,\bar{x})=-p\bar{x}+\tfrac{1}{2}a(\theta)\bar{x}^2+b(\theta)\bar{x}p+\tfrac{1}{2}c(\theta)p^2\ ,
\end{equation}

\noindent where $a(\theta),b(\theta),c(\theta)$ are yet unknown periodic functions. They will be dertermined by requiring that
in the new Hamiltonian $\bar{H}$, the $\mathcal{O}(f)$ oscillating part
is removed. For that purpose we first carry out the transformation up to $\mathcal{O}(f)$. We obtain from the above generating function:

\begin{eqnarray*}
	x &=& -\frac{\partial G_3}{\partial p} = \bar{x}-b\bar{x}-cp\ , \\
	\bar{p} &=& -\frac{\partial G_3}{\partial\bar{x}} = p-a\bar{x}-bp\ , \\
	\frac{\partial G_3}{\partial\theta} &=& +\tfrac{1}{2}\dot{a}\bar{x}^2+\dot{b}\bar{x}p+\tfrac{1}{2}\dot{c}p^2\ .
\end{eqnarray*}

\noindent Up to first order $\mathcal{O}(f)$, we obtain for the new Hamiltonian:

\begin{equation}
	\bar{H}=\tfrac{1}{2}(1+2b+\dot{c})\bar{p}^2+(a-\nu_0^2c+\dot{b})\bar{x}\bar{p}+\tfrac{1}{2}(\nu_0^2-2b\nu_0^2+f+\dot{a})\bar{x}^2\ . 
\end{equation}

\noindent For all first order terms to be zero, the functions $a,b,c$ must obey the following relations:

\begin{eqnarray}
	4\nu_0^2\dot{c}+\dddot{c} &=& -2f(\theta)\ , \label{set1}\\
	b &=& -\tfrac{1}{2}\dot{c}\ , \label{set2}\\
	a &=& \nu_0^2c+\tfrac{1}{2}\ddot{c}\label{set3}\ .
\end{eqnarray}

\noindent We now carry out the transformation Eq~(\ref{transfo1}) up to second order ($\mathcal{O}(f^2)$). In this approximation we get the following relations:

\begin{eqnarray*}
	x &=& (1-b-ac)\bar{x}-c(1+b)\bar{p}\ , \\
	p &=& a(1+b)\bar{x} +(1+b+b^2)\bar{p}\ , \\
	\tfrac{\partial G_3}{d\theta} &=& \tfrac{1}{2}\dot{a}\bar{x}^2+\dot{b}\bar{x}(\bar{p}+a\bar{x}+b\bar{p})+\tfrac{1}{2}\dot{c}(\bar{p}^2+2a\bar{x}\bar{p}+2b\bar{p}^2)\ ,
\end{eqnarray*}

\noindent and find for the new Hamiltonian:

\begin{equation}
 \bar{H}=\tfrac{1}{2}(1+3b^2+\nu_0^2c^2+2b\dot{c})\bar{p}^2+\tfrac{1}{2}\left[a^2+\nu_0^2(1+b^2-2ac)-2bf+2a\dot{b}\right]\bar{x}^2\ . 
\end{equation}

\noindent We bring this Hamiltonian to its normal form using the method explained in Appendix~\ref{normform}. We find:

\begin{equation}
 \bar{H}=\tfrac{1}{2}\bar{p}^2+\tfrac{1}{2}\bar{x}^2\left[\nu_0^2+a^2-2bf+2a\dot{b}+\nu_0^2(b^2-2ac+3b^2+\nu_0^2c^2+2b\dot{c})\right]\ .
\end{equation}

\noindent We insert the expressions for a and b from Eqs.~(\ref{set2},\ref{set3}) and get:

\begin{equation}
 \bar{H}=\tfrac{1}{2}\bar{p}^2+\tfrac{1}{2}(\nu_0^2-\tfrac{1}{4}\ddot{c}^2-\nu_0^2c\ddot{c}+\dot{c}f)\bar{x}^2\ .
\end{equation}

\noindent We use the differential equation for $c$ (Eq.~(\ref{set1})) and apply partial integration to re-write $c\ddot{c}=-\dot{c}^2$ and $\dot{c}\dddot{c}=-\ddot{c}^2$ 
and obtain for the Hamiltonian:

\begin{equation}
 \bar{H}=\tfrac{1}{2}\bar{p}^2+\tfrac{1}{2}\left[\nu_0^2+\tfrac{1}{2} \langle \dot{c}f \rangle \right]
\end{equation}

\noindent Here we only kept the average part of the second order terms and neglected their oscillating parts.

\noindent The function $f$ is periodic in $\theta$ and can be expanded into a Fourier series:

\begin{equation}
f(\theta) = \ssum_{n=1}^{\infty} a_n\cos n\theta +b_n\sin n\theta\ .
\end{equation}

\noindent Inserting this expression in Eq.~(\ref{set1}), we can solve for the periodic solution of the function $c$. For $\dot{c}$ we obtain:

\begin{equation}
\dot{c}(\theta) = 2\ssum_{n=1}^{\infty} \frac{a_n\cos n\theta +b_n\sin n\theta}{n^2-4\nu_0^2}\ .
\end{equation}

\noindent and our final Hamiltonian becomes:

\begin{equation}\label{Hfinal}
 \bar{H}=\tfrac{1}{2}\bar{p}^2+\tfrac{1}{2}[\nu_0^2+\tfrac{1}{2}\ssum_{n=1}^{\infty}\frac{c_n^2}{n^2-4\nu_0^2}]\bar{x}^2\ .
\end{equation}

\noindent where $c_n$ is the amplitude of the n$^{\mbox{th}}$ Fourier component:

\begin{equation}
	c_n=\sqrt{a_n^2+b_n^2}\ .
\end{equation}

\noindent The Hamiltonian does not depend on $\theta$ anymore and therefore the motion can be considered as solved. The square of the tune $\nu_x$ of the motion
is given by:

\begin{equation}\label{tune1}
\boxed{\nu_x^2 = \nu_0^2 +\tfrac{1}{2}\ssum_{n=1}^{\infty}\frac{c_n^2}{n^2-4\nu_0^2}\ .}
\end{equation}

\noindent In paragraphs~\ref{hmot} and \ref{vermot} we use the above results to find the vertical and radial tunes of the isochronous cyclotron.

\noindent  It is seen that if the zero-order tune $\nu_0$ aproaches the value of $n/2$, the tune $\nu_x$ diverges to infinity. This is a case where the motion dynamics is close
to the half-integer resonance. In that case the Hamiltonian of Eq.~(\ref{Hfinal}) does no longer describe the motion correctly. In the next paragraph this special case
will be analyzed in more detail.


\renewcommand{\theequation}{\thesection\arabic{equation}}
\setcounter{equation}{0}

\section{The half-integer resonance}\label{halfint}

We consider again the Hamiltonian of the form as given in Eq.~(\ref{Hnorm}):

\begin{equation}
 H(p,x,\theta) = \tfrac{1}{2}p^2 + \tfrac{1}{2}[\nu_0^2+f(\theta)]x^2\ .
\end{equation}

\noindent where as before the function $f$ is an oscillating function ($\langle f \rangle=0$) with a small oscillating amplitude 
and parameter $\nu_0$ is the zero-order tune of the oscillation. 
\noindent We now study this motion in a different (more general) way such that the result is also valid when the zero-order tune is close to a half-integer $\nu_0\approx n/2$. 
Hereto we introduce action-angle variables $I,\phi$ in a rotating phase space:

\begin{eqnarray}
p &=& \sqrt{2I\nu_0}\sin(\phi-k\theta)\ , \\
x &=& \sqrt{2I/\nu_0}\cos(\phi-k\theta)\ .
\end{eqnarray}

\noindent Here $\phi$ plays the role of new momentum and $I$ the role of new coordinate. The parameter $k$ is an integer or a half-integer. 
We are especially interested in the case $k=N/2$, where $N$ is symmetry number of the periodic function $f$. But for comparisson
with the previous paragraph~\ref{remove} we also allow the values $k=0$ and $k=1$.
The canonical transformation is obtained from the following type-2 generating function:

\begin{align}
 G &= G_3(x,\phi)=\tfrac{1}{2}\nu_0x^2\tan(\phi-k\theta)\ , \\
\frac{\partial G_3}{\partial\theta} &= -\tfrac{1}{2}\frac{k\nu_0x^2}{\cos^2(\phi-k\theta)} = -kI\ ,
\end{align}

\noindent and the new Hamiltonian becomes:

\begin{equation}
 K(\phi,I,\theta) = I[\nu_0-k+\frac{f(\theta)}{\nu_0}\cos^2(\phi-k\theta)]\ .
\end{equation}

\noindent We write this Hamiltonian in the following form:

\begin{equation}
 K(\phi,I,\theta) = I[a(\phi)+f_2(\phi,\theta)]\,.
\end{equation}

\noindent where $a(\phi)$ and $f_2(\phi,\theta)$ are defined as:

\begin{eqnarray}
a(\phi) &=& a_0 +\frac{1}{\nu_0}\mathlarger{\langle} f(\theta)\cos^2(\phi-k\theta)\mathlarger{\rangle}\ , \\
f_2(\phi,\theta) &=& \frac{1}{\nu_0}\mbox{osc}(f(\theta)\cos^2(\phi-k\theta))\ , \\
a_0 &=& \nu_0-k\ . 
\end{eqnarray}

\noindent We want to design a canonical transformation 
which removes the oscillating function $f_2$ from the Hamiltonian up to first order $\mathcal{O}(f)$. The new Hamiltonian may have oscillating terms of $\mathcal{O}(f^2)$, which we
consider small enough to be negligible. We will keep constant ($\theta$-independent) terms up to $\mathcal{O}(f^2)$. When this has been achieved the motion is solved 
(up to $\mathcal{O}(f^2)$) as the Hamiltonian has become a constant. We note that the transformation has been given in the HV-paper\cite{HV-62}, but only for the case where $a$
itself is also a small quantity of $\mathcal{O}(f^2)$. In our cases of interest, this is not true and therefore we generalize the transformation.
We write for the generating function:

\begin{eqnarray*}
G &=& G_3(\phi,\bar{I})=-\bar{I}[\phi+U_2(\phi,\theta)]\ , \\
I &=& -\frac{\partial G}{\partial\phi}=\bar{I}(1+\frac{\partial U_2}{\partial\phi})\ , \\
\bar{\phi} &=& -\frac{\partial G}{\partial\bar{I}}=\phi+U_2(\phi,\theta)\ , \\
\tfrac{\partial G_3}{d\theta} &=& -\frac{\partial U_2}{\partial\theta}\ .
\end{eqnarray*}

\noindent  Here $U_2$ is a  yet unknown periodic function which will be dertermined by requiring that in the new Hamiltonian $\bar{K}$, the $\mathcal{O}(f)$ oscillating part is removed.
We first calculate $\bar{K}$ as afunction of $\bar{I}$ and the old momentum $\phi$:

\begin{equation}\label{barK}
\bar{K}=\bar{I}[a(\phi)+f_2(\phi,\theta)+a\frac{\partial U_2}{\partial\phi}+f_2\frac{\partial U_2}{\partial\phi}-\frac{\partial U_2}{\partial\theta}]\ .
\end{equation}

\noindent So, in order to remove the first order oscillating part $f_2$ from the Hamiltonian, we must define $U_2$ by the following equation:

\begin{equation}\label{Utwo}
 \frac{\partial U_2}{\partial\theta}-a_0\frac{\partial U_2}{\partial\phi}=f_2(\phi,\theta)\ .
\end{equation}

\noindent Note that here we have replaced $a$ by $a_0$, because the difference generates an oscillating term of $\mathcal{O}(f^2)$, which we neglect. With the same reasoning
we can (now that the first order part has been removed) replace in Eq.~(\ref{barK}) $\phi$ by $\bar{\phi}$. We get for the final Hamiltonian the following form:

\begin{equation}\label{barK2}
\bar{K}=\bar{I}[a(\bar{\phi})+\bigsymbol{\langle} f_2\frac{\partial U_2}{\partial\bar{\phi}}\bigsymbol{\rangle}]\ .
\end{equation}

\noindent In order to elaborate this expression furher, we need to find the expressions for $a(\phi)$ and $f_2(\phi,\theta))$ and then solve $U_2$ from Eq.~(\ref{Utwo}).
As we did in Appendix~\ref{remove}, we expand the function $f(\theta)$ in a Fourier series. For the moment however, we represent this function by its cosine components only as:

\begin{equation}
 f(\theta) = \ssum_n a_n\cos\theta\ .
\end{equation}

\noindent Once we have the final result for this simplified case, it can easely be generalized for the full Fourier expansion of $f$. 
We must write expressions for for $a(\phi)$ and $f_2(\phi,\theta))$, but first facilitate the notation as follows:

\begin{eqnarray*}
 S_n^+=\sin(n+2k)\theta,\hspace{0.5cm} C_n^+=\cos(n+2k)\theta,\hspace{0.5cm} S_2=\sin 2\phi,\hspace{0.5cm} C^0 = \cos n\theta\ , \\
 S_n^-=\sin(n-2k)\theta,\hspace{0.5cm} C_n^-=\cos(n-2k)\theta,\hspace{0.5cm} C_2=\cos 2\phi,\hspace{0.5cm} S^0 = \sin n\theta\ . \hspace{0.1cm}
\end{eqnarray*}

\noindent and also define $\bar{a}_n$ as:

\begin{equation}
 \bar{a}_n=\frac{a_n}{4\nu_0}\ . \label{aundn}
\end{equation}

\noindent We now can write:

\begin{eqnarray}
 a(\phi) &=& \nu_0-k+\bar{a}_{2k}C_2\ , \label{aphi}\\
 f_2(\phi,\theta) &=& \ssum_n \bar{a}_n[2C^0 + (C_n^- + C_n^+)C_2 +(-S_n^- + S_n^+)S_2]\ . \label{Ftry}
\end{eqnarray}

\noindent Note here that in the term with $C_n^-=\cos(n-2k)\theta$ we must exclude the case $n=2k$ as this contribution is already included in the expression for $a(\phi)$.

\noindent We try for $U_2$ the following general form:

\begin{equation}\label{Utry}
 U_2 = \ssum_n \bar{a}_n\left[\alpha_nS^0+\beta_nS_n^+C_2+\gamma_nC_n^+S_2+\bar{\beta}_nS_n^-C_2+\bar{\gamma}_nC_n^-S_2\right]\ ,
\end{equation}

\noindent It is easily verified that other contributions to $U_2$, from terms like $C_n^-C_2$, $C_n^+C_2$, $S_n^-C_2$, or $S_n^+C_2$ must be zero, because derivatives of these terms 
(with respect to $\theta$ or $\phi$) do not exist in the function $f_2(\theta,\phi)$.

\noindent Inserting Eqs.~(\ref{Ftry},\ref{Utry}) in Eq.~(\ref{Utwo}), we get the solution for $\alpha_n$ 
and a set of equations for the other unknown parameters  and $\beta_n,\bar{\beta}_n,\gamma_n,\bar{\gamma}_n$:

\begin{eqnarray}
	&\ &\beta_n(n+2k)-2a_0\gamma_n = 1\ , \\
	&\ &\gamma_n(n+2k)-2a_0\beta_n = -1\ , \\
	&\ &\bar{\beta}_n(n-2k)-2a_0\bar{\gamma}_n = 1\ , \\
	&\ &\bar{\gamma}_n(n-2k)-2a_0\bar{\beta}_n = 1\ .
\end{eqnarray}

\noindent The solution of these equations is as follows:

\begin{eqnarray}
	\alpha_n &=& \frac{2}{n}\ ,\\
	\beta_n &=& -\gamma_n = \frac{1}{n+2k+2a_0}=\frac{1}{n+2\nu_0}\ , \label{setb}\\
	\bar{\beta}_n &=& \bar{\gamma}_n = \hspace{0.3cm}\frac{1}{n-2k-2a_0}=\frac{1}{n-2\nu_0}\hspace{2cm}\mbox{for}\ (n\neq2k)\ , \label{setc}\\
	\bar{\beta}_n &=& \bar{\gamma}_n = 0\hspace{6.5cm}\mbox{for}\ (n=2k)\ . \label{setd}
\end{eqnarray}

\noindent For the derivative of $U_2$ with respect to $\phi$ we obtain:

\begin{equation}\label{Udif}
 \frac{\partial U_2}{\partial\phi} = -2\ssum_m \bar{a}_m\left[(\beta_mC_m^+ -\bar{\beta}_mC_m^-)C_2+(\beta_mS_m^+ +\bar{\beta}_mS_m^-)S_2\right]\ .
\end{equation}

\noindent With the expression for $f_2$ in Eq.~(\ref{Ftry}) and the expression for $\partial U_2/\partial\phi$ in Eq.~(\ref{Udif}), we can write for the second term in  Eq.~(\ref{barK2})

\begin{eqnarray}\label{verylong}
\bigsymbol{\langle} f_2\frac{\partial U_2}{\partial\bar{\phi}}\bigsymbol{\rangle} &=& -2\bigsymbol{\langle} \ssum_n\ssum_m \bar{a}_n\bar{a}_m\bigsymbol{[} 2(\beta_mC_m^+C_n^0-\bar{\beta}_mC_m^-C_n^0)C_2 \nonumber \\
&+&\left(\beta_m(C_m^+C_n^- +C_m^+C_n^+)-\bar{\beta}_m(C_m^-C_n^-+C_m^-C_n^+)\right)C_2^2 \nonumber \\
&+&\left(\beta_m(-S_m^+S_n^- +S_m^+S_n^+)+\bar{\beta}_m(-S_m^-S_n^-+S_m^-S_n^+)\right)S_2^2\bigsymbol{]}\bigsymbol{\rangle} \hspace{1cm}
\end{eqnarray}

\noindent Note that here we have already omitted contributions obtained from products between sine-terms and cosine-terms, because their average value is null.\\
\noindent We now will show that all "alternating" terms in Eq.~(\ref{verylong}) do not contribute. 
By this we mean the terms with $C_m^+C_n^-,C_m^-C_n^+,S_m^+S_n^-,S_m^-S_n^+$ and also the terms with $C_m^+C_n^0,C_m^-C_n^0$. 
This can be shown by changing the sign of the summation index $m$ and using the following "symmetry" considerations:

\begin{eqnarray*}
\bar{a}_{-m}&=&\bar{a}_{m}\ , \\
\beta_{-m}&=&-\bar{\beta}_m\ , \\
C_{-m}^+ &=& C_m^- \ ,\\
S_{-m}^+ &=& -S_m^- \ .
\end{eqnarray*}

\noindent Consider for example the term with $C_m^+C_n^-$. For this term we can write:

\begin{eqnarray*}
	\ssum_n\ssum_m \bar{a}_n\bar{a}_m \beta_mC_m^+C_n^- &=& -\ssum_n\ssum_{-m} \bar{a}_n\bar{a}_m \bar{\beta}_mC_m^-C_n^- \\
	&=&-\ssum_n\ssum_{-m} \bar{a}_n\bar{a}_m \bar{\beta}_m\cos(m-2k)\theta\cos(n-2k)\theta\ .
\end{eqnarray*}

\noindent This term will have a non-zero average if $m-2k=n-2k$, so if $m=n$, but this can never happen because $n$ is positive and $m$ is negative. 
The same result is obtained for the tems containing $C_m^-C_n^+,S_m^+S_n^-,S_m^-S_n^+$.
For the term with $C_m^+C_n^0$ we obtain the condition: $m=n+2k$, but also this can never happen because $n$ and $k$ are positive and $m$ is negative.
For the term with $C_m^-C_n^0$ we obtain the condition: $m=n-2k$. In general there could be a solution if $n$ would be any positive integer. However, for cyclotrons
the magnetic field must have $N$-fold symmetry with $N\geq 3$ and  $n\geq N$. Since for our value of $k$ we have $0\leq 2k \leq N$ and $m\leq -N$, 
there are no solutions for this case either.
For the remaining terms in Eq.~(\ref{verylong}) we only will have a contribution to the average if $m=n$. For this we find:

\begin{eqnarray*}\label{lesslong}
\bigsymbol{\langle} f_2\frac{\partial U_2}{\partial\bar{\phi}}\bigsymbol{\rangle} &=& -2\bigsymbol{\langle} \ssum_n \bar{a}_n^2\bigsymbol{[} 
\left(\beta_n C_n^{+^2}-\bar{\beta}_n C_n^{-^2} \right)C_2^2 +\left(\beta_n S_n^{+^2} -\bar{\beta}_n(S_n^{-^2} \right)S_2^2\bigsymbol{]}\bigsymbol{\rangle}\\
&=& -\ssum_n \bar{a}_n^2\left[(\beta_n-\bar{\beta}_n)C_2^2 +(\beta_n-\bar{\beta}_n)S_2^2\right]\\
&=& \ssum_n (\bar{\beta}_n-\beta_n)\bar{a}_n^2
\end{eqnarray*}

\noindent We insert the relations for $\beta_n$ and $\bar{\beta}_n$ as defined in Eqs.~(\ref{setb}-\ref{setd}) and obtain:

\begin{equation}\label{notlong}
\bigsymbol{\langle} f_2\frac{\partial U_2}{\partial\bar{\phi}}\bigsymbol{\rangle}=-\frac{\bar{a}_{2k}^2}{2(k+\nu_0)}+4\nu_0\ssum_{n\neq 2k}\frac{\bar{a}_{n}^2}{n^2-4\nu_0^2}\ .
\end{equation}

\noindent Inserting this expression (Eq.~(\ref{notlong})) and the expression for $a(\phi)$ (Eq.~(\ref{aphi})) and the definition of $\bar{a}_n$ (Eq.~(\ref{aundn}))
in the Hamiltonian given in  (Eq.~(\ref{barK2}), we obtain:

\begin{equation}
\bar{K} = \bar{I}\bigsymbol{[}\nu_0-k+\frac{a_{2k}}{4\nu_0}\cos2\bar{\phi}-\frac{a_{2k}^2}{32\nu_0^2(k+\nu_0)}+\frac{1}{4\nu_0}\ssum_{n\neq 2k}\frac{a_{n}^2}{n^2-4\nu_0^2}\bigsymbol{]}\ .
\end{equation}

\noindent We can now generalize this result for the case that the function $f(\theta)$ not only includes the cosine components but also the sine components:

\begin{equation}
f(\theta) = \ssum_n a_n\cos n\theta + b_n\sin n\theta\ .
\end{equation}

\noindent The general Hamiltonian for this case becomes:

\begin{equation}\label{KIphi}
\bar{K} = \bar{I}\bigsymbol{[}\nu_0-k+\frac{c_{2k}}{4\nu_0}\cos2(\bar{\phi}-k\varphi_{2k})-\frac{c_{2k}^2}{32\nu_0^2(k+\nu_0)}+\frac{1}{4\nu_0}\ssum_{n\neq 2k}\frac{c_{n}^2}{n^2-4\nu_0^2}\bigsymbol{]}\ .
\end{equation}

\noindent Here $c_n$ and $\varphi_n$ are the amplitude and phase of the n$^{\mbox{th}}$ Fourier component of the function $f(\theta)$. They relate to $a_n,b_n$ as follows:

\begin{eqnarray}
	a_n=c_n\cos n\varphi_n\ ,\\
	b_n=c_n\sin n\varphi_n\ . \hspace{0.15cm}
\end{eqnarray}

\noindent Comparing this result with those found in the previous paragraph~\ref{remove}, it is seen that for the cases $k=0$ and $k=1$ both results are the same if applied to a 
cyclotron with $N$-fold symmetry for which $N\geq 3$; for these cases $c_{2k}=0$ and the restriction $n\neq 2k$ in the series summation can be ommited. 
It is seen from  (Eq.~(\ref{KIphi}) that for $k=0$ the tune is given by:

\begin{equation}
                    \nu_x = \nu_0+ \frac{1}{4\nu_0}\ssum_{n}\frac{c_{n}^2}{n^2-4\nu_0^2}\ .
\end{equation}            

\noindent This is (up to $\mathcal{O}(f^2)$) the same as given in Eq.~(\ref{tune1}). For $k=1$ our phase space rotates with frequency 1 and therefore the oscillation
frequency in this phase space should be equal to $\nu_x-1$. This indeed is the case.

\noindent However, in contrast to  the Hamiltonian given in Eq.~(\ref{Hfinal}),  the new Hamiltonian given in Eq.~(\ref{KIphi}) does not have a singularity
at $\nu_0=N/2$ and therefore is valid upto and beyond the half-integer resonance  $\nu_0=\tfrac{N}{2}$. The first singularity now occurs only at the next harmonic $\nu_0=N$.

\noindent Let us consider in more detail the half-integer resonance and take $k=N/2$. 
We now go back to the cartesian description of the phase space and apply the canonical transformation:

\begin{eqnarray*}
X &=& \sqrt{2\bar{I}}\cos(\bar{\phi}-\tfrac{N}{2}\varphi_{N}) \ , \\
P &=&\sqrt{2\bar{I}}\sin(\bar{\phi}-\tfrac{N}{2}\varphi_N) \ .
\end{eqnarray*}

\noindent Note however, that this new cartesian phase space is rotating with frequency $N/2$ relative to the original phase space.

\noindent We also define the parameters $\nu_1,\nu_2,\bar{\nu},\Delta_2$ as follows:

\begin{equation}
\begin{aligned}\label{nu_parameters}
\nu_1 &= \bar{\nu} -\frac{c_N}{4\nu_0}\ , \\
\nu_2 &= \bar{\nu} +\frac{c_N}{4\nu_0}\ , \\
\bar{\nu} &= \nu_0-\frac{N}{2}+\bar{\Delta}_2\ , \\
\bar{\Delta}_2 &=  -\frac{c_N^2}{(4\nu_0)^2(N+2\nu_0)}+\frac{1}{4\nu_0}\ssum_{n>N}\frac{c_{n}^2}{n^2-4\nu_0^2}\ .
\end{aligned}
\end{equation}

\noindent With these definitions the Hamiltonian in cartesian phase space becomes:

\begin{equation}
 \bar{K} =  \tfrac{1}{2}\nu_1 P^2+\tfrac{1}{2}\nu_2 X^2\ ,
\end{equation}

\noindent and the equation of motion for $X$ is given as:

\begin{equation}\label{Hill}
\frac{d^2 X}{d\theta^2} + \nu_1\nu_2 X =0\ .
\end{equation}

\noindent For stable motion of $X$ we must have $ \nu_1\nu_2 >0$. There are two ways to obey this requirement: i) both $\nu_1<0$ and $\nu_2<0$ or
 ii) both $\nu_1>0$ and $\nu_2>0$. The first case i) requires that $\nu_2<0$ and the second case ii) requires that $\nu_1>0$.

\noindent The stable regions are given by:

\begin{equation}
\begin{aligned}\label{stopband}
\nu_0<\bar{\nu}_{1}=\frac{N}{2}-\frac{c_N}{4\bar{\nu}_{1}}-\bar{\Delta}_2\ , \\
\nu_0>\bar{\nu}_{2}=\frac{N}{2}+\frac{c_N}{4\bar{\nu}_{2}}-\bar{\Delta}_2\ .
\end{aligned}
\end{equation}

\noindent These are implicit relations for the limits $\bar{\nu}_{1},\bar{\nu}_{2}$ of the stopband of the resonance. 
We can solve for $\bar{\nu}_{1,2}$ by successive substitution, 
which needs to be carried out up to $\mathcal{O}(f^2)$. One finds:

\begin{equation}\label{stopband2}
\boxed{\nu_0(1,2)=\frac{N}{2} \mp \frac{c_N}{2N} - \Delta_2\ ,}
\end{equation}

\noindent where $\Delta_2$ is defined as:

\begin{equation}\label{delta2}
\boxed{\Delta_2 = \frac{3}{8}\frac{c_N^2}{N^3}+\frac{1}{2N}\sum_{n>N}\frac{c_n^2}{n^2-N^2}\ ,}
\end{equation}

\noindent Here the minus sign applies for the lower limit $\nu_0(1)$ of the stopband and the plus sign for its upper limit $\nu_0(2)$.
\noindent  The width of this stopband is equal to $c_N/N$. Note that center is not exactly positioned at $N/2$ due to the $\mathcal{O}(f^2)$ contributions in Eq.~(\ref{stopband2}).
 In paragraph~\ref{resonancestopband} we use the above results to find the stopband of the isochronous cyclotron.

\noindent In order to illustrate the results, we aproximate Eq.~(\ref{stopband2}) a little bit finer by assuming a hard-edge profile of the function $f(\theta)$ similar to what was done for the azimuthal variation of the magnetic field. In this case  Eq.~(\ref{fcoefs}) applies for the coefficients $c_n$ and the summation in Eq.~(\ref{stopband2}) can be written as
(with the substitution $n=(2k+1)N$):

\begin{align}
\sum_{n>N}\frac{c_n^2}{n^2-N^2} &= \frac{c_N^2}{N^2}\sum_{k>0}\frac{1}{(2k+1)^2((2k+1)^2-1)} \\
   &= \frac{c_N^2}{N^2}\sum_{k>0}\bigsymbol{[}\frac{1}{(2k+1)^2-1}-\frac{1}{(2k+1)^2}\bigsymbol{]}\\
   &=\frac{c_N^2}{N^2}\bigsymbol{(}\tfrac{1}{4}\sum_{k>0}\bigsymbol{[}\frac{1}{k}-\frac{1}{k+1}\bigsymbol{]}-\frac{\pi^2}{8}+1\bigsymbol{)}=\frac{c_N^2}{4N^2}(5-\frac{\pi^2}{2})\ .
\end{align}

\noindent With this assumption we can approximate the limits of the stopband as:

\begin{equation}\label{stopband3}
\boxed{\nu_0(1,2)=\frac{N}{2} \mp \frac{c_N}{2N} - (1-\frac{\pi^2}{16})\frac{c_N^2}{N^3}\ .}
\end{equation}

\noindent Let us now derive the tune of the motion $\bar{\nu}_{x}$ in the stable regions outside of the stopband. From Eq.~(\ref{Hill}) we find:

\begin{equation}\label{tune_x}
\bar{\nu}_x = \frac{N}{2}\mp\sqrt{\nu_1\nu_2} = \frac{N}{2}\mp\sqrt{(\nu_0-\tfrac{N}{2}+\bar{\Delta}_2)^2-\frac{c_N^2}{16\nu_0^2}}\ .
\end{equation}

\begin{figure}[!bht]
   \vspace*{-.5\baselineskip}
   \centering
   \includegraphics[width=9cm]{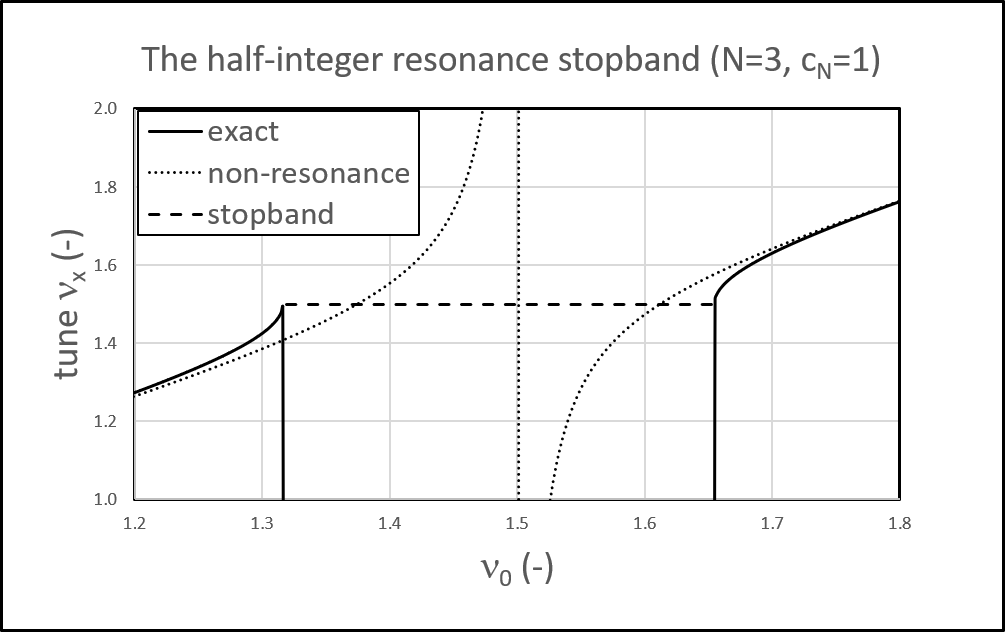}
    \caption[Illustration of the half-integer resonance stopband]{\it Illustration of the half-integer resonance stopband}
    \label{resonance stopband}
\end{figure}

\noindent Here we augment the tune with $N/2$  because we want the tune in a non-rotating frame while Eq.~(\ref{Hill}) applies
for a frame that rotates with frequency $N/2$. Note furher that the $-$ sign in above equation applies for the first stable region and the $+$ sign for the second stable region. 
One could try to develop the square-root in the above equation up to 
$\mathcal{O}(f^2)$ but this will be inaccurate because close to the resonance all three terms ($\nu_0-N/2$, $\Delta_2$ and $c_N^2/16\nu_0^2$) are small and there is no good
way to compare them. As an illustration Figure~\ref{resonance stopband} show the tune $\bar{\nu}_x$ as function of $\nu_0$ for 
the hard-edge profile of $f(\theta)$ with $N=3$ and $c_N=1$. The solid line is calculated from Eq.~(\ref{tune_x}) and the stopband (dashed line) from Eq.~(\ref{stopband3}). The dotted
line is calculated with  Eq.~(\ref{tune1}) which was obtained from the "non-resonance" analysis done in paragraph~\ref{remove}. It is seen that this "non-resonance" approximation is 
good further away from the stopband, but it fails close to the stopband. It is also seen that in the first stable region, due to the resonance, the tune is pushed up 
towards the value of $N/2$. Inside the stopband the tune becomes a complex number with a real and an imagninary part. The real part is  equal to $N/2$; 
the  imagniary part makes that the amplitude of the oscillation increases exponentially.

\renewcommand{\theequation}{\thesection\arabic{equation}}
\setcounter{equation}{0}

\renewcommand{\theequation}{\thesection\arabic{equation}}
\setcounter{equation}{0}

\section{Analytical summation of the series expansions}\label{summation}

\noindent The summations in tune expressions given in Eqs.~(\ref{rtune_coeff},\ref{vtune}) can be done analytically and the coefficients 
$a_N,b_N,c_N,d_N$ can be expressed in elementary mathematical functions.
In order to achieve this, all rational fractions of polynomials (with respect to $k$ since $n=(2k+1)N$)  in the right hand sides of the equations have to be decomposed in a sum of 
simple rational fractions (see partial fraction decomposition\cite{wiki1,Larson}).  For example if we write the coefficient  $\tilde{b}_N$ for the radial tune as:

\begin{equation}
\tilde{b}_N = \sum_{k=0}^\infty F(k)\ ,
\end{equation}

\noindent then we must decompose the function $F(k)$ as follow:

\begin{equation}
F(k) = \sum_{j=1}^m\frac{a_j}{(k+b_j)^{p_j}}\ .
\end{equation}

\noindent Here $p_j=1,2,\dots$ is the power of the linear polynomial in the denominater of the fraction. Such a decomposition can be made for all the rational fractions that
are present in Eqs.~(\ref{rtune_coeff},\ref{vtune}) and they have been derived in appendix~\ref{PFD}. 
For the form as given by $F(k)$, the summation can be carried out analytically and the result is \cite{wiki2,Abram1}:

\begin{equation}
 \sum_{k=0}^\infty F(k)=\sum_{k=0}^\infty\sum_{j=1}^m\frac{a_j}{(k+b_j)^{p_j}}=\sum_{j=1}^m\frac{(-1)^{p_j}}{(p_j-1)!}a_j\psi^{(p_j-1)}(b_j)\ .
\end{equation}

\noindent Here $\psi^{(n})$ is the polygamma function~\cite{wiki3,Abram2} of order $n$. The variables $b_j$ may be imaginary or complex.
In order the series to converge, it is required that the sum of coefficients $a_j$ that correspond to linear powers $p_j=1$ must be equal zero.

\begin{equation}
\sum_{j\rightarrow p_j=1}^m a_j = 0\ .
\end{equation}

\noindent This requirement is met  for all the rational fractions that are present in Eqs.~(\ref{rtune_coeff},\ref{vtune}). We note that the reflection relation for the polygamma function\cite{wiki3,Abram2} must be used in order to express the final results in elementary mathematical functions.
We show as an example the derivation for the coefficient $\tilde{b}_N$ in Eqs.~(\ref{rtune_coeff}). For convenience we define $n=mN=(2k+1)N$ and $\alpha=\sqrt{1+\bmu'}/N$.
We write for $\tilde{b}_N$

\begin{align}
\tilde{b}_N &= \frac{3(1+\bmu')}{N^4}\sum_{k=0}^\infty \frac{1}{(m^2-4\alpha^2)(m^2-\alpha^2)}\ , \nonumber \\
	&= \frac{3\alpha^2}{N^2}\sum_{k=0}^\infty \left( \frac{1}{m^2-4\alpha^2}-\frac{1}{m^2-\alpha^2}\right)\ ,  \nonumber\\
	&= \frac{3\alpha}{4N^2}\sum_{k=0}^\infty \left( \frac{1}{m-2\alpha}-\frac{1}{m+2\alpha}-2\bigsymbol{(}\frac{1}{m-\alpha}-\frac{1}{m+\alpha}\bigsymbol{)}\right)\ ,  \nonumber\\
	&= \frac{3\alpha}{8N^2}\sum_{k=0}^\infty \left( \frac{1}{k+\frac{1-2\alpha}{2}}-\frac{1}{k+\frac{1+2\alpha}{2}}-2\bigsymbol{(}\frac{1}{k+\frac{1-\alpha}{2}}-\frac{1}{k+\frac{1+\alpha}{2}}\bigsymbol{)}\right)\ , \nonumber \\
	&= \frac{3\alpha}{8N^2}\left(-\psi(\frac{1-2\alpha}{2})+\psi(\frac{1+2\alpha}{2}) - 2\bigsymbol{(}-\psi(\frac{1-\alpha}{2})+\psi(\frac{1+\alpha}{2})\bigsymbol{)} \right)\ , \nonumber \\
	&= \frac{3\alpha}{8N^2}\left(-\psi(1-\frac{1+2\alpha}{2})+\psi(\frac{1+2\alpha}{2}) - 2\bigsymbol{(}-\psi(1-\frac{1+\alpha}{2})+\psi(\frac{1+\alpha}{2})\bigsymbol{)} \right)\ , \nonumber \\
&= \frac{3\alpha}{8N^2}\left( -\pi\frac{\cos(\frac{1+2\alpha}{2})\pi}{\sin(\frac{1+2\alpha}{2})\pi}+2\pi\frac{\cos(\frac{1+\alpha}{2})\pi}{\sin(\frac{1+\alpha}{2})\pi}\right)\ .
\nonumber\\
&= \frac{3\alpha\pi}{8N^2}\left(\tan(\pi\alpha)-2\tan(\pi\alpha/2) \right)\ . \nonumber
\end{align}

\noindent Here $\psi$ is the digamma function (the polygamma function of order zero). In the last step we used  the reflection relation for the polygamma function. For the digamma function this relation is:

\begin{equation}
\psi(1-z)-\psi(z)=\pi\cot\pi z\ . \nonumber
\end{equation}

\noindent An analogue but more general reflection relation exist (and has been used in our derivations) for the polygamma function.
\noindent The same method as illustrated here has been applied for all series summations in our derivations (for the radial and vertical tunes and also for the stopband limits
of the half-integer resonance).  For the stopband limits we also needed to use the recurrance relation of the polygamma function. For the digamma function this relation takes the
following form:

\begin{equation}
\psi(1+z)=\psi(z)+\frac{1}{z}\ . \nonumber
\end{equation}


\renewcommand{\theequation}{\thesection\arabic{equation}}
\setcounter{equation}{0}

\section{Partial fraction decomposition}\label{PFD}

We show details of the partial fraction decomposition as needed for the explicit evaluation of the series summations in the expression for the 
radial tune (Eq.~(\ref{rtune_coeff})), the vertical tune (Eq.~(\ref{vtune})) and the stopband limits of the half-integer resonance (Eqs.~(\ref{sblimit})) . In these series summations we first replace $n^2$ by $n^2=N^2m^2$ and define 
$\alpha^2=(1+\bmu')/N^2$ and $\beta^2=4(1+\bmu')/N^2$ (for the radial tune) and $\beta^2=-4\bmu'/N^2$ (for the vertical tune). The decomposition is done in two
steps. In the first (preliminary) step, the parameter $p$ is inserted for $m^2$ and $\alpha^2$ and $\beta^2$ are replaced by $\tilde{\alpha}$ and $\tilde{\beta}$
respectively. The following decompositions are obtained in the first step:

\begin{align}
&\frac{1}{p(p-\balpha)}=\frac{1}{\balpha}(\frac{1}{p-\balpha}-\frac{1}{p}) \nonumber \\
&\frac{p}{(p-\balpha)^2(p-\bbeta)} = \frac{1}{(\balpha-\bbeta)^2}\bigsymbol{(}\frac{\bbeta}{p-\bbeta}-\frac{\bbeta}{p-\balpha}+\frac{\balpha(\balpha-\bbeta)}{(p-\balpha)^2}\bigsymbol{)}\ , \nonumber \\
&\frac{1}{(p-\balpha)^2(p-\bbeta)} = \frac{1}{(\balpha-\bbeta)^2}\bigsymbol{(}\frac{1}{p-\bbeta}-\frac{1}{p-\balpha}+\frac{\balpha-\bbeta}{(p-\balpha)^2}\bigsymbol{)}\ , \nonumber \\
&\frac{1}{p(p-\balpha)^2(p-\bbeta)} = \frac{1}{\balpha^2\bbeta(\balpha-\bbeta)^2}\bigsymbol{(}-\frac{(\balpha-\bbeta)^2}{p}+\frac{\balpha^2}{p-\bbeta}+\frac{(\balpha-\bbeta)^2-\balpha^2}{p-\balpha}+\frac{\balpha\bbeta(\balpha-\bbeta)}{(p-\balpha)^2}\bigsymbol{)}\ , \nonumber \\
&\frac{p}{(p-\balpha)^2} = \frac{1}{p-\balpha}+\frac{\balpha}{(p-\balpha)^2}\ , \nonumber \\
&\frac{1}{p(p-\balpha)^2} = \frac{1}{\balpha^2}\bigsymbol{(}\frac{1}{p}-\frac{1}{p-\balpha}+\frac{\balpha}{(p-\balpha)^2}\bigsymbol{)}\ , \nonumber \\
&\frac{1}{p(p-\balpha)^3} = \frac{1}{\balpha^3}\bigsymbol{(}-\frac{1}{p}+\frac{1}{p-\balpha}-\frac{\balpha}{(p-\balpha)^2}+\frac{\balpha^2}{(p-\balpha)^3}\bigsymbol{)}\ , \nonumber \\
&\frac{p}{(p-\balpha)} = -\frac{1}{\balpha}\bigsymbol{(}\frac{1}{p}-\frac{1}{(p-\balpha)}\bigsymbol{)}\ , \nonumber \\
&\frac{p}{(p-\balpha)^2} = \frac{1}{(p-\balpha)}+\frac{\balpha}{(p-\balpha)^2}\ , \nonumber \\
&\frac{p}{(p-\balpha)^3} = \frac{1}{(p-\balpha)^2}+\frac{\balpha}{(p-\balpha)^3}\ , \nonumber \\
&\frac{1}{(p-\balpha)^3} = \mbox{already fully decomposed form}\ , \nonumber \\
&\frac{1}{(p-\balpha)(p-\bbeta)} =\frac{1}{\balpha-\bbeta}\bigsymbol{(} \frac{1}{(p-\balpha)}-\frac{1}{(p-\bbeta)}\bigsymbol{)}\ . \nonumber
\end{align}

\noindent In the second step each of the terms in the right hand sides of  above expressions are further decomposed by inserting for $p,\tilde{\alpha},\tilde{\beta}$ the original parameters $m^2,\alpha^2,\beta^2$ respectively. We obtain the final expressions below for the radial tune and the vertical tune. 
Note that some of the decompositions for the radial tune also are used for the vertical tune. In the right hand sides of the final expressions we have to substitute $m=2k+1$,
as $k$ is the summation index to be used in the series summations.

\subsection{Radial tune decompositions}

\begin{align}
&\frac{1}{m^2(m^2-\alpha^2)} =\frac{1}{2\alpha^3}\bigsymbol{(}(\frac{1}{m-\alpha}-\frac{1}{m+\alpha})-\frac{2\alpha}{m^2}\bigsymbol{)}\ , \nonumber \\
&\frac{1}{(m^2-\alpha^2)^2} = \frac{1}{4\alpha^3}\bigsymbol{(}-(\frac{1}{m-\alpha}-\frac{1}{m+\alpha})+\alpha(\frac{1}{(m-\alpha)^2}+\frac{1}{(m+\alpha)^2})\bigsymbol{)}\ , \nonumber \\
&\frac{1}{m^2(m^2-\alpha^2)^2} = \frac{1}{4\alpha^5}\bigsymbol{(}\frac{4\alpha}{m^2}-3(\frac{1}{m-\alpha}-\frac{1}{m+\alpha})+\alpha(\frac{1}{(m-\alpha)^2}+\frac{1}{(m+\alpha)^2})\bigsymbol{)}\ , \nonumber \\
&\frac{1}{m^2-\alpha^2}=\frac{1}{2\alpha}\bigsymbol{(}\frac{1}{m-\alpha}-\frac{1}{m+\alpha}\bigsymbol{)}\ , \nonumber \\
&\frac{1}{m^2-4\alpha^2}=\frac{1}{4\alpha}\bigsymbol{(}\frac{1}{m-2\alpha}-\frac{1}{m+2\alpha}\bigsymbol{)}\ , \nonumber \\
&\frac{1}{(m^2-\alpha^2)(m^2-4\alpha^2)}=\frac{1}{12\alpha^3}\bigsymbol{(}(\frac{1}{m-2\alpha}-\frac{1}{m+2\alpha})-2(\frac{1}{m-\alpha}-\frac{1}{m+\alpha})\bigsymbol{)}\ , \nonumber \\
&\frac{1}{m^2(m^2-\alpha^2)(m^2-4\alpha^2)}=\frac{1}{48\alpha^5}\bigsymbol{(}\frac{12\alpha}{m^2}+(\frac{1}{m-2\alpha}-\frac{1}{m+2\alpha})-8(\frac{1}{m-\alpha}-\frac{1}{m+\alpha})\bigsymbol{)}\ , \nonumber \\
&\frac{m^2}{(m^2-\alpha^2)^2(m^2-4\alpha^2)}=\frac{1}{36\alpha^3}\bigsymbol{(}4(\frac{1}{m-2\alpha}-\frac{1}{m+2\alpha})-5(\frac{1}{m-\alpha}-\frac{1}{m+\alpha})\nonumber \\
&\hspace{5cm}-3\alpha(\frac{1}{(m-\alpha)^2}+\frac{1}{(m+\alpha)^2})\bigsymbol{)}\ , \nonumber \\
&\frac{1}{(m^2-\alpha^2)^2(m^2-4\alpha^2)}=\frac{1}{36\alpha^5}\bigsymbol{(}(\frac{1}{m-2\alpha}-\frac{1}{m+2\alpha})+(\frac{1}{m-\alpha}-\frac{1}{m+\alpha})\nonumber \\
&\hspace{5cm}-3\alpha(\frac{1}{(m-\alpha)^2}+\frac{1}{(m+\alpha)^2})\bigsymbol{)}\ , \nonumber \\
&\frac{1}{m^2(m^2-\alpha^2)^2(m^2-4\alpha^2)}=\frac{1}{36\alpha^7}\bigsymbol{(}\frac{1}{4}(\frac{1}{m-2\alpha}-\frac{1}{m+2\alpha})+7(\frac{1}{m-\alpha}-\frac{1}{m+\alpha})  \nonumber \\
&\hspace{5cm} -\frac{9\alpha}{m^2}-3\alpha(\frac{1}{(m-\alpha)^2}+\frac{1}{(m+\alpha)^2})\bigsymbol{)}\ , \nonumber 
\end{align}

\begin{align}
&\frac{1}{m^2(m^2-\alpha^2)^3}=\frac{1}{16\alpha^7}\bigsymbol{(}15(\frac{1}{m-\alpha}-\frac{1}{m+\alpha}) -\frac{16\alpha}{m^2}-7\alpha(\frac{1}{(m-\alpha)^2}+\frac{1}{(m+\alpha)^2}) \nonumber \\
&\hspace{5cm} +2\alpha^2(\frac{1}{(m-\alpha)^3}-\frac{1}{(m+\alpha)^3})\bigsymbol{)}\ . \nonumber
\end{align}

\subsection{Vertical tune decompositions}

\begin{align}
&\frac{m^2}{(m^2-\alpha^2)^3}=\frac{1}{16\alpha^3}\bigsymbol{(}-(\frac{1}{m-\alpha}-\frac{1}{m+\alpha}) +\alpha(\frac{1}{(m-\alpha)^2}+\frac{1}{(m+\alpha)^2}) \nonumber \\
&\hspace{5cm}+2\alpha^2(\frac{1}{(m-\alpha)^3}-\frac{1}{(m+\alpha)^3})\bigsymbol{)}\ , \nonumber \\
&\frac{1}{(m^2-\alpha^2)^3}=\frac{3}{16\alpha^5}\bigsymbol{(}(\frac{1}{m-\alpha}-\frac{1}{m+\alpha}) -\alpha(\frac{1}{(m-\alpha)^2}+\frac{1}{(m+\alpha)^2}) \nonumber \\
&\hspace{5cm}+\frac{2\alpha^2}{3}(\frac{1}{(m-\alpha)^3}-\frac{1}{(m+\alpha)^3})\bigsymbol{)}\ , \nonumber \\
&\frac{m^2}{(m^2-\alpha^2)^2(m^2-\beta^2)}=\frac{1}{4(\alpha^2-\beta^2)^2}\bigsymbol{(}2\beta(\frac{1}{m-\beta}-\frac{1}{m+\beta})  \nonumber \\
&\hspace{2cm}-\frac{\alpha^2+\beta^2}{\alpha}(\frac{1}{m-\alpha}-\frac{1}{m+\alpha})+(\alpha^2-\beta^2)(\frac{1}{(m-\alpha)^2}+\frac{1}{(m+\alpha)^2})\bigsymbol{)}\ , \nonumber \\
&\frac{1}{(m^2-\alpha^2)^2(m^2-\beta^2)}=\frac{1}{4(\alpha^2-\beta^2)^2}\bigsymbol{(}\frac{2}{\beta}(\frac{1}{m-\beta}-\frac{1}{m+\beta})  \nonumber \\
&\hspace{2cm}-(\frac{3\alpha^2-\beta^2}{\alpha^3})(\frac{1}{m-\alpha}-\frac{1}{m+\alpha})+(\frac{\alpha^2-\beta^2}{\alpha^2})(\frac{1}{(m-\alpha)^2}+\frac{1}{(m+\alpha)^2})\bigsymbol{)}\ , \nonumber \\
&\frac{1}{m^2(m^2-\alpha^2)^2(m^2-\beta^2)}=\frac{1}{2\beta^3(\alpha^2-\beta^2)^2}(\frac{1}{m-\beta}-\frac{1}{m+\beta}) -\frac{1}{\alpha^4\beta^2}\frac{1}{m^2} \nonumber \\
&\hspace{1cm}+\frac{3\beta^2-5\alpha^2}{4\alpha^5(\alpha^2-\beta^2)^2}(\frac{1}{m-\alpha}-\frac{1}{m+\alpha})+\frac{1}{4\alpha^4(\alpha^2-\beta^2)}(\frac{1}{(m-\alpha)^2}+\frac{1}{(m+\alpha)^2})\ . \nonumber
\end{align}

\begin{align}
&\frac{m^2}{(m^2-\alpha^2)^2}=\frac{1}{4\alpha}(\frac{1}{m-\alpha}-\frac{1}{m+\alpha}) +\frac{1}{4}(\frac{1}{(m-\alpha)^2}+\frac{1}{(m+\alpha)^2})\ ,  \nonumber \\
&\frac{1}{(m^2-\alpha^2)(m^2-\beta^2)}=\frac{1}{(\alpha^2-\beta^2)}\bigsymbol{(}\frac{1}{2\alpha}(\frac{1}{m-\alpha}-\frac{1}{m+\alpha})-\frac{1}{2\beta}(\frac{1}{m-\beta}-\frac{1}{m+\beta})\bigsymbol{)}\ ,\nonumber \\
&\frac{1}{m^2(m^2-\alpha^2)(m^2-\beta^2)}=\frac{1}{2(\alpha^2-\beta^2)}\bigsymbol{(}\frac{1}{\alpha^3}(\frac{1}{m-\alpha}-\frac{1}{m+\alpha})-\frac{1}{\beta^3}(\frac{1}{m-\beta}-\frac{1}{m+\beta})\bigsymbol{)} \nonumber \\
&\hspace{4cm}+\frac{1}{\alpha^2\beta^2}\frac{1}{m^2}\ ,\nonumber \\
&\frac{1}{m^2(m^2-\alpha^2)}=-\frac{1}{\alpha^2}\bigsymbol{(}\frac{1}{m^2}-\frac{1}{2\alpha}(\frac{1}{m-\alpha}-\frac{1}{m+\alpha})\bigsymbol{)}\ .  \nonumber 
\end{align}

\subsection{Summation of decomposed fractions}

Once the partial fractions in the series expressions have been decomposed, each of the seperate basic contributions have to be analytically calculated. Here we use the method as explained in appendix~\ref{summation}. With $\alpha^2=\gamma^2/N^2$ and $\beta^2=-4(\gamma^2-1)/N^2$ and taking into account that $\alpha^2>0$ and $\beta^2<0$, we obtain
the following basic contributions:

\begin{align}
&\sum_{k=0}^{\infty}\frac{1}{m-\alpha}-\frac{1}{m+\alpha} &&=\frac{\pi}{2}\tan(\frac{\pi\gamma}{2N})\ ,\nonumber \\
&\sum_{k=0}^{\infty}\frac{1}{m-2\alpha}-\frac{1}{m+2\alpha} &&=\frac{\pi}{2}\tan(\frac{\pi\gamma}{N})\ ,\nonumber \\
&\sum_{k=0}^{\infty}\frac{1}{m-\beta}-\frac{1}{m+\beta} &&=\frac{i\pi}{2}\tanh(\frac{\pi\sqrt{\gamma^2-1}}{N})\ ,\nonumber \\
&\sum_{k=0}^{\infty}\frac{1}{m^2} &&=\frac{\pi^2}{8}\ , \nonumber \\
&\sum_{k=0}^{\infty}\frac{1}{(m-\alpha)^2}+\frac{1}{(m+\alpha)^2} &&=\frac{\pi^2}{4}\left(1+\tan^2(\frac{\pi\gamma}{2N})\right)\ ,\nonumber \\
&\sum_{k=0}^{\infty}\frac{1}{(m-\alpha)^3}-\frac{1}{(m+\alpha)^3} &&=\frac{\pi^3}{8}\tan(\frac{\pi\gamma}{2N})\left(1+\tan^2(\frac{\pi\gamma}{2N})\right)\ ,\nonumber\\
&\sum_{k=1}^{\infty}\frac{1}{m-\tfrac{1}{2}}-\frac{1}{m+\tfrac{1}{2}} &&=\frac{\pi}{2}-\frac{4}{3}\  ,\nonumber \\
&\sum_{k=1}^{\infty}\frac{1}{m-1}-\frac{1}{m+1} &&=\frac{1}{2}\ ,\nonumber \\
&\sum_{k=1}^{\infty}\frac{1}{(m-\tfrac{1}{2})^2}-\frac{1}{(m+\tfrac{1}{2})^2} &&=\frac{\pi^2}{2}-\frac{40}{9}\  ,\nonumber \\
&\sum_{k=1}^{\infty}\frac{1}{m^2}&&=\frac{\pi^2}{8}-1\  .\nonumber
\end{align}

\noindent where $i$ is the imaginairy unit and where $m=2k+1$ must be substituted.


\renewcommand{\theequation}{\thesection\arabic{equation}}
\setcounter{equation}{0}

\section[Some properties of canonical systems]{Some properties of canonical systems\protect\footnote{Most part of this  appendix have been copied from\cite{HV-62}}}

\subsection{The Hamiltonian}

Suppose an orbit $x(\theta)$ has to be calculated from the two differential equations:

\begin{eqnarray}
	p^\prime &=& \frac{dp}{d\theta}\ =\ f(p,x,\theta)\ ,  \label{canp} \\
	x^\prime &=& \frac{dx}{d\theta}\ =\ g(p,x,\theta)\ ,  \label{canx}
\end{eqnarray}

\noindent and that $f$ and $g$ obey the relation:

\begin{equation}
	\frac{\partial f}{\partial p} + \frac{\partial g}{\partial x} = 0\ . \nonumber
\end{equation}

\noindent We can define a function $H(p,x,\theta)$, by:

\begin{equation}
        f=-\frac{\partial H}{\partial x}\ ,\ \ \ g=\frac{\partial H}{\partial p}\ , \nonumber
\end{equation}

\noindent so that Eqs.~(\ref{canp}-\ref{canx}) become:

\begin{eqnarray}
	\frac{dp}{d\theta} &=& -\frac{\partial H}{\partial x}\ , \label{Hp}\\
	\frac{dx}{d\theta} &=& +\frac{\partial H}{\partial p}\ . \label{Hx}
\end{eqnarray}

\noindent The variables $p$ and $x$ are called canonical variables and the function $H$ is called the Hamiltonian. This canonical system has the property
that an area occupied by a group of points in the $p,x$-plane (called the phase space) remains constant during the motion. This is the Liouville theorem.
From Eqs.~(\ref{Hp}-\ref{Hx}) one finds that the total derivative of $H$ with respect to $\theta$ is equal to its partial derivative:

\begin{equation}
 \frac{dH}{d\theta}=\frac{\partial H}{\partial \theta}\ . \nonumber
\end{equation}
 
\noindent This means that $H$ is constant of motion if $H$ does not contain the independent variable $\theta$ explicitly. In this case one gets a very useful semi quantative picture 
of the motion in the $p,x$-plane , representing the real motion $x(\theta)$ by observing that the points move on the contours $H$ = constant. An extremum of $H$ gives a stable
stationary position $p$ = constant, $x$ = constant (a stable fixed point). A sadle point in the $H$ surface gives a metastable position (unstable fixed point).   

\subsection{Canonical transformations}

An important property of a canonical system is the possibility to make a transformation from the existing variables $p,x$ to new variables $P,X$:

\begin{eqnarray}
	P &=& P(p,x,\theta)\ , \nonumber \\
	X &=& X(p,x,\theta)\ , \nonumber
\end{eqnarray}

\noindent such that $P$ and $X$ can be derived from a new Hamiltonian $K$ similar to Eqs.~(\ref{Hp},\ref{Hx}):

\begin{eqnarray}
	\frac{dP}{d\theta} &=& -\frac{\partial K}{\partial X}\ , \label{Kp}\\
	\frac{dX}{d\theta} &=& +\frac{\partial K}{\partial P}\ . \label{Kx}
\end{eqnarray}

\noindent A necessary and sufficient condition is that the ratio $R$ of the area of a region in the $p,x$ plane to the area of the corresponding region in the $P,X$ plane 
is independent of $p,x$ and $\theta$.  This means that the determinant of the Jacobian matrix of the transformation must be constant:

\begin{equation}
	R=\begin{vmatrix} \partial P/\partial p & \partial P/\partial x \\ \partial X/\partial p & \partial X/\partial x \end{vmatrix} = \mbox{constant}\ . \label{jacob1}
\end{equation}

\noindent The new Hamiltonian $K$ is obtained from the orginal $H$ as follows:

\begin{equation}\label{jacob2}
	K=R*H+\Xi(P,X,\theta)\ ,
\end{equation}

\noindent where the function $\Xi$ is obtained from:

\begin{eqnarray}
	\frac{\partial\Xi}{\partial P} &=& +\frac{\partial X}{\partial\theta}\ , \nonumber \\
	\frac{\partial\Xi}{\partial X} &=& -\frac{\partial P}{\partial\theta}\ . \nonumber
\end{eqnarray}

Canonical transformations with $R=1$ can be obtained from so called generating functions\cite{Goldstein}. Denote the original variables as $(x,p)$, the new variables as $(X,P)$ and
the independent variable as $\theta$. 

\noindent The first type of generating function $G_1$ depends on the original coordinate $x$ and the new coordinate $X$. The transformation is defined by:

\begin{equation} p=\frac{\partial G_1}{\partial x}\ ,\ \ \ P=-\frac{\partial G_1}{\partial X}\ . \label{gen1}	\end{equation}

\noindent The second type $G_2$ depends on the original coordinate $x$ and the new momentum $P$. The transformation is defined by:

\begin{equation} p=\frac{\partial G_2}{\partial x}\ ,\ \ \ X=\frac{\partial G_2}{\partial P}\ .	\label{gen2} \end{equation}

\noindent The third type $G_3$ depends on the original momentum $p$ and the new coordinate $X$. The transformation is defined by:

\begin{equation} x=-\frac{\partial G_3}{\partial p}\ ,\ \ \ P=-\frac{\partial G_3}{\partial X}\ . \label{gen3}	\end{equation}

\noindent The fourth type $G_4$ depends on the original momentum $p$ and the new momentum $P$. The transformation is defined by:

\begin{equation} x=-\frac{\partial G_4}{\partial p}\ ,\ \ \ X=\frac{\partial G_4}{\partial P}\ . \label{gen4}	\end{equation}

\noindent In all four cases, the new Hamiltonian $K$ is obtained as:

\begin{equation} 
	K=H+\frac{\partial G}{\partial\theta}\ . \nonumber
\end{equation}

\subsection{Orbits in the neighborhood of a known solution}\label{Oneigh}

Let us assume that we have a particular solution $p_e(\theta),x_e(\theta)$ for a given Hamiltonian $H(p,x,\theta)$. 
In this case $p_e$ and $x_e$ obey the equations similar to Eq.~(\ref{Hp},\ref{Hx}):

\begin{eqnarray}
	\frac{dp_e}{d\theta} &=& -\frac{\partial H}{\partial x_e}\ , \label{Hp2}\\
	\frac{dx_e}{d\theta} &=& +\frac{\partial H}{\partial p_e}\ . \label{Hx2}
\end{eqnarray}

\noindent We want to study the motion in the neighborhood of $p_e,x_e$ and therefore introduce the new variables $P,X$ as:

\begin{eqnarray}
 P=p-p_e(\theta)\ , \nonumber\\
 X=x-x_e(\theta)\ . \nonumber
\end{eqnarray}

\noindent This transformation can be obtained from the type 2 generating function (see Eqs.~(\ref{gen2})):

\begin{eqnarray}
 G &=& G_2(P,x)=xP-x_eP+p_ex\ , \nonumber \\
 X &=& \frac{\partial G_2}{\partial P} = x-x_e\ , \nonumber \\
 p &=& \frac{\partial G_2}{\partial x} = P+p_e\ , \nonumber \\
 \frac{\partial G}{\partial\theta} &=& -\dot{x}_e P +\dot{p}_e (X+x_e)\ . \nonumber
\end{eqnarray}

\noindent We now expand the Hamiltonian $H$ around the solution $p_e,x_e$ as follows:

\begin{equation}
H=P\frac{\partial H}{\partial p_e}+X\frac{\partial H}{\partial x_e}+\frac{1}{2}P^2\frac{\partial^2H}{\partial p_e^2}+\dots\ . \nonumber
\end{equation}

\noindent Note that the zero-degree term in this expansion does not contribute to the form of the equations of motion and therefore can be omitted.
The new Hamiltion is obtained as $K=H+\partial G/\partial\theta$ giving:

\begin{equation}
K(P,X,\theta)=P\frac{\partial H}{\partial p_e}+X\frac{\partial H}{\partial x_e}+\frac{1}{2}P^2\frac{\partial^2H}{\partial p_e^2}+\dots-\dot{x}_e P +\dot{p}_e X\ . \nonumber
\end{equation}

\noindent But by virtue of Eqs.~(\ref{Hp2},\ref{Hx2}), the  first degree terms in the above expression for $K$ cancel each other. So, when studying the motion
in the neighborhood of a known solution we can in the expansion of the new Hamiltonian, ignore the first degree terms in $P,X$ and only take into account the quadratic degree 
terms and the higher degree terms:

\begin{equation}
K(P,X,\theta)=\frac{1}{2}P^2\frac{\partial^2H}{\partial p_e^2}+PX\frac{\partial^2H}{\partial p_e\partial x_e}+\frac{1}{2}X^2\frac{\partial^2H}{\partial x_e^2} 
	+\frac{1}{6}P^3\frac{\partial^3H}{\partial p_e^3}+\dots\ . \nonumber
\end{equation}

\noindent The quadratic terms correspond to the linear approximation of the motion with respect to the known solution. The higher degree terms 
must be included when studying nonlinear effects.

\subsection{The normal form of a quadratic Hamiltonian}\label{normform}

Consider a quadratic Hamiltonian of the following form:

\begin{equation}
	H(\pi,\xi,\theta) = \tfrac{1}{2}f\pi^2 +g\pi\xi+\tfrac{1}{2}h\xi^2\ . \nonumber
\end{equation}

\noindent where $f,g$ and $h$ are functions of $\theta$ only and where $f\neq 0$.
We want to reduce this Hamiltonian to its normal form defined as:

\begin{equation}
	K(P,X,\theta) = \tfrac{1}{2}P^2 +\tfrac{1}{2}Q(\theta)X^2\ . \label{knew}
\end{equation}

\noindent We first eliminate the coefficient $f$ in the term with $\pi^2$, using the following type 3 generating function:

\begin{eqnarray}
G&=&G_3(\pi,\bar\xi,\theta) = -\pi\bar{\xi} f^{\tfrac{1}{2}}\ , \nonumber \\
\bar{\pi} &=& \pi f^{\tfrac{1}{2}}\ ,\nonumber \\
\bar{\xi} &=& \xi f^{-\tfrac{1}{2}}\ , \nonumber \\
\partial G/\partial\theta&=&-\tfrac{1}{2}f^{-1}\dot{f}\bar{\pi}\bar{\xi}\ . \nonumber
\end{eqnarray}

\noindent This gives for the new Hamiltonian:

\begin{equation}
 \bar{H}(\bar{\pi},\bar{\xi},\theta)=\tfrac{1}{2}\bar{\pi}^2+(g-\tfrac{1}{2}f^{-1}\dot{f})\bar{\pi}\bar{\xi}+\tfrac{1}{2}fh\ \bar{\xi}^2 \ .  \nonumber
\end{equation}

\noindent With a second transformation (from $\bar{\pi},\bar{\xi}$ to $P,X$) we want to remove the term $\bar{\pi}\bar{\xi}$ in $\bar{H}$. When in the
final Hamiltonian $K$, such a cross-term is not present, we will have $\dot{X}=\partial K/\partial P=P$. From this we can deduce the transformation that will be needed 
by taking $X=\bar{\xi}$ giving

\begin{equation}
  X=\bar{\xi}\ , \hspace{2cm} P=\dot{X}=\dot{\bar{\xi}}\ , \nonumber
\end{equation}

\noindent and where we get $\dot{\bar{\xi}}$ from $\partial \bar{H}/\partial\bar{\pi}$. This leads us to the following transformation:

\begin{eqnarray}
  P&=&\bar{\pi} ~(g-\tfrac{1}{2}f^{-1}\dot{f})\bar{\xi}\ , \nonumber \\
  X&=&\bar{\xi}\ . \nonumber \\
 G&=&G_3(\bar{\pi},X,\theta) = -\bar{\pi}X-(g-\tfrac{1}{2}f^{-1}\dot{f})X^2/2\ . \nonumber
\end{eqnarray}

\noindent  With this the new Hamiltonian $K$ as given in Eq.~(\ref{knew}) is obtained and the function $Q$ is given by:

\begin{equation}\label{Qtheta}
  Q(\theta) = fh-(g-\tfrac{1}{2}f^{-1}\dot{f})^2-\frac{d}{d\theta}(g-\tfrac{1}{2}f^{-1}\dot{f})\ . \nonumber
\end{equation}

\noindent The full transformation (and its inverse) is given by:

\begin{eqnarray}
 P&=&f^{\tfrac{1}{2}}\pi+(g-\tfrac{1}{2}f^{-1}\dot{f})f^{-\tfrac{1}{2}}\xi\ , \hspace{2cm} X=f^{-\tfrac{1}{2}}\xi\ ,	\nonumber \\
 \pi&=&f^{-\tfrac{1}{2}}\left[P-(g-\tfrac{1}{2}f^{-1}\dot{f})X\right]\ , \hspace{1.8cm} \xi=f^{\tfrac{1}{2}} X\ . \nonumber
\end{eqnarray}


\end{document}